\documentclass[a4paper,11pt]{article}

\usepackage{jcappub} 
\usepackage[dvipsnames]{xcolor}
\usepackage[T1]{fontenc}

\usepackage[english]{babel}
\usepackage{xparse}
\usepackage{comment}
\usepackage{multirow}

\title{Sensitivity of the Cherenkov Telescope Array Observatory to Gamma-Ray Signals in Dwarf Irregular Galaxies}
\author[a, b]{Jaume Zuriaga-Puig,}
\author[c]{Viviana Gammaldi}
\author[a,b]{and Miguel Á. Sánchez-Conde}

\affiliation[a]{Instituto de Física Teórica, IFT UAM-CSIC,\\ Calle Nicolás Cabrera 13-15, Campus de Cantoblanco, E-28049 Madrid, Spain.}
\affiliation[b]{Departamento de Física Teórica, Mod. 15, Universidad Autónoma de Madrid,\\ E-28049 Madrid, Spain.}
\affiliation[c]{Department of Information Technology, Escuela Politécnica Superior, Universidad San Pablo-CEU, CEU Universities, Campus Montepríncipe, Boadilla del Monte, Madrid 28668, Spain}
\emailAdd{jaume.zuriaga@csic.es}
\emailAdd{viviana.gammaldi@ceu.es} 
\emailAdd{miguel.sanchezconde@uam.es} 

\abstract{
Dwarf irregular galaxies (dIrrs) are rotationally supported galaxies with a low star formation rate. Thus, their gamma-ray astrophysical emission is expected to be low, making them interesting targets for WIMP dark matter (DM) indirect searches. In this work, we build upon previous work on these objects in this DM context, and identify the best four dIrrs to be observed by the forthcoming Cherenkov Telescope Array Observatory (CTAO). Since dIrrs have not been detected in gamma rays yet, we first explore the prospects for detecting their astrophysical emission with the CTAO. Secondly, we compute the CTAO sensitivity prospects to a DM annihilating signal from these objects, accounting for the presence of DM substructures in them. We do so for both cuspy and cored DM density profiles, as the cusp-core debate remains particularly open for dIrrs. Our best combined limits show the potential to exclude DM annihilation cross-section values around $2\times 10^{-24} \ \mathrm{cm^{3}}\mathrm{s^{-1}}$ for 100 GeV WIMP masses annihilating in the $\tau^+\tau^-$ channel. These prospective results are competitive with and complementary to benchmark targets such as galaxy clusters. We also analyze the case of the velocity-dependent annihilation cross-section (Sommerfeld enhancement), obtaining projected DM constraints that exceed those expected for dwarf spheroidal galaxies, thanks to the contribution of subhalos to the signal. We conclude that dIrrs are compelling targets for the CTAO, not only for DM indirect searches but also as possible astrophysical sources.
}


\begin{document}

\maketitle

\flushbottom

\newpage

\section{Introduction}

One of the foremost candidates for cold dark matter (DM) is the Weakly Interacting Massive Particle (WIMP), which is theorized to be created thermally during the early stages of the Universe \cite{Steigman:2012nb, 2018RvMP...90d5002B}. Being WIMPs their own antiparticle, they can decay or annihilate into Standard Model (SM) particles. Assuming that all DM is composed of self-annihilating WIMPs ($\sim27\%$ of the total energy density budget of the Universe \cite{Aghanim:2018eyx}), the required thermally-averaged self-annihilation cross-section is about $\langle \sigma v \rangle_{\mathrm{th}} \simeq 3 \times 10^{-26} \mathrm{cm}^{3} \mathrm{s}^{-1}$ \cite{Steigman:2012nb, 2018RvMP...90d5002B}. Given the typical mass range of WIMPs ($\mathcal{O}(0.1-100)$ TeV), the primary SM particles produced by these annihilations can be very energetic and, therefore, are expected to produce secondary particles (via decay, hadronization ...) that can reach the Earth. The main cosmic messengers for this emission are cosmic rays (CRs), neutrinos and gamma rays, with the latter being the focus of this work. Efforts aimed at detecting these final particles are referred to as DM indirect detection \cite{2016ConPh..57..496G, Bertone:2004pz, 2018RPPh...81f6201R}.

A considerable amount of work -both theoretical and observational- has been performed on the indirect detection of DM \cite{Fermi-LAT:2016afa, McDaniel:2023bju, 2022PDU....3500912A, 2020PhRvD.102f2001A, 2025MNRAS.tmp.1704A, Di_Mauro_2021, Acharyya_2021, 2012PhRvD..86j3506C, PhysRevLett.129.111101, 2023JCAP...11..063Z, 2023PhRvD.107h3030D, 2012ApJ...750..123A, CTAConsortium:2023yak, 2023MNRAS.520.1348G, 2025arXiv250109789E, 2025ApJ...978L..43C, 2025arXiv250215656F}. Despite this effort, no signal has been detected so far, yet competitive constraints have been set on the thermally-averaged annihilation cross-section $\langle \sigma v \rangle$ vs. the DM mass parameter space. Very diverse astrophysical targets have been observed and used to derive constraints on the DM particle nature, e.g., dwarf spheroidal galaxies (dSphs) \cite{Fermi-LAT:2016afa, McDaniel:2023bju, 2022PDU....3500912A, 2020PhRvD.102f2001A, 2025MNRAS.tmp.1704A}, the Galactic center (GC) \cite{Di_Mauro_2021, Acharyya_2021, 2012PhRvD..86j3506C, PhysRevLett.129.111101, 2023JCAP...11..063Z}, or galaxy clusters \cite{2023PhRvD.107h3030D, 2012ApJ...750..123A, CTAConsortium:2023yak} among other, more exotic objects such as dark subhalos \cite{2023MNRAS.520.1348G, 2025arXiv250109789E}, ultra faint compact stellar systems \cite{2025ApJ...978L..43C} or stellar streams \cite{2025arXiv250215656F}. DSphs have the advantages that they are DM dominated ($10^{5} - 10^{7} \ \mathrm{M}_\odot$), nearby objects ($< 0.5$ Mpc), with a negligible expected astrophysical background, qualities that have made dSphs one of the most constraining targets \cite{McDaniel:2023bju, 2004PhRvD..69l3501E}. Besides, the kinematics of these objects is dominated by pressure and their DM density profile estimation relies on a Jeans analysis \cite{Geringer-Sameth:2014yza, 2004PhRvD..69l3501E}. Another potentially very constraining target is the GC. Its main challenge is the modeling of the so-called Milky Way Galactic diffuse emission (GDE), and the sources present in the region, such as supernova remnants, pulsar wind nebulae, the supermassive black hole Sgr A*, etc. As for galaxy clusters, they are the most massive objects in the Universe ($10^{14} - 10^{15} \ \mathrm{M}_\odot$) located at very large distances ($> 10$ Mpc). They are privileged targets to test decaying DM models, and the expected boost to the annihilation signal from the population of substructures makes them also competitive targets for DM annihilation studies \cite{CTAConsortium:2023yak}. Clusters are also expected to host CR-induced gamma ray emission from $\pi^0$ decays\footnote{This emission has not yet been robustly detected.} and, being also distant objects, gamma-ray propagation effects need to be taken into account \cite{2021JCAP...02..048A}.

Beyond these more standard, benchmark targets, dwarf irregular galaxies (dIrrs) have been proposed as new targets of interest for DM indirect searches \cite{Gammaldi:2017mio}. DIrrs, like the dSphs, are located within the Local Volume, with distances to Earth as small as $0.5-1$ Mpc \cite{Gammaldi:2017mio, 2021PhRvD.104h3026G, HAWC:2023vtl, HESS:2021zzm}, and are DM-dominated objects ($10^{9} - 10^{10} \ \mathrm{M}_\odot$). They are kinematically dominated by rotation, which makes the derivation of their DM density profiles possible thanks to the study of their rotation curves \cite{Gammaldi:2017mio, 2021PhRvD.104h3026G}. DIrrs are star-forming galaxies, thus, in principle, their astrophysical gamma-ray emission could also be important. Yet,  it has been shown that the expected gamma-ray emission in the GeV range is negligible compared to the expected DM annihilation emission, due to their low star formation rate and the small angular size of their star-forming region (SFR)~\cite{Gammaldi:2017mio}. In this work, we revisit this issue and perform an in-depth study on the detectability of their SFR and DM emissions in the TeV energy scale. In particular, we do so for the Cherenkov Telescope Array Observatory (CTAO). The CTAO is the next-generation ground-based imaging atmospheric gamma-ray telescope, with about one order of magnitude better sensitivity than current instruments, an improved angular resolution that can reach up to $\sim 0.02^\circ$ at 100 TeV, and a projected operational range between tens of GeV up to hundreds of TeV \cite{CTAConsortium:2017dvg}. In the Alpha Configuration, the CTAO will be located in two sites, one in the Northern Hemisphere (consisting of 4 Large-Sized Telescopes and 9 Medium-Sized Telescopes in La Palma, Spain) and another in the Southern Hemisphere (14 Medium-Sized Telescopes and 37 Small-Sized Telescopes in the Atacama Desert, Chile), allowing the CTAO to access any point in the sky. Using the latest Instrument Response Function (IRFs) of the CTAO, we discuss which telescope site best suits to our purposes and targets, and also propose the best observational strategy.

Using state-of-the-art models, we divide our analysis into two separate parts, one focused on the astrophysical emission (AE) and another on the DM signal. For the former, we investigate the detectability of both GDE and SFR emission associated with each dIrr. Several works have been published regarding the possible detection of SFR emission in the MeV-GeV energy range (see \cite{Gammaldi:2017mio, 2021PhRvD.104h3026G, 2025A&A...699A..43K, 2012ApJ...755..164A, 2020ApJ...894...88A} for the case of \textit{Fermi}-LAT) and the TeV range (see \cite{HAWC:2023vtl} (HAWC), \cite{2021MNRAS.506.6212S} (CTAO) and \cite{2025A&A...699A..43K} for a more general case). In general, among all the star-forming galaxies considered in the literature, only 13 have been detected by \textit{Fermi}-LAT \cite{2025A&A...699A..43K, 2012ApJ...755..164A, 2020ApJ...894...88A} and 2 in the TeV energy range (M82 by VERITAS \cite{2025ApJ...981..189A} and NGC253 by HESS \cite{2018A&A...617A..73H}). Considering isolated dIrrs in the Local Group (i.e., not satellite galaxies), only flux upper limits have been set for NGC6822 and IC10 with \textit{Fermi}-LAT data \cite{2025A&A...699A..43K, 2021PhRvD.104h3026G}, leaving the analysis of dIrrs a promising prospect for the literature. In our case, we focus on CTAO sensitivity prospects. For the DM annihilation signal, we focus on an extended analysis including all possible gamma-ray emission components (DM, SFR and GDE). We also compare our results after mismodeling the different emissions present, with a point-like (PL) DM-only analysis, an extended analysis with DM-only templates and an explicit mismodeling of the intrinsic SFR emission. We also adopt three different models of the DM distribution in each of the targets, so as to account for current uncertainties, including core/cusp profiles and, also, the inclusion of substructures in the main DM halo. Regarding the DM decay scenario, these galaxies could yield competitive constraints with respect to dSphs thanks to their high DM halo masses \cite{Gammaldi:2017mio}, but we leave this final scenario for a future project. As a final part of the analysis, exploring outside the vanilla WIMP scenario, we include in our analysis the Sommerfeld enhancement case of the DM signal. This enhancement is characterized by a long interaction of a light scalar mediator, in which the DM annihilation signal can be increased by up to 10 orders of magnitude, increasing the detectability prospects of the DM signal significantly.

This work is divided as follows. In Section \ref{sec:candidates}, we perform the selection of the best dIrrs for this analysis, and discuss the optimal CTAO site to observe each of them. In Section \ref{sec:spectral_modeling}, we compute the expected gamma-ray astrophysical and DM emissions and build (spatial and spectral) templates for both components. In Section \ref{sec:setup}, we define and discuss the observation and simulation setup. Section \ref{sec:results_general} is devoted to the main results of our AE sensitivity study and discusses the possibility of either detecting dIrrs as astrophysical emitters with the CTAO, or setting constraints on the total integrated SFR flux needed for a detection. In the same section, the sensitivity of the CTAO to a DM signal in the standard WIMP paradigm is investigated for individual targets and a combined analysis, while in Section \ref{sec:sommerfeld} we show a special case including the Sommerfeld enhancement effect in the DM signal. Finally, in Section \ref{sec:conclusions}, we summarize our work and discuss our conclusions.

\section{Selection of candidates and CTAO Sites}
\label{sec:candidates}

\begin{figure}[t!] 
  \centering 
  \includegraphics[width=0.99\textwidth]{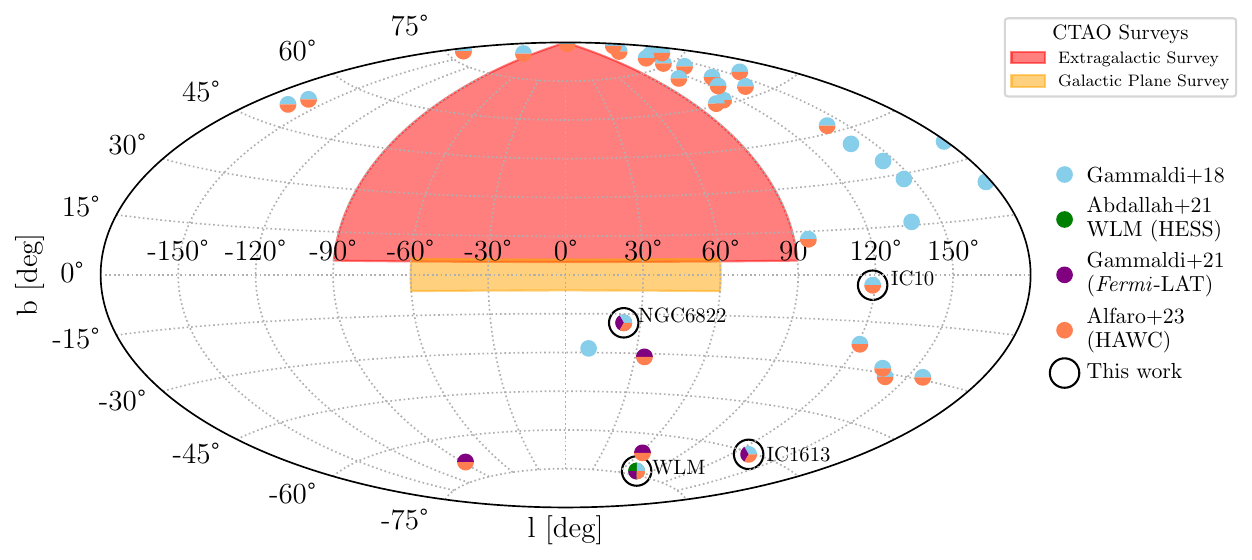}
    \caption{\footnotesize{Location, in Galactic coordinates, of the four selected targets (black circles) and of the rest of the targets analyzed in the literature, i.e., Gammaldi et al. 2018 \cite{Gammaldi:2017mio} (blue), Abdallah et al. 2021 (HESS) \cite{HESS:2021zzm} (green),  Gammaldi et al. 2021 (\textit{Fermi}-LAT) \cite{2021PhRvD.104h3026G} (purple) and  Alfaro et al. 2023 (HAWC) \cite{HAWC:2023vtl} (orange). Red and yellow sky areas refer to the CTAO Extragalactic \cite{CTAConsortium:2017dvg} and GP \cite{CTAConsortium:2023tdz} surveys, respectively. Map in Aitoff projection.}}
\label{fig:dIrr_skymap} 
\end{figure}

In order to prepare the best observation strategy for the CTAO, we first need to select our targets. Several works have already investigated dIrrs in the context of DM searches~\cite{Gammaldi:2017mio, 2021PhRvD.104h3026G, HAWC:2023vtl, HESS:2021zzm}. We closely follow these works as a starting point to select our best targets, and adopt the following criteria: 1) targets with the greatest expected DM flux, which is directly related to their underlying DM density profiles and distances; and 2) targets with a good sky location for the best CTAO site. In addition, the CTAO Consortium is planning extended sky surveys (Extragalactic \cite{CTAConsortium:2017dvg} and Galactic Plane (GP) \cite{CTAConsortium:2023tdz}), so we checked if any of the dIrrs lie within the footprint of either survey. However, this is not the case. Figure~\ref{fig:dIrr_skymap} shows the locations of the dIrrs that have been considered for the selection analysis, together with the footprint of planned CTAO surveys.

In this work, we restrict ourselves to only four targets out of an initial sample of 40 dIrrs: IC10 and IC1613 for the Northern Site, and WLM, NGC6822 and IC1613 for the Southern one. Note that the latter dIrr is visible from the two sites with a sufficiently small zenith angle. Thus, in the following, we work with IC10, WLM, NGC6822, IC1613-North (observed with CTAO-N) and IC1613-South (for CTAO-S). Table~\ref{tab:dIrr_sky_targets} lists the location, distance, zenith angle and best CTAO site for observation, for each of the selected targets.

\begin{table}[t!]
    \begin{center}
        \begin{tabular}{|c|c|c|c|c|c|}
        \hline
        \hline
\textbf{Galaxy}                      & l [deg]        & b [deg]       & $\mathrm{D}_{\odot}$ [Mpc]   & zenith angle [deg] & CTAO Site \\ 
\hline
\hline
\textbf{IC10} & $119.0$ & $-3.3$ & $0.79$ & $30.6$ & North\\
\hline
\textbf{IC1613} & $129.7$ & $-60.6$ & $0.76$ & $26.7 (26.8)$ & North (South) \\
\hline
\textbf{WLM} & $75.9$ & $-73.6$ & $0.97$ & $9.2$ & South \\
\hline
\textbf{NGC6822} & $23.3$ & $-18.4$ & $0.48$ & $8.1$ & South \\
\hline
        \end{tabular}
        \caption{\footnotesize{Parameters of the dIrrs in our sample. Columns refer to location (in Galactic coordinates), distance to Earth $\mathrm{D}_{\odot}$, culminating zenith angle, and best CTAO site for observation. The zenith angle is computed for the best site. In the case of IC1613, both north and south sites are equally suited. Coordinates and distance values are extracted from~\cite{2021PhRvD.104h3026G}.}}
        \label{tab:dIrr_sky_targets}
    \end{center}
\end{table}

\section{Modeling of expected gamma-ray emissions}
\label{sec:spectral_modeling}

In this section, we discuss all the relevant sources of gamma rays that can influence the results, both of astrophysical and DM origin. 

\subsection{Astrophysical emission}

\subsubsection{Intrinsic emission: star-forming region}
\label{sec:SFR_IE}

SFRs are mainly detected through infrared or optical emission, although they are expected to also have an associated, intrinsic and extended gamma-ray emission, originating from the interactions of energetic CRs with the interstellar medium. These energetic CRs are associated with massive stars, mainly via supernova remnants and, with a secondary contribution, pulsar wind nebulae and colliding-wind binaries. The main interactions producing gamma rays are neutral pion decays, inverse Compton (IC) and Bremsstrahlung \cite{Martin:2014nia} (hereafter Martin14).

Our spatial modeling of the expected gamma-ray emission is based on the two-dimensional size of the SFR in the optical band, described as a flat disk with a given inclination, semi-major axis and eccentricity \cite{2012AJ....144....4M, Oh:2015xoa} (for more information on the parameters, see Tables~\ref{tab:dIrr_SFR_1} and \ref{tab:dIrr_SFR_2} in Appendix \ref{ap:SFR_Appendix}). For simplicity, we assume that the SFR-induced gamma-ray emission follows the star formation distribution, with typical sizes of $\theta_\mathrm{opt} \sim 0.1^\circ$. Since the typical angular dimension of the SFRs is only a factor $\sim2-5$ larger than the point spread function (PSF) of the CTAO ($\sim0.05^\circ-0.02^\circ$, corresponding to energies from $\sim 1$ TeV to 100 TeV), we model this emission as homogeneous within this disk.

Similarly to \cite{HAWC:2023vtl}, we model the spectra of this emission by estimating the gamma-ray luminosity, $L_\gamma$, following Martin14 \cite{Martin:2014nia}. To infer $L_\gamma$ for each object, the main parameter we need to determine is the star formation rate ($\mathcal{SFR}$), which is derived from the stellar mass of the galaxy. In this step, we follow Ref.~\cite{2017ApJ...851...22M}. Finally, the spectral shape of the emission is modeled with a power law with photon index $\gamma = 2.5$ in the CTAO energy range, where its normalization is fixed from the normalization of the integral of the power law to the gamma-ray energy flux obtained from $L_\gamma$. For more details, see Appendix \ref{ap:SFR_Appendix}.

\begin{figure}[t!] 
  \centering 
  \includegraphics[width=0.49\textwidth]{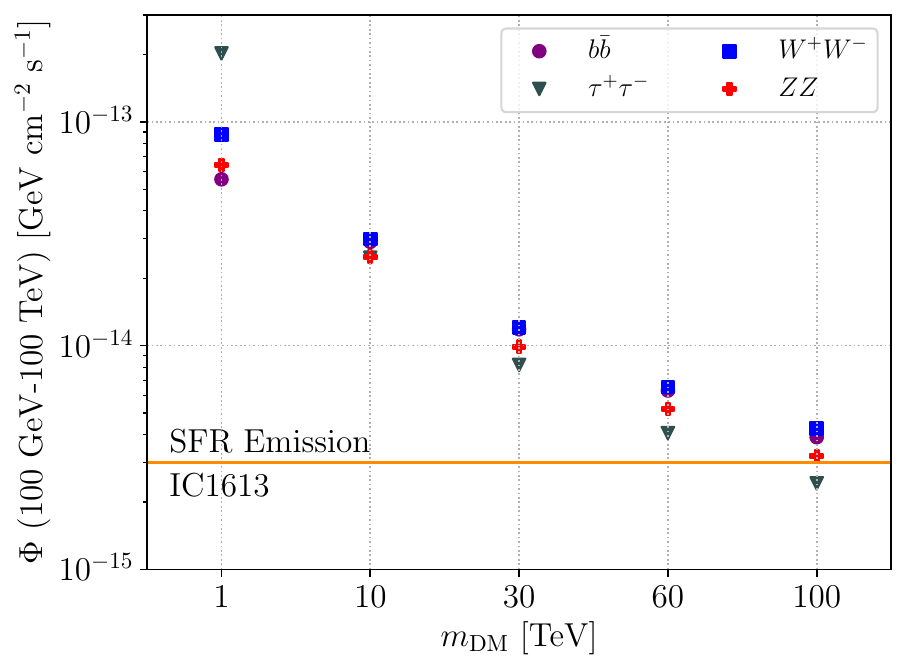}
  \includegraphics[width=0.49\textwidth]{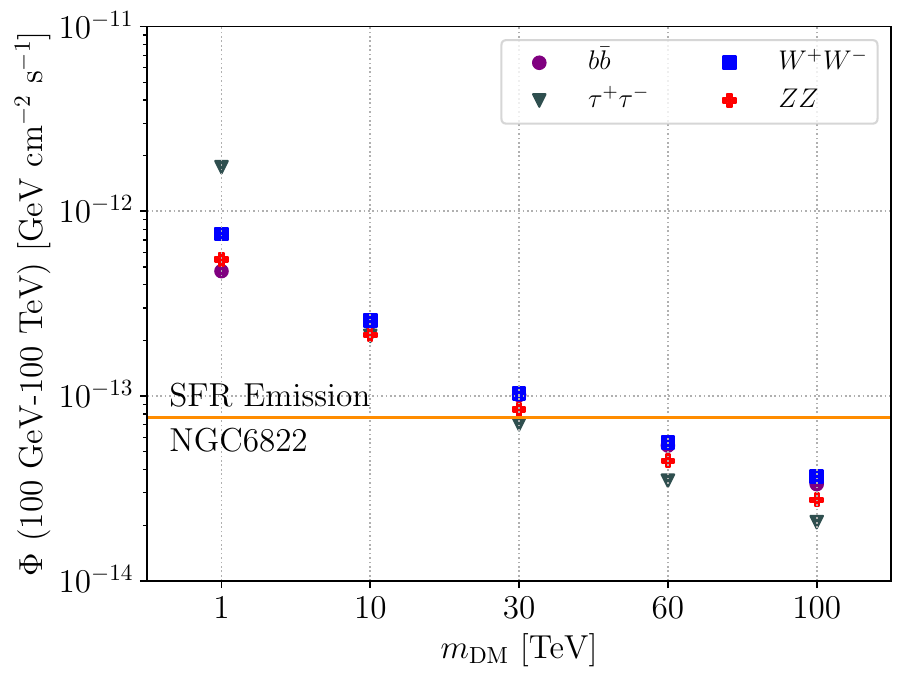}
    \caption{\footnotesize{Expected integrated (from 100 GeV to 100 TeV) gamma-ray SFR emission (left panel IC1613 and right panel NGC6822), computed with the Martin14 model \cite{Martin:2014nia}, compared with the DM integrated energy flux as a function of the DM mass $m_{\mathrm{DM}}$ for different annihilation channels (see Section \ref{sec:WIMP_annih}). For these galaxies, the integrated SFR gamma-ray emission is expected to be of the same order as the DM one for DM masses $m_{\mathrm{DM}}> 60$ and $30 \,\text{TeV}$, respectively. Instead, the integrated SFR gamma-ray emission is negligible for DM masses below those values. The Burkert-MED DM case has been adopted here (see next Section \ref{sec:DM_spatial_modeling}), with the corresponding fluxes integrated up to the virial radius $R_{200}$ (Table~\ref{tab:dIrr_Burkert_params}). See Appendix \ref{ap:rest_of_figs}, Figure~\ref{fig:dIrr_DMFlux_SFR_appendix}, for the Figures corresponding to WLM and IC10.}}
\label{fig:dIrr_DMFlux_SFR} 
\end{figure}

In Figure~\ref{fig:dIrr_DMFlux_SFR}, we show the expected benchmark gamma-ray energy flux from the SFR emission for IC1613 (left panel), a galaxy with an expected low emission, and NGC6822 (right panel), one of the brightest galaxies, integrated over the 100 GeV -- 100 TeV energy range. We compare it with the expected DM annihilation flux via the $\tau^+\tau^-$, $b \bar b$, $W^+W^-$ and $ZZ$ channels, assuming the range of DM masses $m_{\mathrm{DM}}$ considered in this work (see Section~\ref{sec:results_DM}). For these galaxies, the integrated SFR gamma-ray emission is expected to be of the same order as the DM one for DM masses $m_{\mathrm{DM}}> 60$ and 30 TeV, respectively. Instead, the integrated SFR gamma-ray emission begins to be negligible for DM masses below $m_{\mathrm{DM}} \sim 30 \ \mathrm{TeV}$. We have checked that similar cases occur for the rest of the dIrrs considered. Yet, since the normalization of the SFR emission flux is affected by a large uncertainty of 2 orders of magnitude (see Figure~\ref{fig:dIrr_All_spectral_fluxes} and Appendix \ref{ap:SFR_Appendix}), we decide to perform an extended analysis to determine its possible detection with the CTAO in Section \ref{sec:results_SFR}. These uncertainties are driven by the modeling of the $\mathcal{SFR}$ of the galaxy from its stellar mass and the estimation of the total SFR luminosity from this value (Appendix \ref{ap:SFR_Appendix}). Note that, considering isolated dIrrs in the Local Group (i.e., not satellite galaxies), only flux upper limits have been set for NGC6822 and IC10 with \textit{Fermi}-LAT data \cite{2025A&A...699A..43K, 2021PhRvD.104h3026G}. Besides, in the TeV range only two star-forming galaxies have been detected \cite{2025ApJ...981..189A, 2018A&A...617A..73H}. Therefore, detecting the SFR emission would mean detecting isolated dIrrs, for the first time, as astrophysical sources in gamma rays.

\subsubsection{Galactic diffuse emission}
\label{sec:GDE}

\begin{figure}[t!] 
  \centering 
  \includegraphics[width=0.49\textwidth]{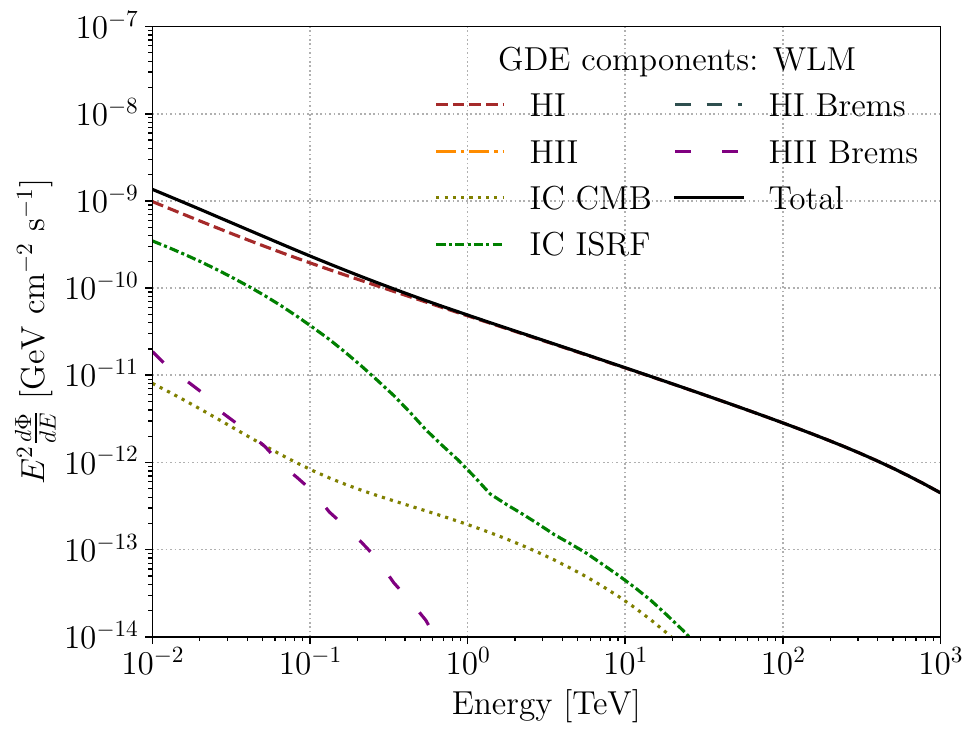}
  \includegraphics[width=0.49\textwidth]{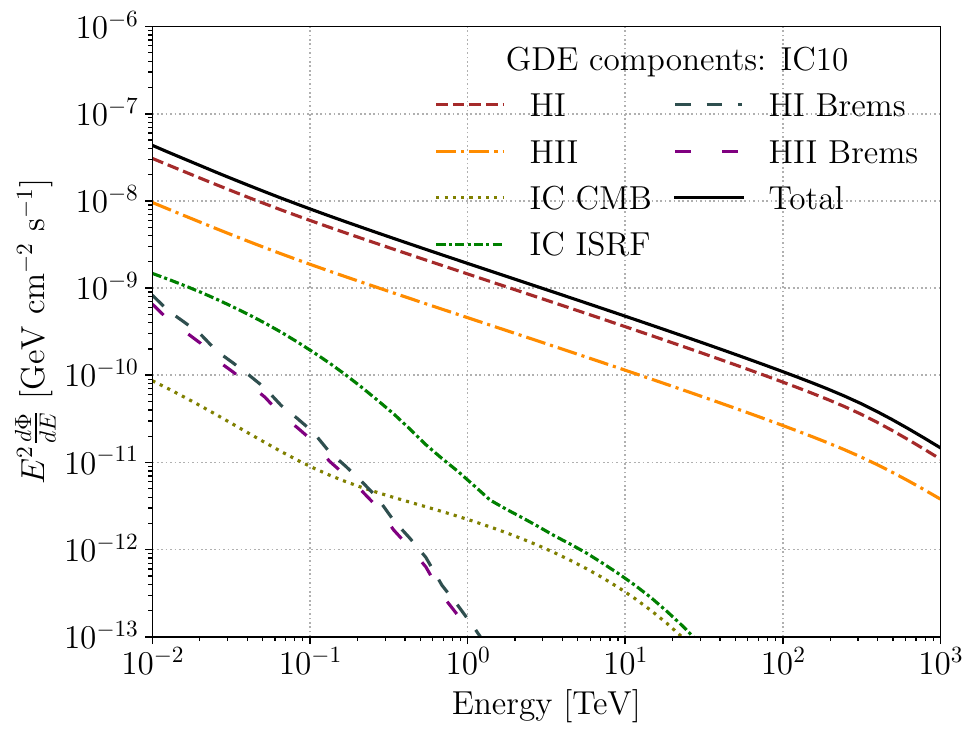}
    \caption{\footnotesize{Spectral energy distribution of the different components used to compute the GDE emission for the direction of the galaxies WLM (left panel) and IC10 (right panel), based on the $``\gamma$-optimized Max'' model~\cite{Luque:2022buq}. The different contributions to the flux from the HI and HII molecular clouds, as well as the associated IC and Bremsstrahlung (Brems), are shown.See Appendix \ref{ap:rest_of_figs}, Figure~\ref{fig:GDE_components_appendix}, for the Figures corresponding to IC1613 and NGC6822.}}
\label{fig:GDE_components} 
\end{figure}

Another potentially important contribution to the gamma-ray emission in the regions of interest is the GDE. The GDE is a diffuse emission created by the interactions between CRs and the interstellar medium, similar to the interactions described for the SFR, but originated within the Galaxy. We model the GDE using the ``$\gamma-$optimized Max'' model in Ref.~\cite{Luque:2022buq}, computed with the \texttt{DRAGON2} code \cite{Evoli_2017,Evoli:2017vim}. This model features a radial dependence (taking the origin at the GC) on the diffusion coefficient which well describes both the \textit{Fermi}-LAT data observed in the GC \cite{Gaggero:2014xla} and the Ridge emission observed by HESS \cite{Gaggero:2017jts}\footnote{More precisely, two models are presented in Ref.~\cite{Luque:2022buq}. We choose the `$\gamma-$optimized Max'' model because it gives a slightly larger flux from the Galaxy in this energy range.}. In Figure~\ref{fig:GDE_components}, we show the spectral flux of the different components of the GDE, computed for the simulated observation of the WLM galaxy (right panel), located far from the GP, and IC10 galaxy (left panel), located closer to the GP, where the emission should be significant. In general, the contribution to the flux coming from the HI molecular cloud is the most dominant one, while the IC and Bremsstrahlung emissions are of second order. Additionally, we observe that the HII contribution is also relevant only in galaxies close to the GP (IC10 and NGC6822). In Figure~\ref{fig:dIrr_All_spectral_fluxes}, we compare the total GDE spectral emission, which dominates the rest of the fluxes modeled, with the SFR and DM fluxes considered for the WLM (left panel) and IC10 (right panel) dIrr galaxies. Note that the GDE flux corresponds to the Galactic emission present in the region when observing the target.

\subsection{Gamma-ray flux from WIMP annihilation}
\label{sec:WIMP_annih}

As stated in the Introduction, dIrrs are DM-dominated galaxies, with extended DM halos. The differential gamma-ray flux expected by the annihilation of Majorana DM particles in the halo is given by \cite{Bertone:2004pz}:

\begin{equation}
  \frac{d \Phi_{\text{DM}}}{d E}= \frac{\langle\sigma v\rangle}{8 \pi m_{\mathrm{DM}}^{2}} \sum_{i}^{\mathrm{channels}} \mathrm{BR}_{i}  \frac{d N_{i}}{d E} J(\Delta \Omega),
  \label{eq:flux_annih} 
\end{equation}

\noindent where $\langle\sigma v\rangle$ is the thermally averaged annihilation cross-section, and $\mathrm{BR}_i$ are the branching ratios of the WIMP annihilation process, with $\sum_{i}^{\mathrm{channels}} \mathrm{BR}_i = 1$. $\frac{d N_{i}}{d E}$ is the differential number of gamma-rays created by the DM annihilation at the target, which in this case is computed using \texttt{PPPC4DMID} \cite{Cirelli_2011, Ciafaloni2010}. The DM annihilation happens producing pairs of particle - antiparticle of the SM, such as fermions, quarks and bosons. Once produced, these highly energetic particles can propagate and subsequently create a set of secondary particles that can reach detectors in the form of gamma rays, CR and neutrinos. In our case, we focus on the gamma rays produced by a model-independent approach (i.e., $\mathrm{BR}=1$ for each channel $i$). We study the $\tau^+\tau^-$, $b\bar{b}$, $W^+W^-$ and $ZZ$ annihilation channels independently, assuming that the final spectrum is very similar to one of (or in between) those.

\begin{figure}[t!] 
  \centering 
  \includegraphics[width=0.6\textwidth]{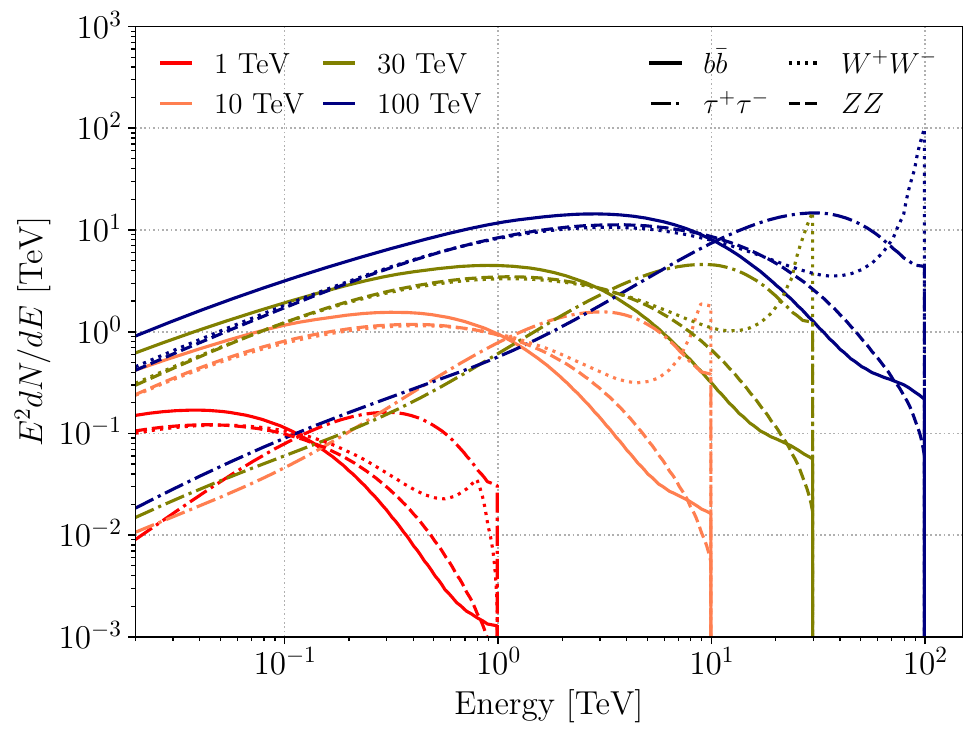}
    \caption{\footnotesize{Differential gamma-rays counts $\frac{d N_{i}}{d E}$ produced in the source, for different DM masses $m_{\mathrm{DM}}$ ($1$ (red), $10$ (orange), $30$ (green) and $100$ (blue) TeV) and channels ($b\bar{b}$ (solid line), $\tau^+\tau^-$ (dot dashed line), $W^+W^-$ (dotted line) and $ZZ$ (dashed line)). These fluxes have been computed using \texttt{PPPC4DMID} \cite{Cirelli_2011, Ciafaloni2010}.}}
\label{fig:DM_prompt_flux} 
\end{figure}

In Figure~\ref{fig:DM_prompt_flux}, we show how the expected gamma-ray flux $\frac{dN}{dE}$ depends on the DM mass $m_{\mathrm{DM}}$ and the annihilation channel, and in Figure~\ref{fig:dIrr_All_spectral_fluxes} we show, compared with the rest of the modeled fluxes, the particular case of WLM (left panel) and IC10 (right panel) with $m_\mathrm{DM} = 10$ TeV. Furthermore, depending on the specific DM particle model, the annihilation might result in a combination of several channels with different branching ratios, keeping the summation over the channels equal to 1 in Equation~\ref{eq:flux_annih}. Here, no gamma-ray propagation effects (e.g., attenuation by the extragalactic background light) are considered. In fact, for the nearby objects ($D_\odot < 1$ Mpc) considered in this work, these effects are not significant (see \cite{2021MNRAS.507.5144S} and references therein). Also, secondary gamma rays can be produced via IC, Bremsstrahlung and synchrotron interactions. However, these effects show a secondary, smaller imprint on the gamma-ray flux   \cite{Djuvsland2022, Buch2015}, so for simplicity we also neglect them.

In the same Equation~\ref{eq:flux_annih}, the J-factor $J(\Delta \Omega)$ is the astrophysical parameter for WIMP annihilation, computed from the DM density profile in the target of interest. It is defined as the integral of the DM density profile $\rho_{\mathrm{DM}}$ squared over the line of sight $l(\hat\theta)$ in the direction of the target $\hat\theta$:

\begin{equation}
    J(\Delta \Omega)= \int_{\Delta \Omega(\alpha_{\mathrm{int}})} \mathrm{d} \Omega \int_{l(\hat{\theta})_{\min }}^{l(\hat{\theta})_{\max }} \rho^{2}_{\mathrm{DM}}[r(l)] \mathrm{d} l(\hat{\theta})
    \label{eq:JFac_formula}
\end{equation}

Note that, since our targets are extended, both the integration angle $\alpha_\mathrm{int}$ and the solid angle $\Delta \Omega(\alpha_{\mathrm{int}}) = 2 \pi (1 - \cos(\alpha_{int}))$ change with the target.

\subsubsection{Spatial DM modeling and DM density profiles}
\label{sec:DM_spatial_modeling}

We have treated the DM spatial modeling of each dIrr within three possible scenarios, defined by its two halo components: the functional form of the DM density profile in the halo and the description of the subhalo population, following the same modeling adopted by  \cite{2021PhRvD.104h3026G}. Several works discuss whether the DM density profile in the halo of such objects is cuspy or cored \cite{Burkert:1995yz, Karukes:2017kne}. This debate arises  because numerical simulations seem to prefer cuspy profiles, such as the Navarro-Frenk-White (NFW) \cite{Navarro:1995iw}:

\begin{equation}
    \rho_{\mathrm{NFW}} (r) = \frac{\rho_0}{ (\frac{r}{r_{\mathrm{s}}}) (1 + \frac{r}{r_{\mathrm{s}}})^2},
    \label{eq:NFW_profile}
\end{equation}

\noindent where $r_{\mathrm{s}}$ is the scale radius and $\rho_0$ is the characteristic density, the normalization factor of the profile. On the other hand, rotation-curve measurements seem to prefer cored profiles for dIrrs and objects of similar mass, like the Burkert profile \cite{Burkert:1995yz}:

\begin{equation}
    \rho_{\mathrm{Burkert}} (r) = \frac{\rho_0}{ (1 + \frac{r}{r_{\mathrm{s}}}) (1 + (\frac{r}{r_{\mathrm{s}}})^2)}
    \label{eq:Burkert_profile}
\end{equation}

Because of this, although the rotation curve fits prefer the Burkert profile over the NFW \cite{2021PhRvD.104h3026G, Karukes:2017kne}, we keep an agnostic point of view and consider both profiles in our work, with the same scale radius $r_{\mathrm{s}}$ and central density $\rho_0$ parameters adopted in \cite{2021PhRvD.104h3026G}. In the case of the NFW profile, one of the key parameters to estimate $\rho_0$ and $r_s$ is the concentration $c$, which in Ref.~\cite{2021PhRvD.104h3026G} is computed following the concentration-mass $c-M$ relation in \cite{2014MNRAS.442.2271S}. From the definition $c = R_{200} / r_{\mathrm{s}}$, $r_s$ is estimated, where $R_{200}$ is the virial radius of the galaxy, defined where the DM halo has an average density 200 times the critical density of the Universe ($\rho_\mathrm{crit} = 137$ $\mathrm{M}_{\odot} \mathrm{kpc}^{-3}$). To give an idea of the sizes of the dIrrs in our sample, we can also define the angle subtended by the DM halos and their cusps as $\theta_{200} = \arctan{(R_{200}/D_\odot)}$ and $\theta_\mathrm{s} = \arctan{(r_\mathrm{s}/D_\odot)}$, respectively, with $D_\odot$ being the distance of the dIrrs to Earth. In Tables~\ref{tab:dIrr_NFW_params} and \ref{tab:dIrr_Burkert_params}, we summarize the different parameters of the Burkert and NFW DM density profiles for the considered targets\footnote{It can be noted that the $R_{200}$ values in Tables~\ref{tab:dIrr_NFW_params} and \ref{tab:dIrr_Burkert_params} are different. The values shown for the Burkert profile are directly obtained from the fit to the rotation curves in each galaxy performed in \cite{2021PhRvD.104h3026G}, whereas the NFW $R_{200}$ values are directly computed with the overdensity equation as explained in the text. Both values are consistent, as the NFW values are within the error bars.}.

\begin{table}[b!]
    \begin{center}
        \begin{tabular}{|c|c|c|c|c|c|c|c|}
        \hline
        \hline
                      &  \multicolumn{7}{c|}{\textbf{NFW}} \\ 
\hline
\hline
\textbf{Galaxy}    &  $\rho_0$ [$\mathrm{M}_{\odot}$/$\mathrm{kpc}^3$] &  $r_\mathrm{s}$ [kpc]  &  $\theta_\mathrm{s}$ [deg]  &   $R_{200}$ [kpc] &   $\theta_{200}$ [deg]  & $M_{200}$ [$\mathrm{M}_{\odot}$] & $c$ \\ 
\hline
\textbf{IC10}      & $6.31\times 10^6$                                          & $6.8$                   & $0.49$                     & $70.3$            &  $5.09$          &    $3.78\times 10^{10}$   & 10.3 \\
\hline
\textbf{IC1613}    & $7.94\times 10^6$                                          & $4.0$                   & $0.30$                     & $45.7$            &  $3.44$           &    $1.02\times 10^{10}$  & 11.4 \\
\hline
\textbf{WLM}       & $1.00\times 10^7$                                          & $2.8$                   & $0.17$                     & $33.6$            &  $1.98$            &   $4.53\times 10^{9}$  & 12.0 \\
\hline
\textbf{NGC6822}   & $7.94\times 10^6$                                          & $5.9$                   & $0.70$                     & $62.6$            &  $7.43$             & $3.15\times 10^{10}$    & 10.6 \\
\hline
\hline

        \end{tabular}
        \caption{\footnotesize{Parameters of the NFW DM density profile (Equation~\ref{eq:NFW_profile}) used to define the main DM halo, virial radius ($R_{200}$), subtended angles ($\theta_\mathrm{s}$, $\theta_{200}$), virial masses ($M_{200}$) and concentrations ($c$) of the four dIrrs studied in this work. $\rho_0$, $r_\mathrm{s}$ and $R_{200}$ are extracted from \cite{2021PhRvD.104h3026G}.}}
        \label{tab:dIrr_NFW_params}
    \end{center}
\end{table}

\begin{table}[b!]
    \begin{center}
        \begin{tabular}{|c|c|c|c|c|c|c|}
        \hline
        \hline
                      &  \multicolumn{6}{c|}{\textbf{Burkert}} \\ 
\hline
\hline
\textbf{Galaxy}    &  $\rho_0$ [$\mathrm{M}_{\odot}$/$\mathrm{kpc}^3$] &  $r_\mathrm{s}$ [kpc]  &  $\theta_\mathrm{s}$ [deg]  &   $R_{200}$ [kpc] &   $\theta_{200}$ [deg] &  $M_{200}$ [$\mathrm{M}_{\odot}$] \\ 
\hline
\textbf{IC10}      & $1.58\times 10^8$                                          & $2.0$                   & $0.15$                     & $71.3^{+7.1}_{-47.6}$           &  $5.16$            & $4.47\times10^{10}$      \\
\hline
\textbf{IC1613}    & $2.00\times 10^6$                                          & $7.0$                   & $0.53$                     & $45.7^{+1.7}_{-6.9}$           &  $3.44$              &   $1.07\times10^{10}$  \\
\hline
\textbf{WLM}       & $6.31\times 10^7$                                          & $1.3$                   & $0.08$                     & $33.3^{+2.0}_{-1.5}$           &  $1.97$              &  $4.35\times10^{9}$  \\
\hline
\textbf{NGC6822}   & $3.16\times 10^7$                                          & $3.3$                   & $0.39$                     & $62.9^{+8.4}_{-6.7}$           &  $7.47$              & $3.16\times10^{10}$    \\
\hline
\hline

        \end{tabular}
        \caption{\footnotesize{Parameters of the Burkert DM density profile (Equation~\ref{eq:Burkert_profile}) used to define the main DM halo, virial radius ($R_{200}$), subtended angles ($\theta_\mathrm{s}$, $\theta_{200}$) and virial masses ($M_{200}$) of the four dIrrs studied in this work. $\rho_0$, $r_\mathrm{s}$ and $R_{200}$ are extracted from \cite{2021PhRvD.104h3026G}.}}

        \label{tab:dIrr_Burkert_params}
    \end{center}
\end{table}

Since the J-factor (Equation~\ref{eq:JFac_formula}) is the integral of the total DM density profile squared, any modification in the profile can significantly change the J-factor value. One of the predictions of $\Lambda$CDM is the existence of smaller DM subhalos within the main hosting halo. With the inclusion of the subhalos, the total DM density profile is then given by $\rho_{\mathrm{DM}} = \rho_{\mathrm{main}} + \rho_{\mathrm{sub}}$, where $\rho_{\mathrm{main}}$ is the DM density profile of the host and $\rho_{\mathrm{sub}}$ is the subhalo contribution. The overall effect is an increase of the expected annihilation signal, also known as ``subhalo boost factor'' in the literature (see e.g., \cite{2019Galax...7...68A} and references therein).

To model the subhalo population, we use the following prescription (e.g., \cite{2024MNRAS.530.2496A}):

\begin{equation}
    \frac{d^3 N}{d V d M d c}=N_{\mathrm{tot}} \frac{d P_V}{d V}(V) \frac{d P_M}{d M}(M) \frac{d P_c}{d c}(M, c),
    \label{eq:subhalo_distribution_formula}
\end{equation}

\noindent where $N_{\mathrm{tot}}$ is the total number of subhalos, the subhalo radial distribution $dP_V/dV$ is the probability distribution of having a subhalo in a certain volume $dV$, the subhalo mass function $dP_M/dM$ is the probability distribution of finding a subhalo of mass between $M$ and $M + dM$, and the subhalo concentration distribution $dP_c/dc$ is referred to the probability of a subhalo having a certain concentration. We adopt the subhalo boost model first presented in Ref.~\cite{2017MNRAS.466.4974M}. This model assumes that the internal structure of galactic subhalos is well described by NFW profiles that have been heavily truncated in the outskirts by tidal stripping, nevertheless keeping intact their inner cusps. Whether subhalos are able to retain their inner NFW cusps or not due to tidal stripping is still a subject of debate, and is expected to also depend strongly on the particular accretion history of every subhalo, orbital configuration, etc. (see e.g., Refs.~\cite{2023MNRAS.518...93A,2025arXiv250601152A} and references therein). On the other hand, the smallest of these subhalos could have even steeper inner profiles, so-called ``prompt cusps'' as a result of the way structure formation works in $\Lambda$CDM \cite{2025arXiv250601152A, 10.1093/mnras/stx1658, 10.1093/mnras/stac3373}. We note that the survival of subhalos, as well as the exact shape of their inner cusps, will directly impact the computation of the subhalo boost. Beyond these considerations, the model also implicitly assumes the inclusion of two levels of halo substructure (i.e., subhalos within subhalos) \cite{2016CoPhC.200..336B}. For each of the probability distributions in Equation~\ref{eq:subhalo_distribution_formula}, we have followed those adopted in~\cite{2021PhRvD.104h3026G, CTAConsortium:2023yak}. 

\begin{figure}[t!]
  \centering 
  \includegraphics[width=0.99\textwidth]{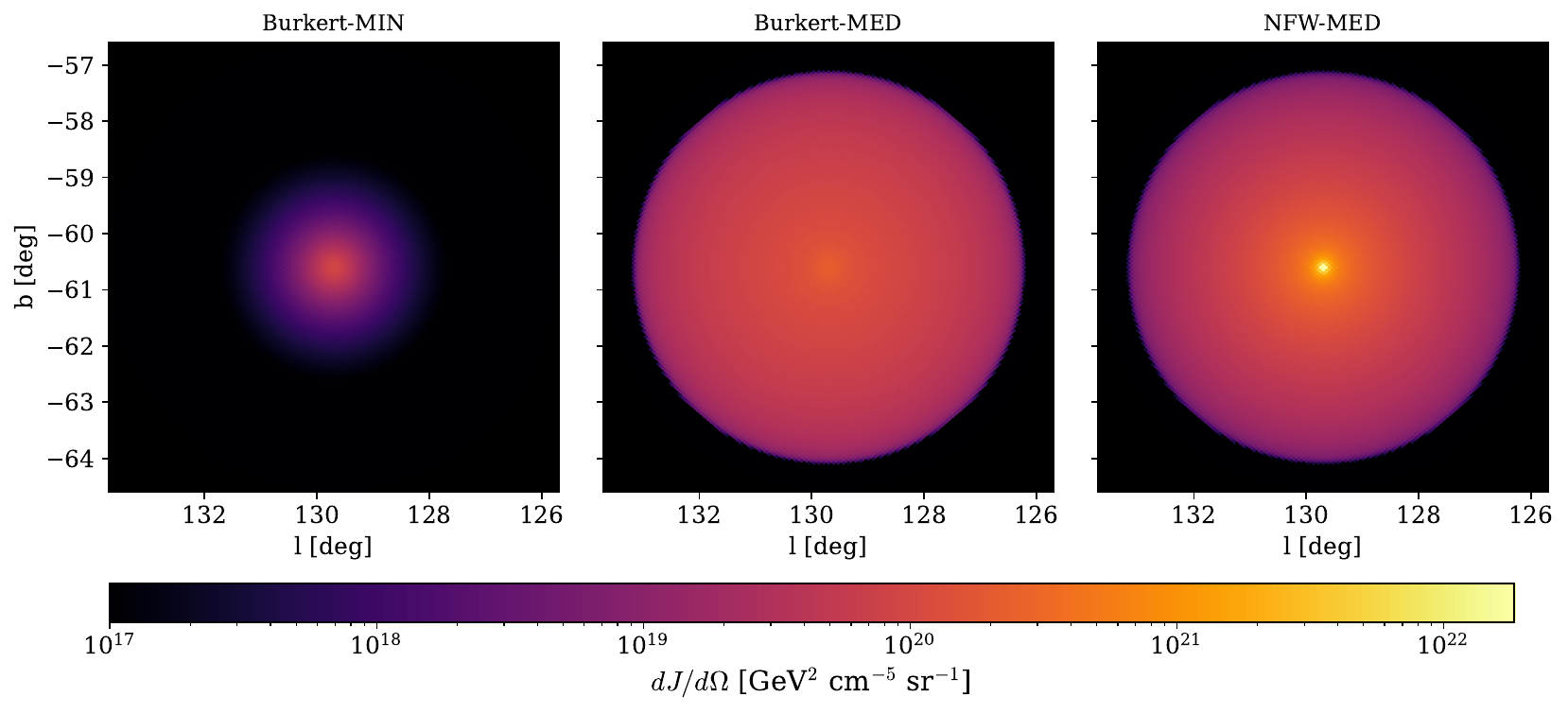}
    \caption{\footnotesize{Differential J-factor $dJ/d\Omega$ of the IC1613 dIrr, for the three DM density profiles adopted in this work: from left to right the Burkert-MIN (no subhalos), Burkert-MED and NFW-MED; see text for details. These maps have been created with \texttt{CLUMPY}~\cite{2012CoPhC.183..656C, 2016CoPhC.200..336B, 2019CoPhC.235..336H}.}}
\label{fig:clumpy_maps} 
\end{figure}

\begin{figure}[t!] 
  \centering 
  \includegraphics[width=0.75\textwidth]{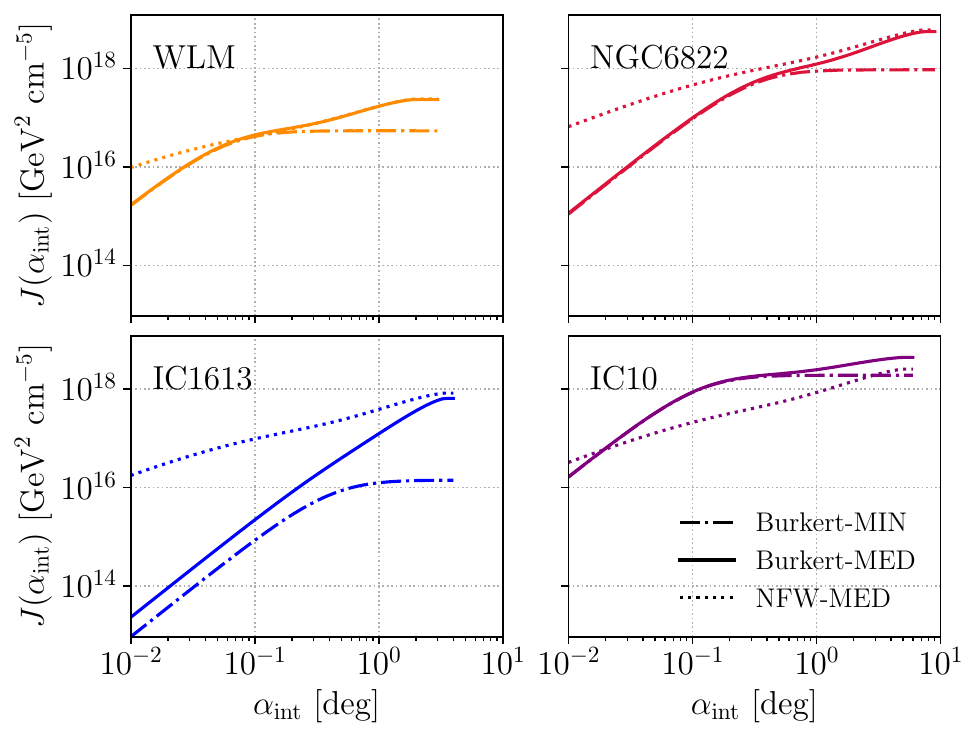}
    \caption{\footnotesize{Integrated J-factors versus the integration angle $\alpha_\mathrm{int}$, for the four different targets in our sample (from top to bottom and left to right, WLM, NGC6822, IC1613, and IC10) and three considered DM density profiles (dot-dashed for Burkert-MIN, solid for Burkert-MED, and dotted for NFW-MED). The integral is computed up to $\theta_{200}$ of each dIrr. These values have been computed with \texttt{CLUMPY} \cite{2012CoPhC.183..656C, 2016CoPhC.200..336B, 2019CoPhC.235..336H}.}}
\label{fig:All_Jfac_integrated} 
\end{figure}

\begin{figure}[t!] 
  \centering 
  \includegraphics[width=0.49\textwidth]{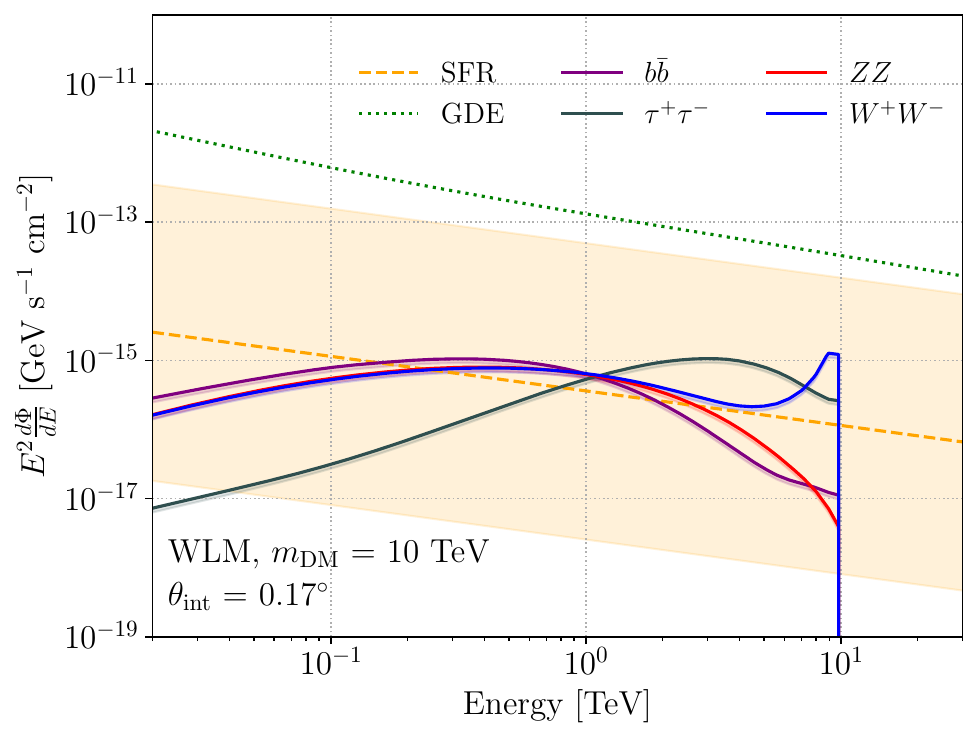}
  \includegraphics[width=0.49\textwidth]{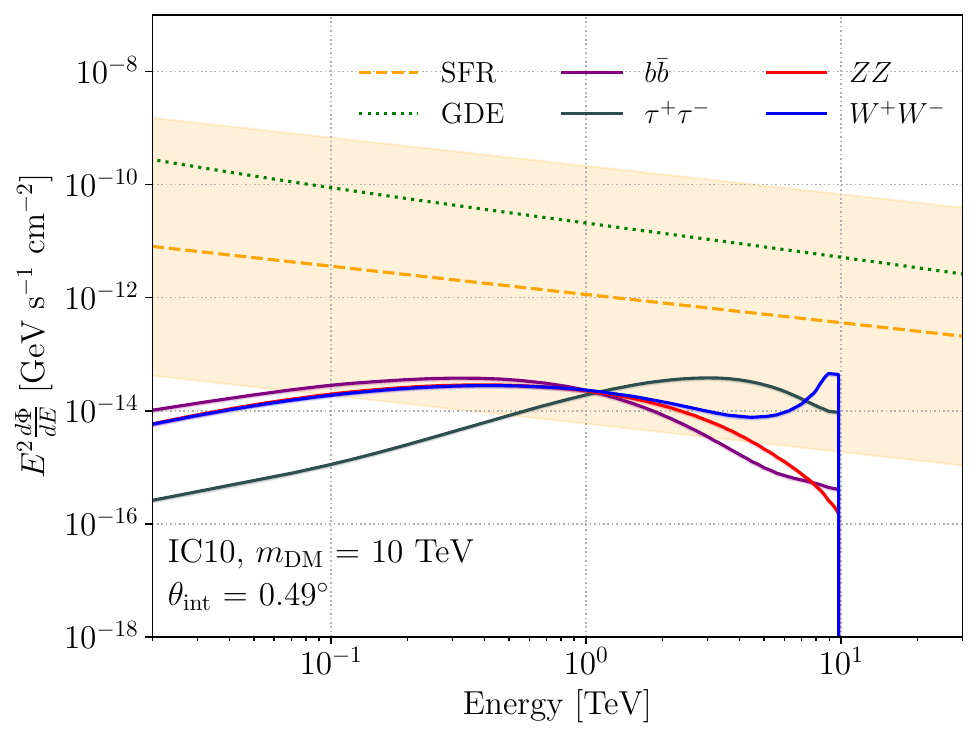}
    \caption{\footnotesize{Spectral energy distribution of the different components simulated for the dIrrs WLM (left panel) and IC10 (right panel), integrated within its NFW scale radius ($\theta_{\mathrm{int}} = \theta_{\mathrm{s}} = 0.17^\circ$ and $0.49^\circ$, respectively). The solid line shows the DM annihilation flux corresponding to the case of a WIMP mass of $m_{\mathrm{DM}} = 10$ TeV, for the annihilation channels considered in this work. As a reference, the solid lines represent the Burkert-MED modeling, whereas the uncertainty bands represent the possible scatter coming from the Burkert-MIN and NFW-MED (see Section \ref{sec:WIMP_annih}). The yellow dashed line refers to the benchmark SFR intrinsic emission, with the yellow band being its associated uncertainty, following the Martin14 model \cite{Martin:2014nia} as explained in Appendix \ref{ap:SFR_Appendix}. The green dotted line shows the expected GDE emission (see Section \ref{sec:GDE}), modeled with the $\gamma$-optimized Max model \cite{Luque:2022buq} in the direction of the target. See Appendix \ref{ap:rest_of_figs}, Figure~\ref{fig:dIrr_All_spectral_fluxes_appendix}, for the Figures corresponding to IC1613 and NGC6822.}}
\label{fig:dIrr_All_spectral_fluxes} 
\end{figure}

Given the possible uncertainties on the tidal evolution and baryonic effects, to model the DM distribution, we closely follow \cite{2021PhRvD.104h3026G, CTAConsortium:2023yak}, and adopt the Burkert-MIN, Burkert-MED and NFW-MED DM density profiles, where we use their nomenclature as well. In the Burkert-MIN case, the main halo is described by a Burkert profile with no subhalos; for the Burkert-MED, our benchmark case, we adopt the same profile for the host but now including subhalos following the recipes given above; finally, the NFW-MED corresponds to an NFW main halo profile with substructures. With these choices, we are left with a wide range that covers all possible scenarios: a case with no subhalo boost on the signal as a lower limit on the constraints Burkert-MIN, the benchmark state-of-the-art case Burkert-MED and the cuspy case with subhalos NFW-MED. In Table~\ref{tab:dIrr_Jfactors}, we show the integrated J-factors up to the virial radius $R_{200}$ for each galaxy and DM density profile. In Figure~\ref{fig:clumpy_maps}, we show the corresponding DM emission templates for the galaxy IC1613, created with the publicly available code \texttt{CLUMPY} \cite{2012CoPhC.183..656C, 2016CoPhC.200..336B, 2019CoPhC.235..336H}. In the Figure, the differences in the expected signal can be seen: the inclusion of subhalos makes the signal brighter in the outskirts, while the NFW profile creates a more concentrated signal in the center. Finally, Figure~\ref{fig:All_Jfac_integrated} shows the different integrated J-factors as a function of the integration angle $\alpha_{int}$. Our J-factors have also been computed using \texttt{CLUMPY}.

\begin{table}[t!]
    \begin{center}
        \begin{tabular}{|c|c|c|c|}
        \hline
        \hline

\multirow{2}{*}{\textbf{Galaxy}} & Burkert-MIN & Burkert-MED & NFW-MED\\ 
& [$\mathrm{GeV}^{2}\,\mathrm{cm}^{-5}$] & [$\mathrm{GeV}^{2}\,\mathrm{cm}^{-5}$] & [$\mathrm{GeV}^{2}\,\mathrm{cm}^{-5}$]\\ 
\hline
\hline
\textbf{IC10} & $1.92\times 10^{18}$ & $4.43\times 10^{18}$ & $2.55\times 10^{18}$ \\
\hline
\textbf{IC1613}  & $1.41\times 10^{16}$ & $6.49\times 10^{17}$ & $8.32\times 10^{17}$ \\
\hline
\textbf{WLM}  & $5.39\times 10^{16}$ & $2.32\times 10^{17}$ & $2.38\times 10^{17}$ \\
\hline
\textbf{NGC6822}   & $9.34\times 10^{17}$ & $5.57\times 10^{18}$ & $6.05\times 10^{18}$ \\
\hline
\hline

        \end{tabular}
        \caption{\footnotesize{J-factor values for the three different DM density profiles adopted in this work, integrated up to $\theta_{200}$ (see Tables~\ref{tab:dIrr_NFW_params} and \ref{tab:dIrr_Burkert_params}).}}
        \label{tab:dIrr_Jfactors}
    \end{center}
\end{table}

\section{Simulations setup and mock data analysis}
\label{sec:setup}

In the following, for each of the four targets, i.e., IC10, IC1613, WLM and NGC6822, we consider 50 hours of observation time on axis, i.e., with each dIrr located at the center of the field of view\footnote{Note that the adopted observation strategy is an on axis pointing centered on the target, as opposed to the 'wobble' method commonly used by IACTs. A detailed study of the optimal observation strategy is left for future work/proposals.}.  The simulated observations are all performed with the public code \texttt{gammapy} (version 1.2) \cite{2023A&A...678A.157D}. Instrumental parameters such as the effective area, PSF, energy dispersion and CR instrumental background are all encoded in the IRFs of the CTAO, publicly available on the CTAO website \cite{cherenkov_telescope_array_observatory_2021_5499840}. More specifically, we make use of the prod5-IRFs (v0.1), based on the full modelization of the Alpha configuration of the CTAO for both the North and South sites\footnote{The chosen IRFs are \texttt{Prod5-North-20deg-AverageAz-4LSTs09MSTs.180000s-v0.1} for the North site and \texttt{Prod5-South-20deg-AverageAz-14MSTs37SSTs.180000s-v0.1} for the South site, corresponding to a zenith angle of $20^\circ$. For more information on the Alpha Configuration expected performance, see \href{https://www.ctao.org/for-scientists/performance/}{https://www.ctao.org/for-scientists/performance/}}.

\begin{figure}[t!] 
  \centering 
  \includegraphics[width=0.99\textwidth]{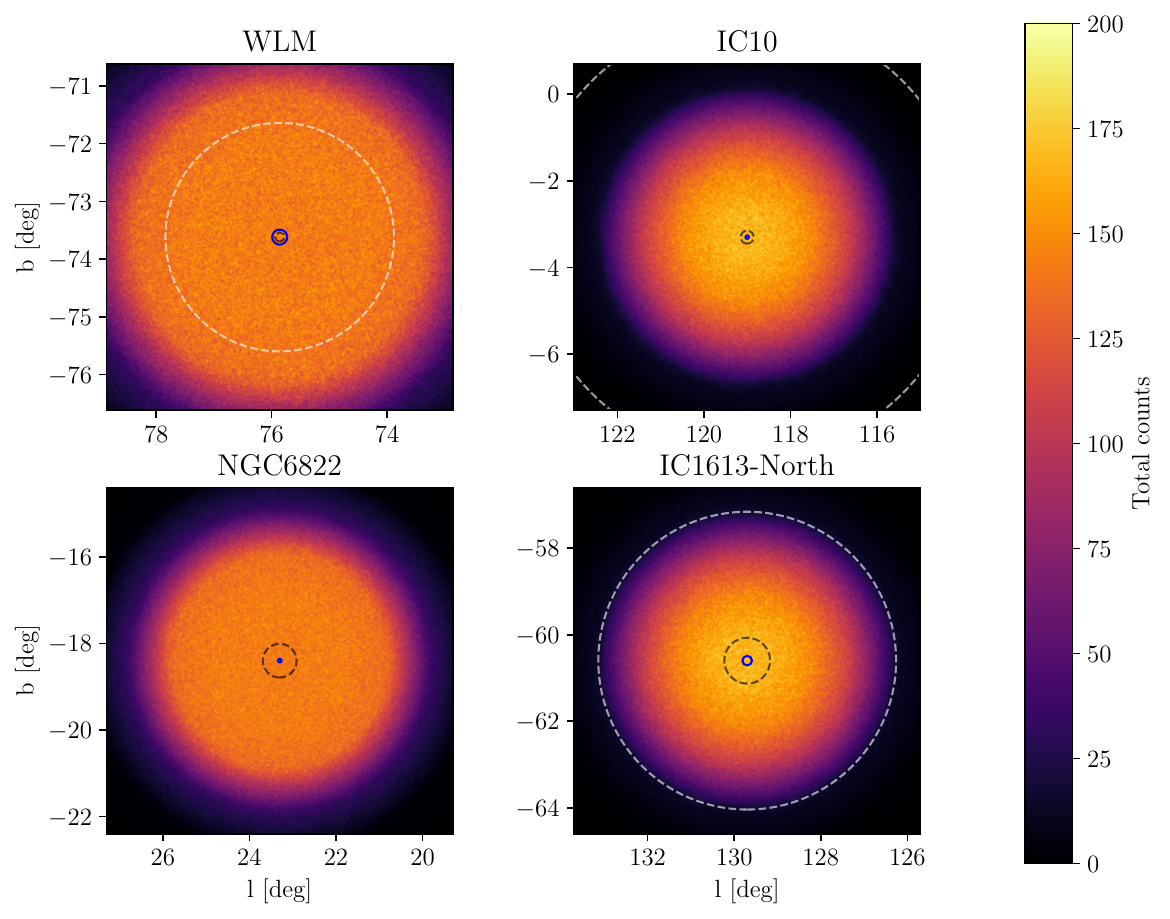}
    \caption{\footnotesize{Simulated gamma-ray counts of the four targets, i.e., IC10, IC1613, WLM and NGC6822 (for IC1613, we only show the North case). The figure shows the expected spatial distribution of total counts (gamma-ray emission and instrumental CR background) for the case of Burkert-MED, and $m_{\mathrm{DM}} = 10$ TeV annihilating into the $b\bar{b}$ channel. The grey dashed circle shows the $\theta_{200}$ value of each dIrr, the black dashed circle represents $\theta_{s}$, and the solid blue circle represents the size of the modeled SFR. The simulated regions are defined by their halo size $\theta_{200}$. Note that, for the case of NGC6822 and IC10, $\theta_{200}$ is larger than the FoV of the CTAO, so we limit the simulations up to a total width of $8^\circ$ instead.}}
\label{fig:dIrr_sky_counts_simulation} 
\end{figure}

Following the discussion so far, the full modeling of these targets is a complex task. Firstly, the GDE spectral component dominates both the expected DM flux and the SFR emission (Figure~\ref{fig:dIrr_All_spectral_fluxes}). Secondly, the SFR emission can be of the same order of magnitude as the DM flux (depending on the DM mass), although the spatial distribution and extension of the signals are completely different (Figures~\ref{fig:All_Jfac_integrated} and \ref{fig:dIrr_sky_counts_simulation}). In fact, the GDE is a diffuse foreground emission which originates from our own galaxy; the DM signal is a moderate extended emission of several degrees in the sky, with a peak at the center of the targeted dIrr, and, finally, the SFR is nearly a PL signal for CTAO, indeed with an extension at most a few times greater than the PSF of the CTAO. This, in turn, will help break degeneracies and disentangle the different components in the template analysis, as discussed later below.

\begin{figure}[t!] 
  \centering 
  \includegraphics[width=0.99\textwidth]{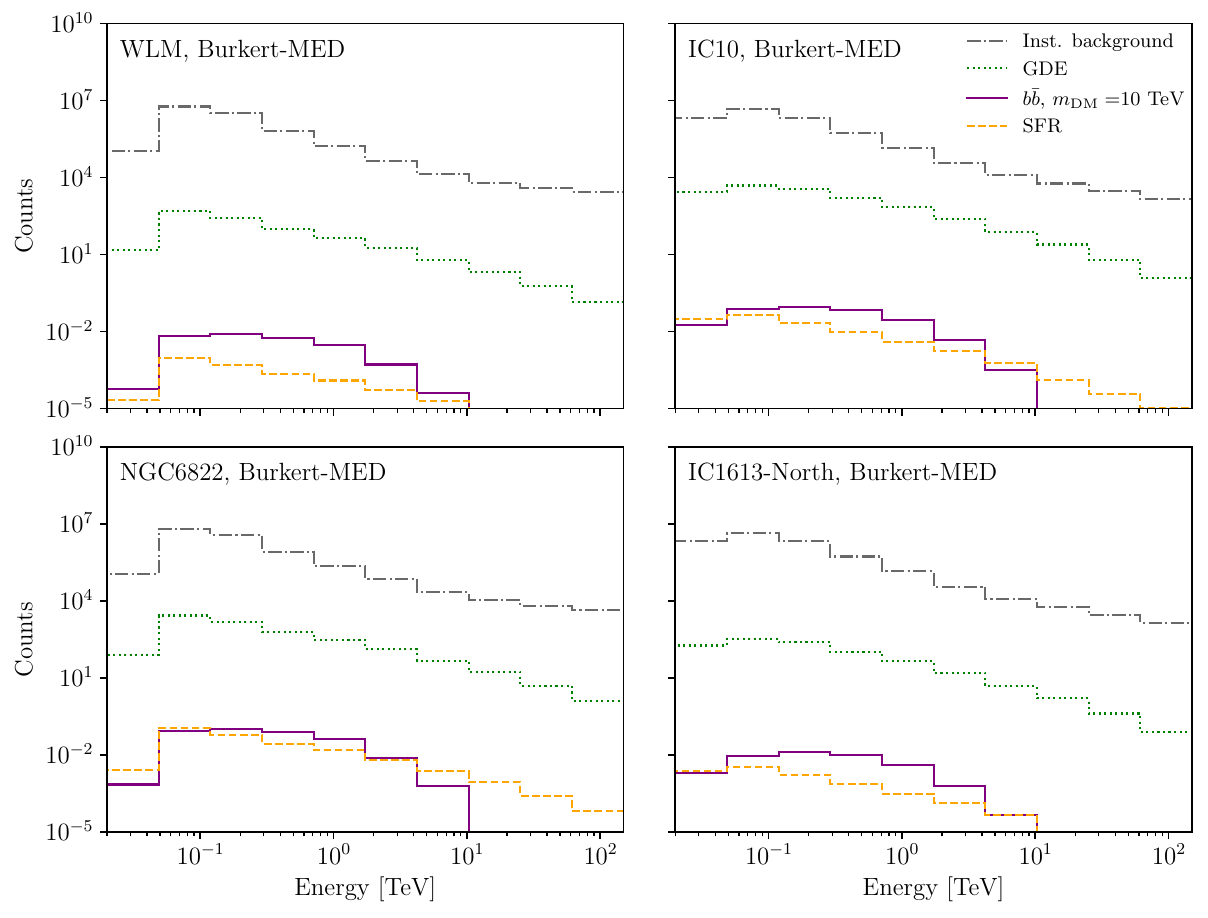}
    \caption{\footnotesize{Simulated observed spectral counts of the four dIrrs, i.e., WLM, IC10, NGC6822 and IC1613 (for IC1613, we only show the North case). The figure shows the expected spectral distribution of counts (spatially integrated in the whole ROI, as shown in Figure~\ref{fig:dIrr_sky_counts_simulation}) for all the components of our analysis: the instrumental background counts (dash-dotted grey line), GDE (dotted green line), SFR counts (dashed yellow line) and DM annihilation ($b\bar{b}$ channel, Burkert-MED modeling and $m_{\mathrm{DM}} = 10$ TeV) (solid purple line).
    }}
\label{fig:dIrr_spectral_counts_simulation} 
\end{figure}

Finally, we simulate the sky observation of all the expected counts (gamma-ray sky emission and instrumental background counts) as shown in Figure~\ref{fig:dIrr_sky_counts_simulation}, with a spatial pixel binning of $0.02^\circ$ up to the virial radius of the targets $\theta_{200}$, which is the maximum relevant size of this analysis. In the cases of IC10 and NGC6822, the corresponding values of $\theta_{200}$ are greater than the field of view (FoV) of the CTAO, so we limit these simulations to a size of $8^\circ\times8^\circ$. As for the energy binning, we define it with 10 bins logarithmically spaced between 20 GeV and 150 TeV. In Figure~\ref{fig:dIrr_spectral_counts_simulation}, we show the spectral counts (for each energy bin, spatially integrated over the full simulated map) of the four galaxies studied in this work. As expected, the instrumental background counts dominate over all of the models considered\footnote{This does not represent a problem, since a template fitting analysis allows us to recover the actual gamma-ray counts from the total detected counts.}.

\subsection{Analysis procedure and considerations}

In our work, we adopt a template fitting analysis, with the following likelihood:

\begin{equation}
    \mathcal{L}(\mu | n) = \prod_{i,j} e^{-\mu_{i,j}} \frac{\mu_{i,j}^{n_{i,j}}}{n_{i,j}!} 
    \label{eq:likelihood_formula}
\end{equation}

Where the indexes $i$ and $j$ run over the energy and spatial bins, respectively; $\mu = \{ \mu_{i,j} \}$ are the predicted counts, created from the templates of the theoretical models that we want to test; and $n = \{ n_{i,j} \}$ are the observed (mock) data counts. The model prediction $\mu$ is defined by a set of signal templates (depending on the case, the DM template described in Section \ref{sec:WIMP_annih}, $\mu_{i,j}^\mathrm{DM}$, or the SFR emission in Section \ref{sec:SFR_IE}, $\mu_{i,j}^\mathrm{SFR}$) and a set of background components $\{ \mu_{i,j}^\mathrm{bkg} \}$ (the instrumental CR background and the rest of AE templates not considered as signal, like the GDE). Once the theoretical models are chosen, we need to convolve them with the IRFs of the CTAO and then compare with the mock data, $n$. We allow a renormalization $\theta^X$ of each of the templates such that $\mu_{i,j} = \sum_{X} \theta^X \mu_{i,j}^X$. These normalization parameters are the ones that will be fitted to the observed (mock) counts. Note that, when $\theta^X=1$ for all the models, the expected emission presented in Section \ref{sec:spectral_modeling} is recovered. For the case of the DM, as it can be seen in Equation~\ref{eq:flux_annih}, $\theta^\mathrm{DM}$ is directly proportional to the annihilation cross-section $\langle \sigma v \rangle = \theta^\mathrm{DM} \langle \sigma v \rangle_\mathrm{th}$, so from now on we will refer to this parameter instead. Finally, as the CTAO is still in its construction phase, the observed (mock) data $n^{X}_{i,j}$ is created by extracting the counts from a Poissonian distribution with mean $\mu^{X}_{i,j}$ for each of the simulated models  (background and also signal whenever necessary), with the final counts $n_{i,j}$ being a result of a summation of all the mock data $n^X_{i,j}$ of each model $X$ considered. As before, the counts are also convolved with the IRFs of the CTAO.

In general, to evaluate a possible detection scenario, we define the following Test Statistics (TS), in which we can compare the best fit of the normalization parameters to the signal+background case ($\theta^{\mathrm{signal}}$ and $\{ \theta^\mathrm{bkg} \}$) with the null hypothesis given only by the corresponding background \cite{2015APh....62..165C}:

\begin{equation}
   \mathrm{TS} = -2 \log{\frac{\mathcal{L}_{\mathrm{null}}(\mu(\theta^{\mathrm{sig}}=0,  \hat{\hat{\theta}}^\mathrm{bkg})| \mathrm{n})}{\mathcal{L}_{\mathrm{best-fit}}(\mu(\hat{\theta}^{\mathrm{sig}},  \hat{\theta}^\mathrm{bkg})| \mathrm{n})}} \simeq \sigma^2,
    \label{eq:TS_formula_general}
\end{equation}

\noindent where $\sigma$ is the detection significance\footnote{For a nested signal+background model with one additional parameter relative to the null hypothesis, following Wilks theorem it can be approximated $\mathrm{TS} \simeq \sigma^2$ \cite{2015APh....62..165C}.}, $\mathcal{L}_{\mathrm{null}}(\mu(\theta^{\mathrm{sig}}=0,  \hat{\hat{\theta}}^\mathrm{bkg}))$ is the maximized likelihood corresponding to the null hypothesis, with $\hat{\hat{\theta}}^\mathrm{bkg}$ being the best-fit parameters where no signal is assumed in the model. In the denominator, $\mathcal{L}_{\mathrm{best-fit}}(\mu(\hat{\theta}^{\mathrm{sig}},  \hat{\theta}^\mathrm{bkg})| \mathrm{n})$ is the maximized likelihood in the signal and background scenario, with its corresponding best-fit parameters $\hat{\theta}^{\mathrm{sig}}$ and $\hat{\theta}^\mathrm{bkg}$. Being interested in a possible detection scenario of the signal, the mock data $n$ is simulated including both the background and signal templates. To study the detection prospects for dIrrs as astrophysical emitters, the signal is identified with the SFR emission of each galaxy ($\hat{\theta}^{\mathrm{sig}} = \hat{\theta}^{\mathrm{SFR}}$) and the background as the GDE and instrumental CR background. On the other hand, for the DM flux detection case, the signal is defined by the DM template ($\hat{\theta}^{\mathrm{sig}} = \hat{\langle \sigma v \rangle}$) and the background by the instrumental CR background and a subset of AE templates.

In case of no detection expected ($\mathrm{TS}<25$), we proceed and set upper limits to the normalization parameter of the corresponding signal template. For the SFR emission, we will investigate the value of the total integrated flux needed to have a $5\sigma$ detection of the dIrr as an astrophysical source (Section \ref{sec:results_SFR}) and the underlying $\mathcal{SFR}$ needed for such emission. For the case of the DM signal (Sections \ref{sec:results_DM} and \ref{sec:sommerfeld}), the projected upper limits at the $95\%$ confidence level (C.L.) on the annihilation cross-section are estimated instead. To investigate the latter, following the Wilks theorem we use that the TS of the annihilation cross-section $\langle \sigma v \rangle$ follows a $\chi^2$ distribution with one degree of freedom \cite{2015APh....62..165C}\footnote{Since our parameter of interest is the annihilation cross-section $\langle \sigma v \rangle$, to compute its upper limits we need to maximize the likelihood over the rest of the nuisance parameters $\theta^\mathrm{bkg}$ when sampling the TS profile.}. Assuming that the distribution is one-sided, the $95\%$ C.L. upper limits on $\langle \sigma v \rangle$ correspond to an increase in the log-likelihood ratio over the best fit by 2.71 \cite{2015APh....62..165C}:

\begin{equation}
    -2 \log{\frac{\mathcal{L} (\mu(\langle \sigma v \rangle_{95\%},  \theta^\mathrm{bkg}_{95\%})| \mathrm{n})}{\mathcal{L}_{\mathrm{best-fit}}(\mu(\hat{\langle \sigma v \rangle},  \hat{\theta}^\mathrm{bkg})| \mathrm{n})}} = 2.71,
    \label{eq:TS_upper_limits}
\end{equation}

\noindent where the set of parameters $\langle \sigma v \rangle_{95\%}$ and $\theta^\mathrm{bkg}_{95\%}$ define the constraints that enclose the $95\%$
C.L. region over $\langle \sigma v \rangle$, and $\hat{\langle \sigma v \rangle}$ and $\hat{\theta}^\mathrm{bkg}$ are the best-fit parameters of the templates to the mock data $n$. As we are focusing on setting upper limits to the signal, note that the mock data is simulated only from the background templates and assuming no DM signal. We present an analysis that includes the full AE modelization of the GDE and SFR emission (Section \ref{sec:results_DM}), but a deeper discussion on different analysis set-ups is presented in Appendix \ref{ap:mismodeling_constraints}, where we employ three different approaches: a PL DM-only case in which the AE is not included; a second case where we include the spatial distribution of the DM-only signal; and a final one where we explore the implications of not including the SFR emission as a background template affects the constraints, by simulating it in the mock data and neglecting it in the fitting templates.

Finally, given the Poissonian randomness of the simulated counts, $n$, for each sky configuration (i.e., each subset of templates from which $n$ is simulated), we create 100 simulations to ensure convergence of the results, as in \cite{CTAConsortium:2023yak}. In Appendix \ref{ap:poisson_uncertainties}, we show how the computation of the individual upper limits on the annihilation cross-section $\langle \sigma v \rangle$ is performed, and how the expected Poissonian uncertainties are inferred. Considering that we have a total of five case-galaxies (WLM, IC10, NGC6822, IC1613-North and IC1613-South), we can also perform a combined analysis that represents the final constraints on the annihilation cross-section $\langle \sigma v \rangle$, where the combined likelihood is computed by means of the addition of the individual log-likelihood profiles. Given that we have two different prospective observations for IC1613, in Section \ref{sec:results_DM} we discuss how the combined results are affected by the selection of the IC1613-North or South case. In Appendix \ref{ap:combined_analysis}, we discuss the convergence of the individual TS profiles and describe the computation of the combined results.

\section{Results}
\label{sec:results_general}

In this section, we present our template fitting analysis results for the four considered dIrrs, and discuss the prospects to detect with the CTAO any of the different gamma-ray components considered in this work.

\subsection{Detection of the SFR emission}
\label{sec:results_SFR}

Among all star-forming galaxies, only 13 have been detected by \textit{Fermi}-LAT \cite{2025A&A...699A..43K, 2012ApJ...755..164A, 2020ApJ...894...88A} and 2 at TeV energies (M82 by VERITAS \cite{2025ApJ...981..189A} and NGC253 by HESS \cite{2018A&A...617A..73H}) \cite{2025A&A...699A..43K}. More specifically, for isolated dIrrs, only with NGC6822 and IC10 upper limits on the flux have been computed with \textit{Fermi}-LAT data \cite{2025A&A...699A..43K, 2021PhRvD.104h3026G}. Given that no gamma-ray emission has been detected from isolated dIrrs, we put to the test the possibility of having a CTAO detection of the SFR component in dIrrs. In this case, focusing on the AE, we neglect in our simulations the contribution due to the DM annihilation, thus the simulation only includes the instrumental CR background, the GDE, and the intrinsic astrophysical emission in the dIrr (i.e., the SFR). We also vary such SFR contribution to the overall flux to account for the SFR uncertainties, which are driven by the modeling of the $\mathcal{SFR}$ of the galaxy from its stellar mass and the estimation of the total SFR luminosity from this value (see Appendix \ref{ap:SFR_Appendix}). 

\begin{figure}[t!] 
  \centering 
  \includegraphics[width=0.7\textwidth]{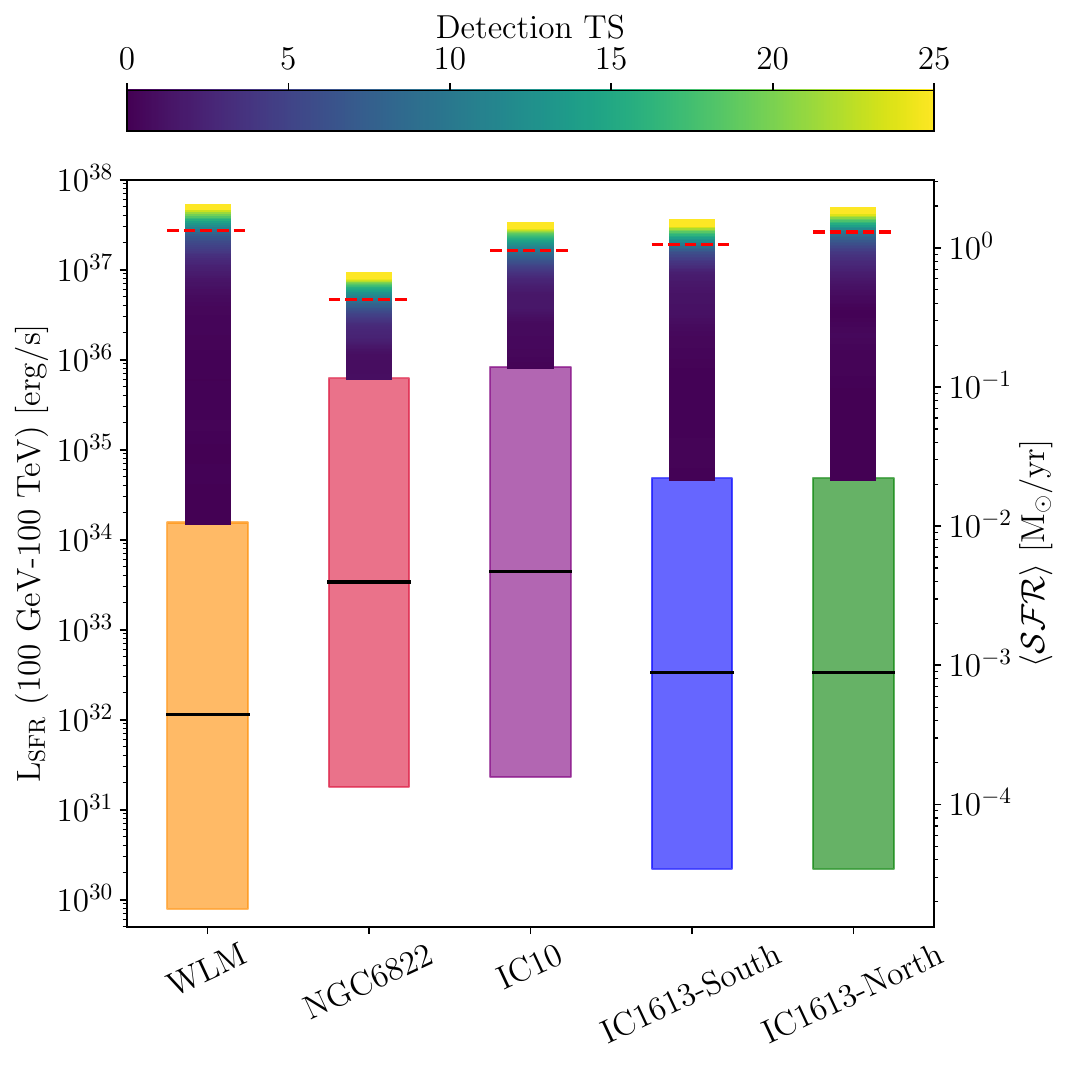}
    \caption{\footnotesize{TS of the SFR emission produced in each of the different galaxy cases considered in this work and an observation time of 50 h per target, as a function of the integrated SFR luminosity (from 100 GeV to 100 TeV) and the corresponding $\langle \mathcal{SFR} \rangle$ representative value related to the emission (see text for more information). The horizontal black line represents the mean, benchmark value of the SFR luminosity computed with the Martin14 model \cite{Martin:2014nia}, while colored bands refer to the uncertainty of the SFR modeling. For each case, the red dashed horizontal line represents the luminosity needed to have a $3\sigma$ hint of the galaxy. Also, we show the integrated luminosity needed to have a $5\sigma$ detection (TS $\simeq 25$), i.e., top of each bar. The SFR emission is modeled as explained in Appendix \ref{ap:SFR_Appendix}.}}
\label{fig:Flux_TS_SFR_detection} 
\end{figure}

The results of this study are presented in Figure~\ref{fig:Flux_TS_SFR_detection}, where the black horizontal line represents the benchmark SFR luminosity value used throughout this work, and the colored bands refer to the uncertainty of our modeling. In order to reach a $3\sigma$ (TS $\simeq 9$, marked by the red dashed horizontal line) or a $5\sigma$ (TS $\simeq 25$) detection, the luminosity must be from 3 to 5 orders of magnitude greater than the expected benchmark SFR luminosity, although in some galaxies (IC10 and NGC6822), a detection is possible if the luminosity is greater than about ten times the upper limit of the SFR luminosity, whereas for a $3\sigma$ hint, the needed factor is reduced to $\sim 7$ for NGC6822. Given that the main parameter that regulates this emission is the $\mathcal{SFR}$, we also show in the Figure the expected mean values corresponding to such a luminosity. Note that, because of the scatter given by the four models of the SFR-luminosity relation (Equation~\ref{eq:PCR_to_Lgamma} in Appendix \ref{ap:SFR_Appendix}), there is a small range of $\mathcal{SFR}$ values that can yield the same luminosity. Therefore, we choose $\langle \mathcal{SFR} \rangle$ as the representative value for this parameter. Our results show that an individual SFR detection for these four galaxies is unlikely with the CTAO, even considering large variations of the SFR flux within the range allowed by current uncertainties (see Table~\ref{tab:dIrr_SFR_1}). In general, we find that an overall $\mathcal{SFR}$ of order $\sim 1 \ \mathrm{M_\odot/yr}$ is needed to have a detection of a SFR, with NGC6822 and IC10 being the best galaxy candidates among the four targets. To ensure statistical consistency, this detection TS is computed with 100 simulated datasets per SFR luminosity, and we show in the Figure its median. 

Regarding a combined analysis, we have checked that all of the likelihoods are dominated by NGC6822, so we do not show the combined results as they are very similar to NGC6822 individual constraints. For an in-depth discussion of all the possible correlations between the fitted parameters, in Appendix \ref{ap:Correlation_matrix} we show the correlation matrices of the full analysis (including also the DM-induced emission, not only the AE models used in this Section), where we list the possible degeneracies between the SFR emission and the GDE and CR instrumental background. As a final remark, we perform in Appendix \ref{ap:SFR_bin_by_bin_lklh} a bin-by-bin likelihood fit, showing the expected observed spectral flux and upper limits for the best target NGC6822. In this Appendix, we show the cases of when the total SFR flux corresponds to the SFR-MAX case (see Table~\ref{tab:dIrr_SFR_1}) and when we set the normalization of the flux to have a $5\sigma$ detection. Besides, we compare the expected flux with the \textit{Fermi}-LAT upper limits presented in \cite{2025A&A...699A..43K,2021PhRvD.104h3026G}.

\subsection{Galactic diffuse emission detection}
\label{GDE_detection}

In this part of the analysis, we investigate the possibility of detecting the GDE in the direction of the targets. To approach this exercise, like the SFR analysis, we simulate a sky model composed of the GDE, the instrumental CR background, and the SFR (no DM template is included). As usual, we run the simulations 100 times to ensure consistency with the statistics. We have checked that for all 100 simulations made, the CTAO will consistently detect the GDE only in those cases where the dIrrs lie close to the GP, i.e., where the GDE emission is expected to be significantly higher (Figure~\ref{fig:dIrr_skymap}). Indeed, in the direction of IC10 and NGC6822, the detection of the GDE is expected with a high TS ($\mathrm{TS} > 50$ in most simulations). For the rest of the galaxies, no GDE detection is expected instead (for more information, see Appendix \ref{ap:GDE_detection}). Finally, in Appendix \ref{ap:Correlation_matrix}, we list the degeneracies between the SFR emission and the GDE and CR instrumental background by quantifying the correlation between the free parameters in the fit. The correlation matrices also include the DM emission to illustrate the correlations of the full analysis.

\subsection{Constraints on the DM annihilation signal}
\label{sec:results_DM}

We focus on four annihilation channels ($\tau^+\tau^-$, $b\bar{b}$, $W^+W^-$ and $ZZ$), sampling the constraints into 13 DM mass values $m_{\mathrm{DM}}$ from $50$ GeV to $100$ TeV logarithmically spaced (note that, for kinematic reasons, in the $ZZ$ and $W^+W^-$ channels the first value for $m_{\mathrm{DM}}$ is $94$ GeV, not $50$ GeV). No detection of DM signal with a significance $\sigma > 5$ has been found consistently throughout the 100 iterations of each sky realization of the DM models under consideration, and independently of including AE as a background in the simulations. Thus, in the following, we provide the projected $95\%$ C.L. upper limits on the annihilation cross-section, $\langle \sigma v \rangle$, found for each dIrr (individual limits) and for the statistical combination of the whole sample (combined analysis). These projected constraints correspond to the full modeling of the scenario, with all the components already introduced (GDE, SFR emission, DM annihilation flux and the instrumental CR background). For comparison, and in order to quantify the level of uncertainty coming from not using the full modelization in these targets, we show in Appendix \ref{ap:mismodeling_constraints} the projected upper limits purposely mismodeling the signal for three different analyses: an ideal DM-only scenario in which the DM is PL (PL analysis); another DM-only in which its emission is extended; and a final where we explore the impact of not including the SFR emission as a background template, by simulating it in the mock data and neglecting it in the fitting templates. This helps in understanding how the DM constraints vary by explicitly not considering the AE in the analysis or the extended nature of the DM emission. In the following, we present the full extended analysis (DM and AE), in which, considering all the models, we are left with four free normalization parameters in the fits: the amplitudes of the instrumental CR background, the GDE, the SFR emission, and the DM annihilation cross-section $\langle \sigma v \rangle$.

\begin{figure}[t!] 
  \centering 
  \includegraphics[width=0.99\textwidth]{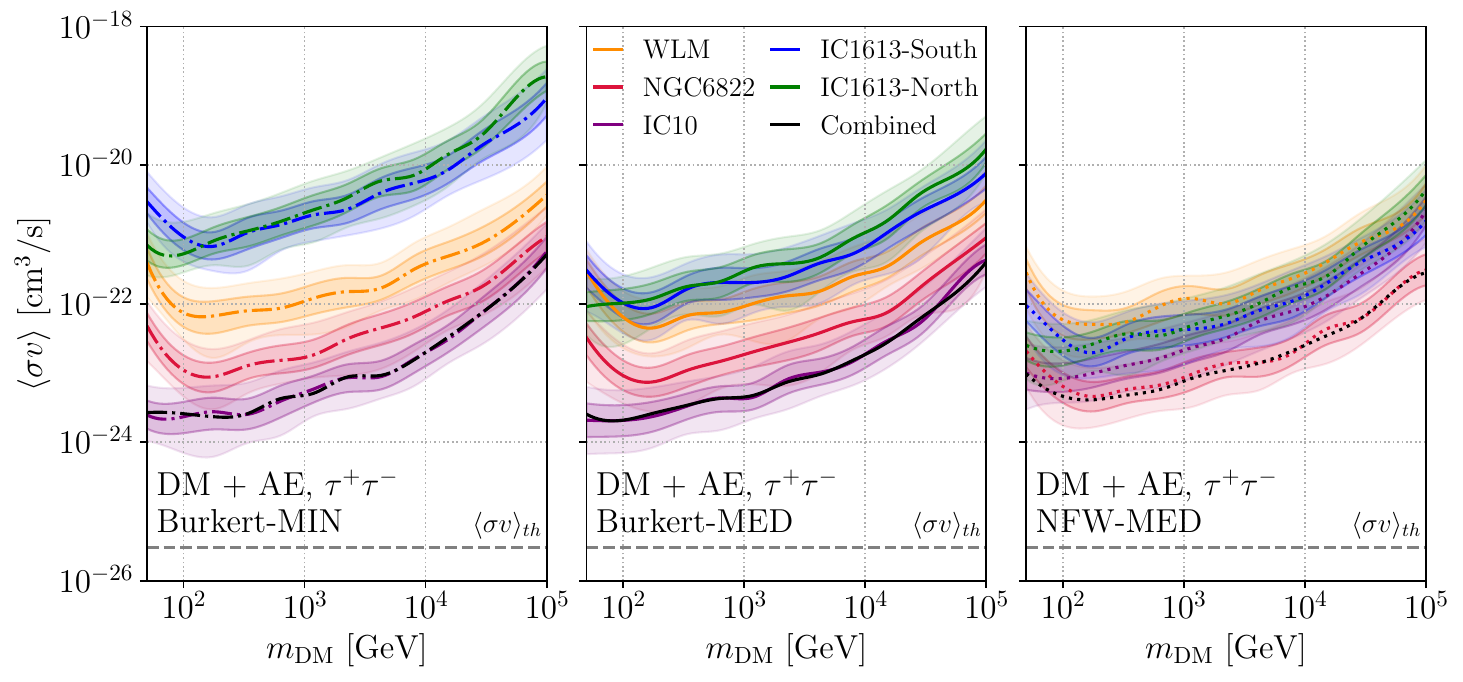}
    \caption{\footnotesize{DM annihilation cross-section $\langle \sigma v \rangle$ projected upper limits for individual dIrrs (colored lines) and the combined sample (black lines, see Appendix \ref{ap:combined_analysis}), in the case of the full DM + AE modeling and in the $\tau^+\tau^-$ annihilation channel. Each column refers to a different DM density profile: Burkert-MIN (first column), Burkert-MED (second column) and NFW-MED (third column); as described in Section \ref{sec:DM_spatial_modeling}. We also show the 1$\sigma$ and 2$\sigma$ uncertainties (colored bands) for each of the individual upper limits due to the Poissonian noise related to the simulation of the counts (Appendix \ref{ap:poisson_uncertainties}). The combined limits are computed including four galaxies (WLM, NGC6822, IC10 and IC1613-South), totaling up to 200h of observation time. As a comparison, the horizontal grey dashed line represents the thermal relic cross-section $\langle \sigma v \rangle_{\mathrm{th}}$.}}
\label{fig:tau_channel_UL_all} 
\end{figure}

We show in Figure~\ref{fig:tau_channel_UL_all} the individual (colored bands) and combined (black lines) projected upper limits for the case of DM annihilating into the $\tau^+\tau^-$ channel, and the corresponding $1\sigma$ and $2\sigma$ uncertainties estimated from the Poissonian noise of our simulations. For more information on the computation of the uncertainties, see Appendix \ref{ap:poisson_uncertainties}, and, for the computation of the combined analyses, see Appendix \ref{ap:combined_analysis}. Regarding the individual projected constraints, IC10 dominates the Burkert-MIN and Burkert-MED results, with NGC6822 dominating the NFW-MED case. Besides this, their respective individual constraints almost coincide with the combined results, making them also the best dIrr targets for DM indirect searches. As for the galaxy cases IC1613-North and South, we find that in both cases the individual constraints are similar and among the weakest, with the South case yielding slightly better constraints at higher masses (thanks to the SSTs in CTAO-South) than the North and the opposite at lower masses (because of the LSTs planned in CTAO-North). Being also one of the weakest constraints, when computing the combined analysis, no difference is found when including IC1613-North or South (see Figure~\ref{fig:combined_individual_likelihoods} in Appendix \ref{ap:combined_analysis}), totaling the combined observation time up to 200h (50 h per galaxy, as IC1613 is not considered twice). Therefore, hereafter, no distinction will be made when considering the final combined results.

Among all channels considered, the best constraints reach $\sim 2 \times 10^{-24} \ \mathrm{cm^{3}/s}$ at $m_\mathrm{DM} \simeq 100 $ GeV in the Burkert-MED case and $\tau^+\tau^-$ channel. We find that including all possible astrophysical gamma-ray emissions does not change the projected upper limits compared to the extended DM-only analysis and simulation (Appendix \ref{ap:DM_only_results}) or without including the SFR emission as a fitted template, even when including in the mock dataset the maximum SFR flux allowed by the uncertainties (Appendix \ref{ap:UL_no_SFR_fitted}). In other words, the astrophysical gamma-ray background in dIrrs, the SFR emission (see Appendix \ref{ap:SFR_Appendix}), is completely negligible for DM searches for the four targets, as in the case of dSphs, although dIrrs presenting higher $\mathcal{SFR}$ might require an explicit modeling. As for the GDE, in those two galaxies where the GDE contribution is more relevant compared to the total counts (IC10 and NGC6822), the template fitting analysis is able to correctly isolate/identify that part of the emission from the total, and, therefore, it does not affect the DM sensitivity prospects. In the case of the other two galaxies, located far from the GP (WLM and IC1613), since the GDE does not play an important role in the total counts, the DM constraints are also unaffected by the non-detection of this emission.

\begin{figure}[t!] 
  \centering 
  \includegraphics[width=0.99\textwidth]{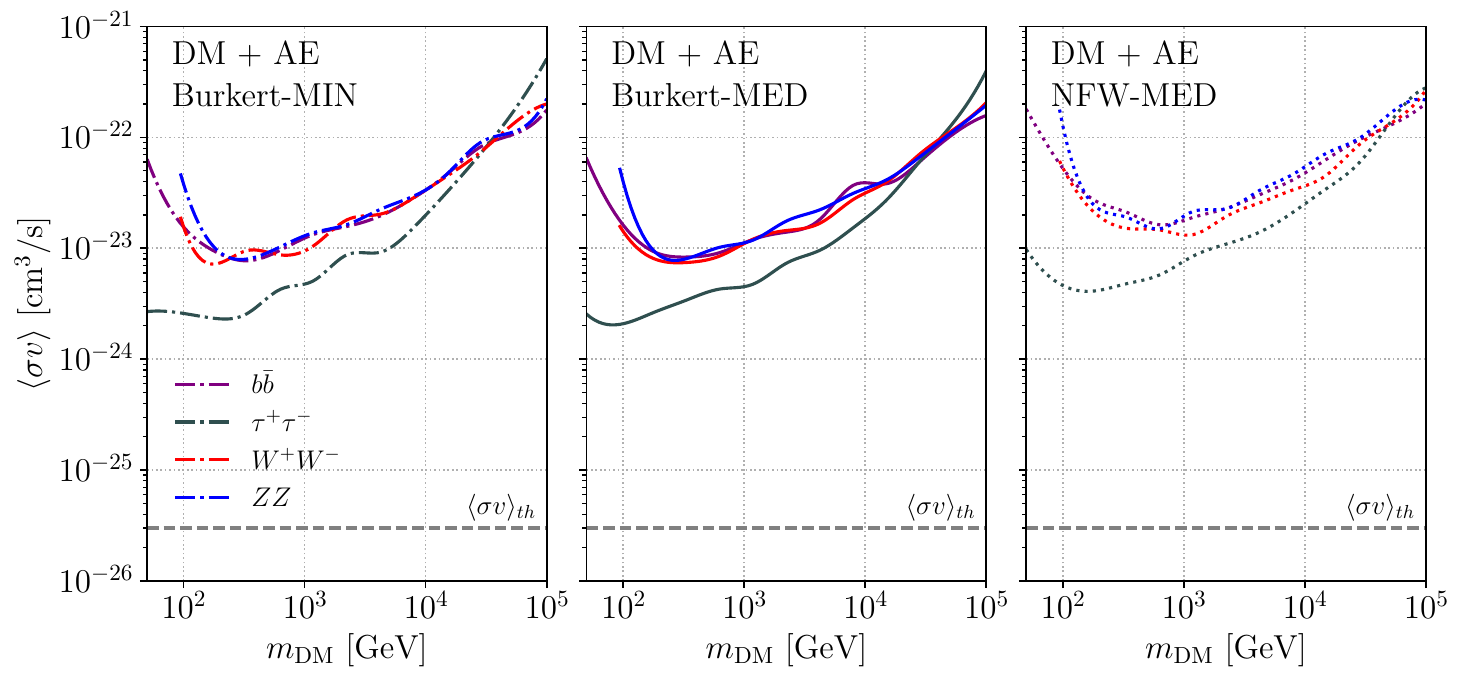}
    \caption{\footnotesize{Projected combined upper limits to the annihilation cross-section $\langle \sigma v \rangle$, for the DM + AE analysis (Section \ref{sec:results_DM}) and assuming the Burkert-MIN (first panel), Burkert-MED (second panel), NFW-MED (last panel) as the DM density profiles. We show the limits for the four different annihilation channels used in this work: $b\bar{b}$ (in purple), $\tau^+\tau^-$ (grey), $W^+W^-$ channel (red) and $ZZ$ channel (blue). As a comparison, the horizontal grey dashed line represents the thermal relic cross-section $\langle \sigma v \rangle_{\mathrm{th}}$.}
    }
\label{fig:all_channel_combined} 
\end{figure}

In Figure~\ref{fig:all_channel_combined}, we show all the combined annihilation cross-section $\langle \sigma v \rangle$ projected upper limits, for the three DM density profiles used -- Burkert-MIN (first panel), Burkert-MED (second panel), NFW-MED (last panel)-- and for the four different SM annihilation channels used in this work: $b\bar{b}$ (in purple), $\tau^+\tau^-$ (grey), $ZZ$ (blue) and $W^+W^-$ channel (red). Interestingly, among the combined results, we find that the best constraints are given by the Burkert-MED and Burkert-MIN, with NFW-MED being worse by a factor of a few. The reason is in the two dominating galaxies. The Burkert limits are dominated by IC10, whose J-factors exceed the NFW-MED in the inner region (Figure~\ref{fig:All_Jfac_integrated}) and are very similar in the outskirts, making it a brighter source than NGC6822 in the center. For the NFW-MED case, NGC6822 has the greatest J-factor of all the targets, even greater than the second best galaxy, IC10, by a factor of $\sim 2.5$ (Table~\ref{tab:dIrr_Jfactors}).

\begin{figure}[t!] 
  \centering 
    \includegraphics[width=0.85\textwidth]{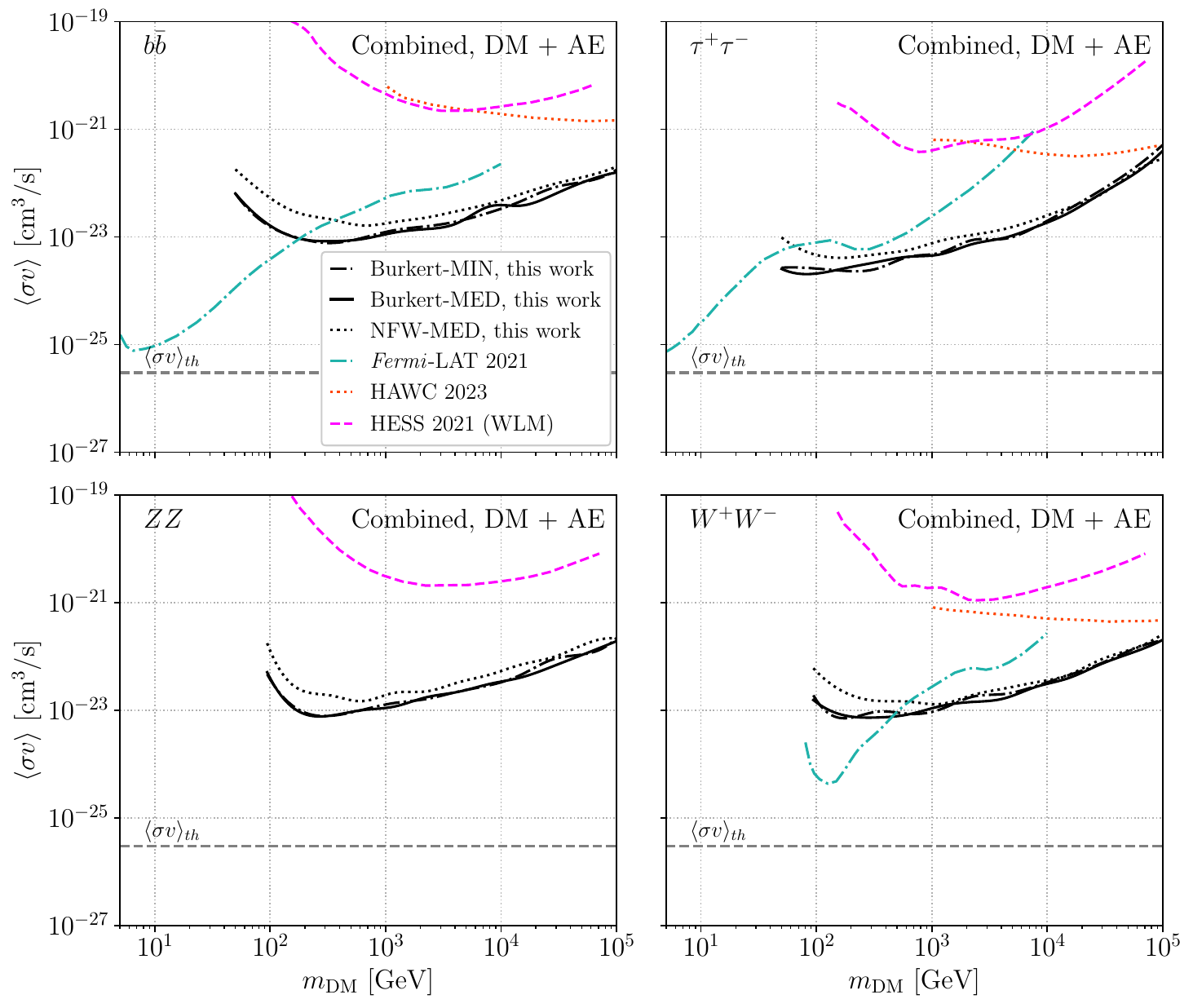}
    \caption{\footnotesize{Projected upper limits on the annihilation cross-section $\langle \sigma v \rangle$ as a function of the DM mass $m_\mathrm{DM}$ for the four different annihilation channels studied in this work (first panel $b\bar{b}$, second $\tau^+\tau^-$, third $ZZ$, and last panel $W^+W^-$ channel). In black are shown our combined results (IC10, NGC6822, WLM, IC1613-North and IC1613-South) for the different DM density profiles (dash-dotted line for Burkert-MIN, solid for Burkert-MED and dotted lines for NFW-MED). We also show the same results by other experiments with dIrrs as targets: ``\textit{Fermi}-LAT 2021'' \cite{2021PhRvD.104h3026G} (blue dash-dotted line), ``HAWC 2023'' \cite{HAWC:2023vtl} (red dotted line) and ``HESS 2021'' \cite{HESS:2021zzm} (pink dashed line). Note that the HESS results are only for one target (WLM), while the \textit{Fermi}-LAT and HAWC upper limits come from a combined analysis of 7 and 31 targets, respectively. As a final comparison, the horizontal grey dashed line represents the thermal relic cross-section $\langle \sigma v \rangle_{\mathrm{th}}$.}}
\label{fig:UL_other_experiments_comparison} 
\end{figure}

In Figure~\ref{fig:UL_other_experiments_comparison}, we compare the CTAO projected upper limits with actual upper limits derived in the literature so far for dIrrs using data from different instruments: ``\textit{Fermi}-LAT 2021''~\cite{2021PhRvD.104h3026G} (blue dash-dotted line), ``HAWC 2023'' \cite{HAWC:2023vtl} (red dotted line), and ``HESS 2021'' \cite{HESS:2021zzm} (pink dashed line). The black lines show our DM+AE combined results (dash-dotted line for Burkert-MIN, solid line for Burkert-MED and dotted line for NFW-MED). We can see that our predictions have the potential to yield the most stringent constraints in the mass range between $\sim100$ GeV and 100 TeV. For example, in the $\tau^+\tau^-$ channel, with respect to ``\textit{Fermi}-LAT 2021''~\cite{2021PhRvD.104h3026G}, we have obtained an improvement of $\sim 2$ at $m_\mathrm{DM} \sim 100$ GeV, but increasing up to a factor of $\sim 50$ at $m_\mathrm{DM} \sim 10$ TeV. At higher masses, our results are a factor $\sim 100$ more constraining than ``HESS 2021'' \cite{HESS:2021zzm} and between $\sim 100$ and a factor of a few with respect to ``HAWC 2023'' \cite{HAWC:2023vtl}. This demonstrates the great potential of this class of targets and the CTAO compared to current-generation GeV-TeV observatories. 

\begin{figure}[t!] 
  \centering 
    \includegraphics[width=0.99\textwidth]{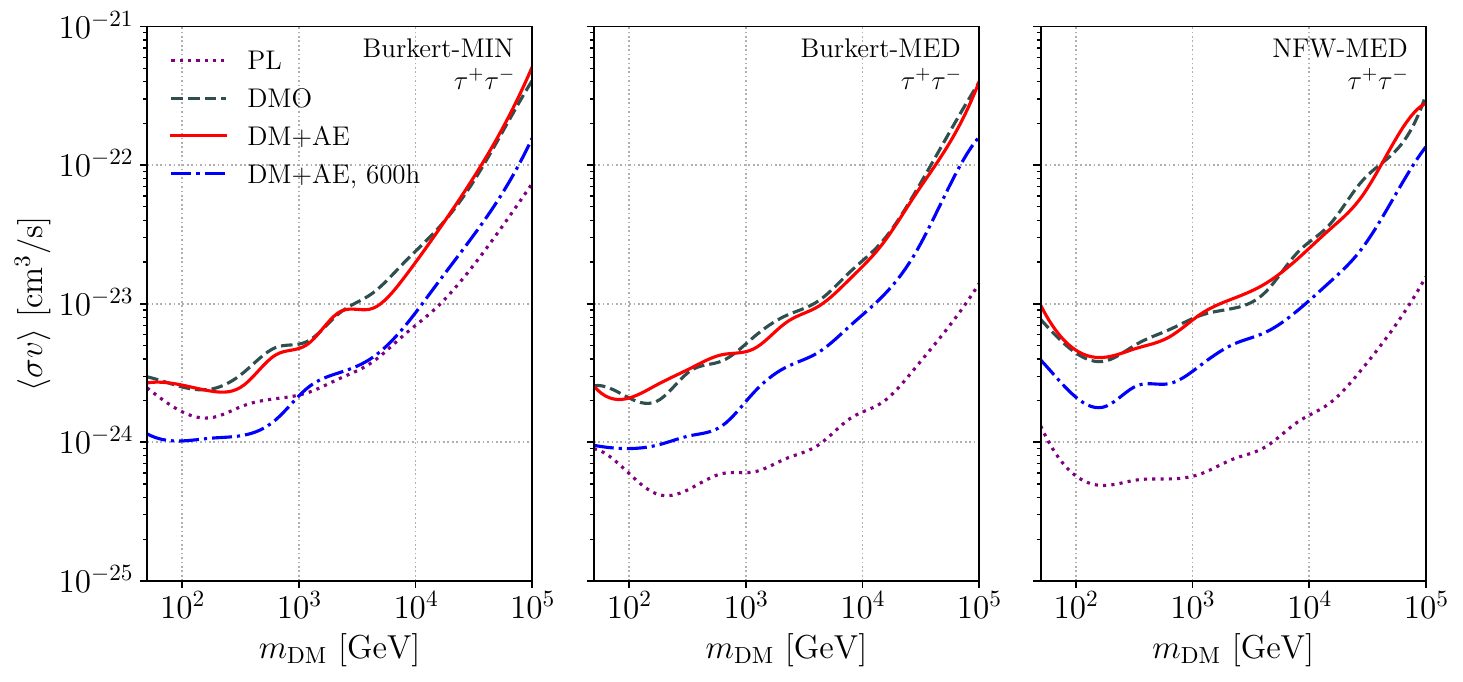}

    \caption{\footnotesize{Projected combined upper limits on the annihilation cross-section $\langle \sigma v \rangle$ as a function of the DM mass $m_\mathrm{DM}$ for the $\tau^+\tau^-$ annihilation, assuming the Burkert-MIN (first panel), Burkert-MED (second panel) and NFW-MED (last panel). We show the quantified error that can arise from mismodeling the gamma-ray flux in the case of a PL analysis (dotted purple line, see Appendix \ref{ap:Point_Like_results}) and a DM-only extended analysis (DMO, dashed grey line, see Appendix \ref{ap:DM_only_results}) from the full DM+AE extended analysis (solid red line). Besides this, we show the improvements of the projected combined constraints when extending the observation up to 600h, evenly distributed between the two best targets (IC10 and NGC6822) (dash-dotted blue line).}}
\label{fig:UL_different_analysis} 
\end{figure}

To quantify the impact from mismodeling the gamma-ray flux in these targets (Appendix \ref{ap:mismodeling_constraints}), we show in Figure~\ref{fig:UL_different_analysis} how the projected combined upper limits change from the full DM + AE modeling (solid red line) compared to a DM-only PL (PL, dotted purple line, see Appendix \ref{ap:Point_Like_results}) and a DM-only extended analysis (DMO, dashed grey line, see Appendix \ref{ap:DM_only_results}) for the $\tau^+\tau^-$ annihilation channel. We can see that, since the whole signal is concentrated in a PL source, using a DM-only PL analysis overestimates the projected constraints by a factor of a few, while performing a DM-only extended analysis yields results almost identical to those of the full DM+AE analysis. We also show with the dash-dotted blue line the improvements of the constraints when extending the observation up to 600h, evenly distributed between the two best targets (IC10 and NGC6822). In this last case, given that we are increasing the observation time per target from 50 h to 300 h, the projected upper limits increase by a factor of $\sim 2-3$. For a final check, in Appendix \ref{ap:Correlation_matrix} we list the possible correlations for all the free parameters in the full analysis: the GDE normalization, the SFR normalization, the DM normalization $\langle \sigma v \rangle$ and the CR instrumental background normalization.

\section{Velocity-dependent cross-section: Sommerfeld enhancement}
\label{sec:sommerfeld}

So far, we have determined the projected upper limits on the DM annihilation cross-section vs DM mass parameter space, assuming that the DM annihilation process happens through a constant annihilation cross-section $\langle \sigma v \rangle$, usually of the order of $\mathcal{O} (10^{-26}) \, \mathrm{cm}^{3} \mathrm{s}^{-1}$. However, the latter is a simplistic approach and can be easily generalized. Indeed, a more complete description for the WIMP vanilla scenario is to extend the annihilation cross-section $\sigma v$ into a power series, from which the thermally averaged cross-section has the following form (in natural units):

\begin{equation}
   \langle \sigma v\rangle = \sum_{i} c_i x^i= c_0 + c_1{x^{-1}} + \mathcal{O}(x^{-2}),
    \label{eq:freeze_out_thermal}
\end{equation}

\noindent where $\langle \sigma v\rangle_\mathrm{{s-wave}}=c_0$ and $\langle \sigma v\rangle_\mathrm{p-wave}=c_1/x$ correspond to the s-wave and p-wave terms of the power series and $x = m_\mathrm{DM}/T$, being $T$ the temperature of the primordial plasma when the DM decouples. Indeed, the coefficients $c_i$ are constant parameters that explicitly depend on the theoretical DM model of interest \cite{2003PhRvD..68j3505C}. Within the zero-order approximation, assuming the DM freeze-out mechanism described by the Boltzmann equation, with the benchmark value for $c_0=\langle \sigma v\rangle _\mathrm{s-wave} \simeq \langle \sigma v\rangle_\mathrm{th} = 3 \times 10^{-26} \, \mathrm{cm^3s^{-1}}$ (for a unique DM candidate with a mass in the GeV-TeV range) the thermal relic abundance $\Omega_{\mathrm{DM}}\simeq 0.27$ \cite{Aghanim:2018eyx} can be correctly recovered \cite{Bertone:2004pz}.

Besides the effective interaction scenario, interesting phenomenology can arise if a WIMP annihilation is produced with a mediator between the DM particles and the SM. We focus on the case of a light scalar mediator  $\phi$, in which the long interaction given by the light mediator leads to the Sommerfeld enhancement $\mathcal{S}$ of the annihilation cross-section, leading to a possible boost of the gamma-ray signal up to several orders of magnitude \cite{2023JCAP...02..004F, 2022JCAP...10..021L, 2010PhRvD..82h3525F}. This enhancement appears in the non-relativistic regime: indeed, following the expansion in velocities of $\sigma v$, the generalization for the Sommerfeld enhancement case is given by \cite{2022JCAP...10..021L, 2019PhRvD..99f1302B}:

\begin{equation}
   \sigma v_{\mathrm{rel}} = \mathcal{S}_\mathrm{s-wave}(\sigma c)_0 + \mathcal{S}_\mathrm{p-wave} (\sigma c)_1 (v_{\mathrm{rel}} / c)^2 + \mathcal{O} ((v_{\mathrm{rel}} / c)^4),
    \label{eq:sommerfeld_enhancement_sigma}
\end{equation}

\noindent where $\mathcal{S}_\mathrm{i} = \mathcal{S}_\mathrm{i}(v_\mathrm{rel})$ are the Sommerfeld enhancement factors, explicitly dependent on the relative velocity of the annihilating particles $v_\mathrm{rel}$, and $(\sigma c)_i$ are constant parameters on which the constraints will be set. As a first approximation, we focus only on the s-wave term, assuming that $\mathcal{S}_\mathrm{s-wave}(\sigma c)_0 \gg     \mathcal{S}_\mathrm{p-wave} (\sigma c)_1 (v_{\mathrm{rel}} / c)^2  $.

In this formalism, the Sommerfeld enhancement factor has the following form for the s-wave approximation \cite{2022JCAP...10..021L}:

\begin{equation}
    \mathcal{S}_\mathrm{s-wave}\left(v_{\mathrm {rel}}\right) \approx \begin{cases} \frac{\pi}{\epsilon_v}\frac{\sinh \left(\frac{2 \pi \epsilon_v}{\epsilon_\phi^*}\right)}{{\cosh \left(\frac{2 \pi \epsilon_v}{\epsilon_\phi^*}\right)-\cos \left(2 \pi \sqrt{\frac{1}{\epsilon_\phi^*}-\frac{\epsilon_v^2}{\epsilon_\phi^{* 2}}}\right)}} & \mathrm { if } \quad \epsilon_v \leqslant \sqrt{\epsilon_\phi^*} \\ \frac{\pi / \epsilon_v}{1-\mathrm{e}^{-\pi / \epsilon_v}} & \mathrm { otherwise,}\end{cases}
\end{equation}

\noindent where the dependency on the velocity appears in the parameter $\epsilon_v = \frac{v_\mathrm{rel}}{\alpha_\mathrm{D}c}$, the dependency on the scalar mediator mass $m_\phi$ (with respect to the DM mass) is given by the parameters $\epsilon_\phi = \frac{m_\phi}{\alpha_\mathrm{D}m_\mathrm{DM}}$ and $\epsilon_\phi^* = \epsilon_\phi\pi^2/6$. The $\alpha_\mathrm{D}$ term is the dark fine-structure constant of the interaction potential. Following \cite{2022JCAP...10..021L}, we fix $\alpha_\mathrm{D} = 0.01$ as a benchmark value. The value of the Sommerfeld enhancement depends directly on the value of the relative velocity $v_\mathrm{rel}$ and the parameter $\epsilon_\phi$. Intuitively, the enhancement only appears in the case of a light scalar; as in the case of a massive mediator ($\epsilon_\phi \gg 1$), the vanilla WIMP case is recovered $\mathcal{S}_\mathrm{s-wave} \simeq 1$. In the intermediate regime of $\epsilon_\phi \ll \epsilon_v \ll 1$, the interaction can be approximated by a Coulomb potential, with the Sommerfeld factor approximated by $\mathcal{S}_\mathrm{s-wave} \simeq \pi / \epsilon_v \propto 1/v$. Following this behavior on $1/v$, in the case of large velocities $\epsilon_v \gg 1$, it can be shown that there is no enhancement. Finally, at lower velocities $\epsilon_v \ll \epsilon_\phi \ll 1$, it is where the resonances appear in the Sommerfeld factor and the DM signal is enhanced the most. This resonances appear at specific values of $\epsilon_\phi = \frac{6}{\pi^2n^2}$ (where $n$ is an integer), giving $\mathcal{S} _\mathrm{s-wave} \simeq 1/(\epsilon_v n)^{2} \propto 1/v^2$ \cite{2022JCAP...10..021L}.

Since the total s-wave annihilation cross-section $\mathcal{S}_\mathrm{s-wave}(\sigma c)_0$ depends on the relative velocity $v_{\mathrm{rel}}$ of the annihilating particles via the Sommerfeld enhancement factor, this directly translates to a modification in the computation of the J-factor. In fact, the integration of the phase-space density $f(\vec{r}, \vec{v_i})$ of the two DM annihilating particles does not give directly the DM density profile squared as in Equation~\ref{eq:JFac_formula}, but a more general J-factor:

\begin{equation}
    J_{\mathrm{S}}(\Delta \Omega)=\int_{\Delta \Omega} \mathrm{d} \Omega \int_{\mathrm {l.o.s. }} \mathrm{d}l \int \mathrm{~d}^3 \vec{v}_1 \int \mathrm{~d}^3 \vec{v}_2 f\left(r(l, \Omega), \vec{v}_1\right) f\left(r(l, \Omega), \vec{v}_2\right) \mathcal{S}_\mathrm{s-wave}\left(\frac{v_{\mathrm {rel }}}{2}\right)
    \label{eq:sommerfeld_enhancement_Jfactor}
\end{equation}

Note that, in the absence of the Sommerfeld enhancement ($\mathcal{S}_\mathrm{s}\left(\frac{v_{\mathrm {rel }}}{2}\right) = 1$), or a non-dependency on the velocities in general, the integral of the phase spaces is again the DM density profile over the mass squared, recovering the standard J-factor formula (Equation~\ref{eq:JFac_formula}): 

\begin{equation}
\int \mathrm{~d}^3 \vec{v}_1 \int \mathrm{~d}^3 \vec{v}_2 f\left(r(s, \Omega), \vec{v}_1\right) f\left(r(s, \Omega), \vec{v}_2\right) = \rho_{\mathrm{DM}}(r)^2
\end{equation}

It is relevant to stress that a velocity-dependent cross-section does not necessarily imply the need for the annihilation cross-section normalization $(\sigma c)_0$ to be of the order of $\langle \sigma v \rangle_{\mathrm{th}} = 3 \times 10^{-26} \ \mathrm{cm}^3 \mathrm{s}^{-1}$. Generally, modulo $\mathcal{O}(1)$ factors, the tree-level cross-section of the Sommerfeld case has the following form $(\sigma c)_0 \simeq \frac{\pi\alpha_\mathrm{D}^2}{m_\mathrm{DM}^2}$ \cite{2010PhRvD..82h3525F}. Here, we choose to use the standard value $\langle \sigma v \rangle_{\mathrm{th}}$ as a reference, allowing for comparison with the literature \cite{Abazajian:2011ak,Slatyer:2011kg, Lu:2017jrh}. During freeze-out in early times, the Sommerfeld enhancement is negligible, and with $(\sigma c)_0 \simeq 3 \times 10^{-26} \ \mathrm{cm}^3 \mathrm{s}^{-1}$ the correct value of  $\Omega_{\mathrm{DM}}h^2$ can be recovered \cite{2010PhRvD..82h3525F}. However, given the dependency of $(\sigma c)_0$ on $\alpha_\mathrm{D}$ and $m_\mathrm{DM}$, this only holds for certain values of those parameters. Taking this into account, the correct values of $\alpha_\mathrm{D}$ that yield the correct thermal relic abundance can be of the order of $\sim 0.01$ or different, depending on the decoupling temperatures from the primordial plasma and masses of the DM and scalar particles $\phi$ \cite{2010PhRvD..82h3525F}. For simplicity and to ease the comparison with the literature, we keep the value $\alpha_\mathrm{D} = 0.01$ fixed as stated before.

The two main parameters that we constrain are, therefore, $\epsilon_\phi$ and the normalization parameter $(\sigma c)_0$. For high values of $\epsilon_\phi > 10$ (massive mediators), the Sommerfeld effect is negligible, but resonances appear in $J_{\mathrm{S}}(\Delta \Omega)$ for lower values of $\epsilon_\phi$ \cite{2022JCAP...10..021L}. Very similarly to Equation~(\ref{eq:flux_annih}), with these definitions, the gamma-ray flux is given by the expression: 

\begin{equation}
  \frac{d \Phi_{\text{DM}}}{d E}= \frac{(\sigma v)_0}{8 \pi m_{\mathrm{DM}}^{2}} \sum_{i}^{\mathrm{channels}} \mathrm{BR}_{i}  \frac{d N_{i}}{d E} J_\mathrm{S}(\Delta \Omega)
  \label{eq:flux_annih_v_dependent} 
\end{equation}

An explicit computation of this process is beyond the scope of this paper (for a more comprehensive review, see \cite{2022JCAP...10..021L}, and for the explicit computation of the J-factor see \cite{2023JCAP...02..004F}). For our purposes, we use the results of the phase-space integration shown in \cite{2022JCAP...10..021L} for the galaxies WLM, NGC6822 and IC10 (Figure 4 of the same Reference). Their corresponding enhanced J-factors are determined for a $0.5^\circ$ integration angle, including the main halo and the subhalos' contributions. The inclusion of subhalos can give even greater enhancements on the J-factors, in which case the dIrrs can exceed the dSphs' J-factors \cite{2022JCAP...10..021L}. Finally, we can use our previous results (Figure~\ref{fig:tau_channel_UL_all}, Burkert-MED) to rescale the limits and rule out a part of the parameter space ($\epsilon_\phi$, $m_{\mathrm{DM}}$ and $(\sigma c)_0$) for the three mentioned galaxies. 

\begin{figure}[t!] 
  \centering 
    \includegraphics[width=0.85\textwidth]{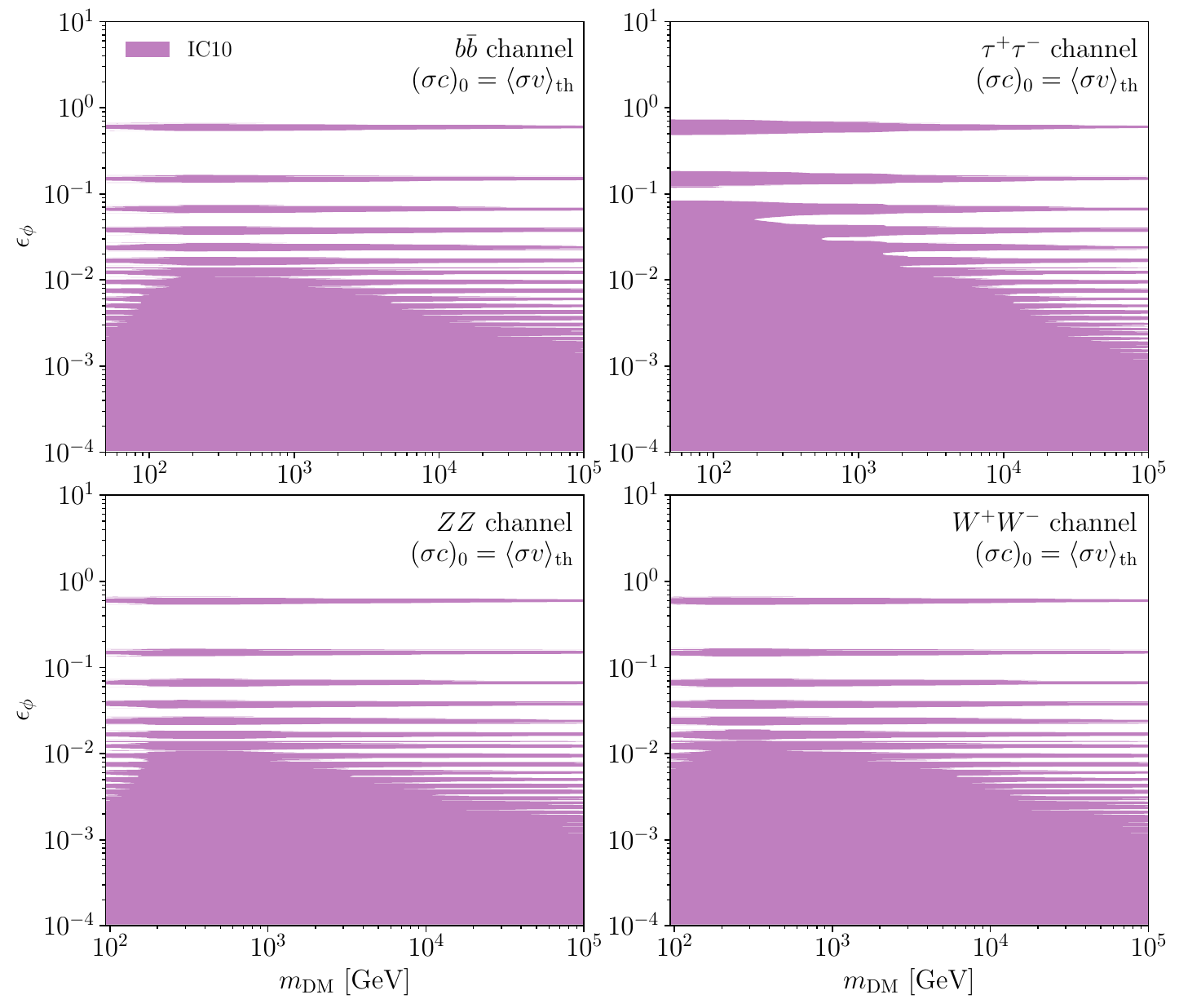}
  
    \caption{\footnotesize{CTAO sensitivity predictions for the $\epsilon_\phi - m_\mathrm{DM}$ parameter space for the IC10 galaxy, the most constraining galaxy. The value of the thermal relic cross-section $\langle \sigma v \rangle_{\mathrm{th}} = 3 \times 10^{-26} \ \mathrm{cm}^3 \mathrm{s}^{-1}$ has been adopted as the normalization parameter $(\sigma c)_0$. We show results for the $b\bar{b}$ (upper left panel), $\tau^+\tau^-$ (upper right panel), $ZZ$ (lower left) and $W^-W^+$ (lower right) annihilation channel. See Section \ref{sec:sommerfeld} for details.}}
\label{fig:velocity_dependent_epsilonlimits} 
\end{figure}

\begin{figure}[t!] 
  \centering 
    \includegraphics[width=0.85\textwidth]{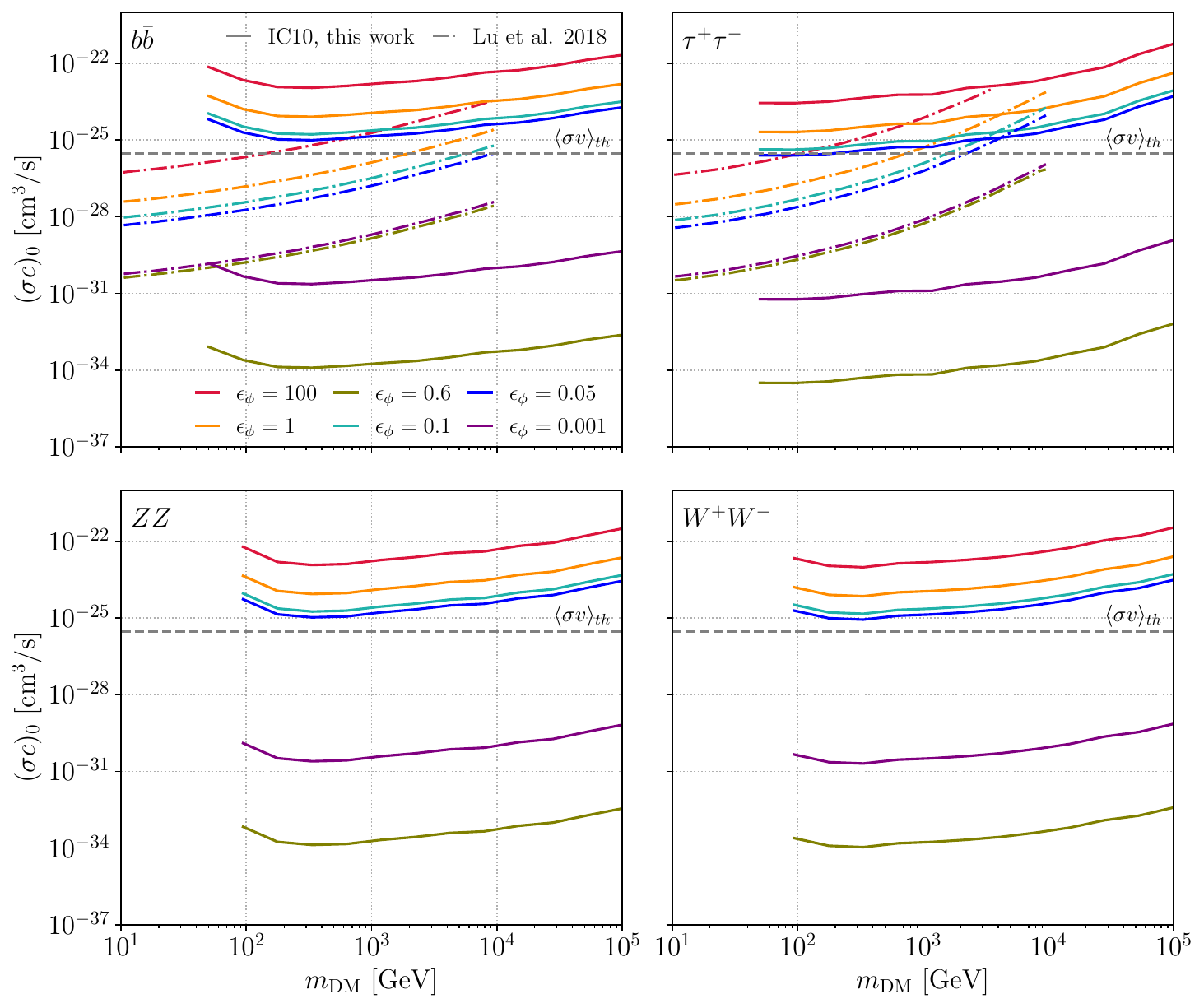}
    \caption{\footnotesize{CTAO predicted upper limits on the $(\sigma c)_0$ parameter (see Section~\ref{sec:sommerfeld}), adopting IC10 as the target, and for different values of $\epsilon_\phi$: 100 (red), 1 (orange), 0.6 (green), 0.1 (light blue), 0.05 (blue) and 0.001 (purple). The dot-dashed lines in the top panels represent the upper limits computed with \textit{Fermi}-LAT dSphs for the same values of $\epsilon_\phi$ \cite{Lu:2017jrh}. The results are for the $b\bar{b}$ (left panel), $\tau^+\tau^-$ (right panel), $ZZ$ (lower left) and $W^-W^+$ (lower right) annihilation channels. Note that the cases $\epsilon_\phi = 0.6$ and $0.001$ correspond to resonances in the Sommerfeld factor. For a benchmark value, the horizontal grey dashed line represents the thermal relic cross-section $\langle \sigma v \rangle_{\mathrm{th}}$.}}
\label{fig:velocity_dependent_sigmalimits} 
\end{figure}

Figure~\ref{fig:velocity_dependent_epsilonlimits} shows the $\epsilon_\phi - m_\mathrm{DM}$ parameter space that is projecetd to be ruled out at a $95\%$ C.L. by the CTAO for the IC10 galaxy for the case of the $\tau^+\tau^-$ (upper left panel), $b\bar{b}$ (upper right), $ZZ$ (lower left) and $W^-W^+$ (lower right) annihilation channels, fixing $(\sigma c)_0 = \langle \sigma v \rangle_{\mathrm{th}} = 3 \times 10^{-26} \mathrm{cm}^3 \mathrm{s}^{-1}$. Furthermore, we note that in the Sommerfeld enhancement case, IC10 also yields the best constraints, followed by NGC6822. In Appendix \ref{ap:Sommerfeld_rest_galaxies} we show how these projected constraints compare with NGC6822 and WLM. In Figure~\ref{fig:velocity_dependent_sigmalimits}, we show with the solid lines the $(\sigma c)_0 - m_\mathrm{DM}$ $95\%$ C.L. projected upper limits for six different values of $\epsilon_\phi$: 100 (red), 1 (orange), 0.6 (green), 0.1 (light blue), 0.05 (blue) and 0.001 (purple). The dot-dashed lines represent the upper limits computed with \textit{Fermi}-LAT data from dSphs~\cite{Lu:2017jrh} for the same values of $\epsilon_\phi$. Note that some of these particular values of $\epsilon_\phi$ correspond to resonances in the J-factor computation \cite{2022JCAP...10..021L}, hence the highly constraining curves of $\epsilon_\phi = 0.6$ or $0.001$. We can see that at the resonances our constraints are better than those from dSphs \cite{Lu:2017jrh}, whereas when $\epsilon_\phi \gg 1$ (red lines) there is no enhancement and the constraints from Section \ref{sec:results_DM} are recovered. In general, at resonances, the projected constraints form dIrrs improve the dSphs limits \cite{Lu:2017jrh}; but outside the resonances, as the J-factor given by dSphs is greater in this regime, the corresponding constraints improve the ones from dIrrs. We note, though, that the J-factors we are using include the contribution from subhalos,  while no substructure boost was included in \cite{Lu:2017jrh} for dSphs. As a comparison, in the case of the vanilla WIMP DM, substructures can enhance the J-factors up to a factor 10 for dIrrs and an order of $\sim20-40\%$ for the case of dSphs~\cite{2017MNRAS.466.4974M}. A similar effect happens in the Sommerfeld enhancement case: the total J-factor value for IC10 or NGC6822 is higher than that of the classical and ultra-faint dSphs (e.g., Draco, Sculptor or Reticulum II). The reason for this is that the boost factor created by the inclusion of subhalos is, in general, more than 1 order of magnitude greater for dIrrs than dSphs \cite{2022JCAP...10..021L}, therefore yileding greater J-factors in total.

\section{Conclusions}
\label{sec:conclusions}

In this work, we have identified and performed simulations of the four best dIrrs for observation with the CTAO. The main goal was to investigate whether a gamma-ray signal detection in CTAO data should be expected from these objects or not (either of astrophysical origin or DM-related). DIrrs are rotation-supported galaxies located in the Local Group, with masses in the range $\sim 10^{10} \ \mathrm{M_\odot}$ that are expected to host not only large reservoirs of DM but also SFRs. Because of this, we included in our DM analyses not only the former but also the contribution of the latter to the total expected gamma-ray flux as given by state-of-the-art models. Since these models are currently suffering from large uncertainties, we include these in the analysis as well by adopting different, yet realistic levels of SFR-induced flux. As for the modeling of the DM component, we have considered different DM density profiles, namely Burkert and NFW, representative of cored and cuspy profiles, respectively. We do so given the open debate on the precise slope of the DM profile in the center of these targets \cite{Burkert:1995yz, Karukes:2017kne, 2021PhRvD.104h3026G}. In addition to the smooth DM component, we included DM halo substructure in the DM flux calculations as well, since subhalos are expected to enhance the annihilation signal significantly in dIrrs~\cite{2017MNRAS.466.4974M}. 

Although the four candidates have not been selected for their SFR-induced emission, we showed that a detection of dIrrs as conventional astrophysical emitters, i.e., purely SFR-induced, is not expected. Indeed, we find that the corresponding $\mathcal{SFR}$ needed for a detection with the CTAO is of the order of $\sim 1 \ \mathrm{M_\odot/yr}$, with the best candidates being NGC6822 and IC10. The expected luminosity value is from 3 to 5 orders of magnitude greater than the benchmark from current SFR models, as shown in Figure~\ref{fig:Flux_TS_SFR_detection}. For the best candidate, NGC6822, it corresponds to a factor of $\sim 10$ above the uncertainties, whereas for a $3\sigma$ hint the needed factor is reduced to $\sim 7$. Besides SFR detection, we have also studied the expected sensitivity of the CTAO to the foreground gamma-ray flux associated with the GDE within the ROI of each dIrr. Interestingly, and as shown in Figure~\ref{fig:GDE_TS_histograms}, the GDE can be detected consistently in each of the 100 performed simulations with a high TS (up to 1000) for the cases of IC10 and NGC6822 (the two galaxies in our sample that lie closer to the GP). No GDE detection is expected in the other two dIrrs. 

We evaluated the sensitivity of the CTAO to DM-induced signals: our analysis consistently yields no detection from any of the considered dIrrs across all performed simulations. 
In the absence of an expected DM signal, we have computed predicted limits on the WIMP parameter space under different observational setups. We include as background in our analysis the intrinsic emission of the galaxies due to the SFR. Spatially very concentrated in a small region of $\sim 0.1^\circ$, our SFR benchmark emission is of the same order of magnitude as the DM emission, although the uncertainties can make the flux up to $\sim2$ orders of magnitude higher (see Figure~\ref{fig:dIrr_All_spectral_fluxes}). Also, considering that two target galaxies (IC10 and NGC6822) are close to the GP, we have also included the modeling of the GDE. We show the results in Figure~\ref{fig:tau_channel_UL_all} for the individual projected constraints in the $b\bar{b}$ channel, and a comparison of the combined analysis with the four channels considered in Figure~\ref{fig:all_channel_combined}. As a final step, we have checked the dependance of the results on the mismodeling of the emission, concluding that the astrophysical intrinsic emission of dIrrs, in the form of the SFR emission, does not play an important role in the $\langle \sigma v \rangle$ upper limits even when considering the maximum flux allowed by the uncertainties of our modeling (Appendix \ref{ap:UL_no_SFR_fitted}, Figure~\ref{fig:UL_no_SFR_fitted}), as in the case of dSphs. Regarding the GDE, in the two galaxies where the GDE contribution is relevant compared to the total counts (IC10 and NGC6822), the template analysis can extract that part of the emission from the total, and therefore, it should not interfere with the DM sensitivity prospects. However, in the other two galaxies far from the GP (WLM and IC1613), since the GDE does not play an important role in the total counts, the DM constraints are also unaffected by the non-detection of this emission.

\begin{figure}[t!] 
  \centering 
    \includegraphics[width=0.6\textwidth]{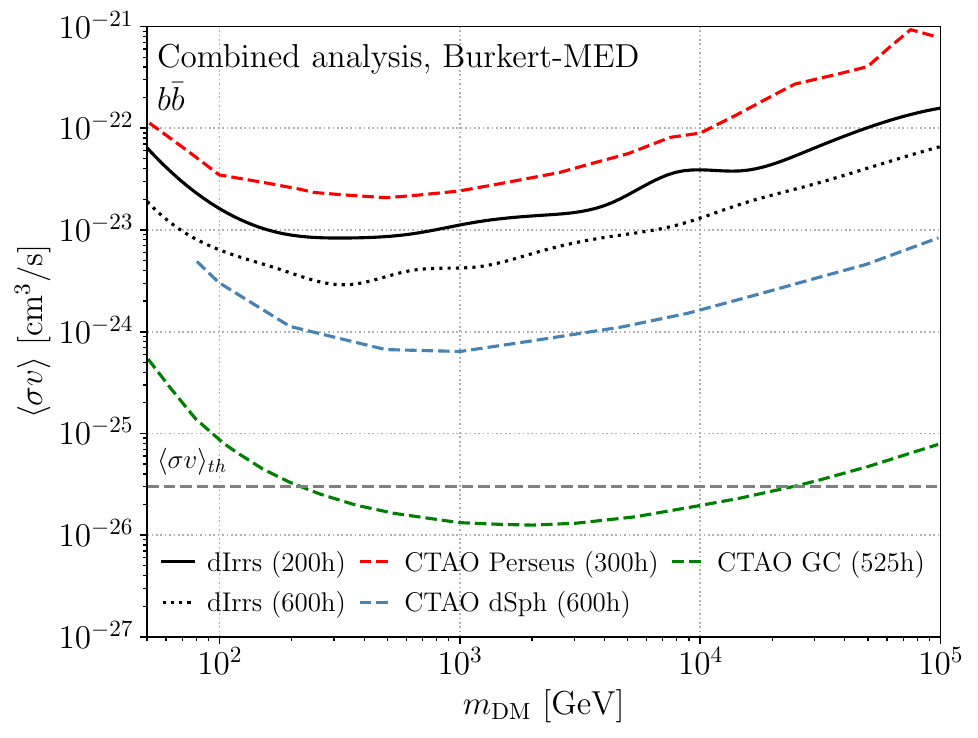}
    \caption{\footnotesize{Projected combined upper limits on the annihilation cross-section $\langle \sigma v \rangle$ as a function of the DM mass $m_\mathrm{DM}$ for the $b\bar{b}$ channel. We show, in black, the projected combined upper limits of dIrrs for our benchmark DM density profile model Burkert-MED, totaling up to 200h of observation time among all the galaxies (solid line), and increasing the total observational simulated time to 600 hours and restricting the observation to the two dominant galaxies (IC10 and NGC6822, 300 hours per target). The dashed lines represent the limits by other CTAO simulations: the Perseus galaxy cluster \cite{CTAConsortium:2023yak} (red), dSphs \cite{2025MNRAS.tmp.1704A} (blue) and the GC \cite{Acharyya_2021} (green). As a final comparison, the horizontal grey dashed line represents the thermal relic cross-section $\langle \sigma v \rangle_{\mathrm{th}}$.}}
\label{fig:UL_CTAO_comparison} 
\end{figure}

In order to put our results into a more general context, in the Figure~\ref{fig:UL_CTAO_comparison}, we show with the solid black line the DM sensitivity prospects for the $b\bar{b}$ channel and benchmark DM density profile model Burkert-MED, compared with other DM predictions recently performed within the CTAO Consortium (dashed lines), namely for the Perseus galaxy cluster\footnote{With a total of 300h of observation time, we take the projected constraints computed with their benchmark DM density profile modeling Burkert-MED.} \cite{CTAConsortium:2023yak} (red), dwarf spheroidal galaxies \cite{2025MNRAS.tmp.1704A} (blue)\footnote{For dSphs, we show the best projected constraints derived in \cite{2025MNRAS.tmp.1704A}, computed from their two best candidates with a total of 600h of observation time evenly distributed between them assuming an Einasto DM density profile.}, and the GC\footnote{With a total of 525h of observation time, we take their projected constraints computed with the Einasto DM density profile.} \cite{Acharyya_2021} (green). The dIrrs projected upper limits are 2-5 times more stringent than those expected from observations of Perseus. This is so despite the fact that for Perseus, a total of 300 hours observation time was assumed in \cite{CTAConsortium:2023yak}, while in this work only 50 hours per target is considered. Indeed, it is worth noting that the two most competitive dIrrs alone (IC10 for the Burkert-MIN and Burkert-MED profiles and NGC6822 for the NFW-MED profile) provide projected limits that are already more constraining than those derived for Perseus. This is a direct consequence of two factors: the comparison of the J-factor values and the distance to Earth, being the dIrrs closer targets. The best projected constraints are from the GC, reaching the thermal relic they are between 2 and 3 orders of magnitude better than our results. Finally, dIrrs projected upper limits are about one order of magnitude worse than those expected from dSphs (at $m_\mathrm{DM} \sim 1$ TeV) \cite{2025MNRAS.tmp.1704A}. However, it must be noted that the dSphs results were obtained assuming an observation time of 600h. Thus, in order to make a fairer comparison between these results, we show with the dotted black line the dIrrs projected limits simulating an observation time of 600 hours as well, evenly distributed among IC10 and NGC6822 alone. For this exercise, we expect the dIrrs predicted upper limits to scale roughly as $\propto t_{\mathrm{obs}}^{1/2}$. Indeed, the 600 hours combined results (300 hours per galaxy) improve the results obtained for 50 hours by approximately a factor $\sim (300/50)^{1/2} \simeq 2.5$. In this new observation scenario, the results get more competitive at lower masses, with differences of about a factor $\sim 2$ at $m_\mathrm{DM} \sim 100 $ GeV, yet at higher DM masses, the difference is still almost one order of magnitude. We recall that the $t_{\mathrm{obs}}^{1/2}$ scaling implicitly assumes the typical behavior of the signal-to-noise. At higher masses, though, where the limits may be statistics-limited for the CTAO, a significantly larger photon statistics may lead to more significant improvements of the limits, even $\propto t_{\mathrm{obs}}$; see e.g.~\cite{2016PhR...636....1C} for \textit{Fermi}-LAT.

Finally, we have also investigated the expected sensitivity of the CTAO to DM annihilation in dIrrs in the case that a light scalar mediator between the DM and SM particles exists. In such a scenario, an interesting phenomenology arises due to the Sommerfeld enhancement, in which the J-factor can be enhanced up to eight orders of magnitude in some cases \cite{2023JCAP...02..004F, 2022JCAP...10..021L, 2010PhRvD..82h3525F}. A new parameter $\epsilon_\phi$ appears, which controls the impact of the effect. We show the fraction of the $\epsilon_\phi - m_\mathrm{DM}$ parameter space that could be ruled out after performing the proposed dIrrs DM analysis in Figure~\ref{fig:velocity_dependent_epsilonlimits}. Interestingly, as it can be seen in Figure~\ref{fig:velocity_dependent_sigmalimits}, we found that the predicted upper limits can become more constraining than those from dSphs when the Sommerfeld enhancement is taken into account, due to its important dependence on the properties of the subhalo population within hosts (subhalos being much more relevant in the case of dIrrs compared to dSphs). Besides this, it must be noted that the dSphs analysis is performed at energies covered by \textit{Fermi}-LAT, lower than those of this work.

This work shows the potential of dIrrs as targets for gamma-ray analysis. Since only very few dwarf galaxies have been detected with gamma rays, it is imperative to study the sensitivity prospects to these targets. We propose IC10 and NGC6822 for the best dIrr galaxy targets, both as astrophysical sources and as DM indirect searches targets. Although we do not expect the CTAO to detect dIrrs as astrophysical sources, we have found that the $\mathcal{SFR}$ needed is about $\sim 1 \ \mathrm{M_\odot/yr}$, setting a benchmark value for other SFR targets. Finally, we have shown that the expected sensitivity to a DM signal is complementary and competitive with other targets, such as galaxy clusters, as these two galaxies alone are expected to yield better constraints than the planned Perseus survey. Furthermore, in the Sommerfeld enhancement case, dIrrs constraints can also surpass dSphs constraints.

\acknowledgments

The authors would like to thank Judit Pérez-Romero, Stela Adduci Faria, Tomohiro Inada, Jonathan Biteau, Sergio Hernández-Cadena, Pedro de la Torre Luque, Daniele Gaggero, Thomas Lacroix and all the DAMASCO group\footnote{\href{https://projects.ift.uam-csic.es/damasco/}{https://projects.ift.uam-csic.es/damasco/}} for fruitful discussions. 
The work of JZP, VG and MASC was supported by the grants PID2024-155874NB-C21, PID2022-139841NB-I00, PID2021-125331NB-I00 and CEX2020-001007-S, all funded by MCIN\slash AEI\slash10.13039\slash501100011033 and by ``ERDF A way of making Europe''. Authors also acknowledge the MultiDark Network, ref. RED2022-134411-T. JZP's contribution to this work has been supported by \textit{FPI Severo Ochoa} PRE2021-099137 grant.  The simulations and numerical computations made in this paper have been performed in the Hydra HPC cluster at the Instituto de Física Teórica (IFT UAM-CSIC) and the Centro de Computación Científica-Universidad Autónoma de Madrid (CCC-UAM). VG thanks the University San Pablo CEU travel grant for the dissemination of this science. This research has made use of the CTAO instrument response functions provided by the CTA Consortium and Observatory\footnote{\href{https://www.ctao-observatory.org/science/cta-performance/}{https://www.ctao-observatory.org/science/cta-performance/}}  (version prod5 v0.1 \cite{cherenkov_telescope_array_observatory_2021_5499840}).

\appendix
\section{SFR emission modeling}
\label{ap:SFR_Appendix}

To compute the SFR emission, similarly to what has been done in HAWC 2023 \cite{HAWC:2023vtl}, we first compute the $\mathcal{SFR}$ of the galaxies from their stellar mass $M_*$ following \cite{2017ApJ...851...22M}:

\begin{equation}    
    \log (\mathcal{SFR} \, \mathrm{[M_\odot/yr]}) = (-10.75\pm0.53) + (1.04\pm0.06) \log(M_{*}/\mathrm{M_\odot}) \pm 0.34,
    \label{eq:M_to_SFR}
\end{equation}

\noindent where the stellar mass values have been extracted from \cite{Oh:2015xoa} (WLM, IC10, IC1613) and \cite{2012AJ....144....4M} (NGC6822).

From this value, we can now compute the CR Power ($P_{\mathrm{CR}} = P_{\mathrm{CR \, p}} + P_{\mathrm{CR \,e}}$, Equation~\ref{eq:SFR_to_PCR}), and, finally, obtain the gamma-ray luminosity $L_\gamma$ (between 100 GeV and 100 TeV) given by Martin14 \cite{Martin:2014nia}:

\begin{align}
 P_{\mathrm{CR \, p,e}} = Q_{\mathrm{MW \, p,e}} \frac{\mathcal{SFR}}{1.9 \mathrm{M}_\odot \, \mathrm{yr}^{-1}} \nonumber \\
    \mathrm{where} \, Q_{\mathrm{MW \, p}} = 7.10 \times 10^{40} \mathrm{erg} \, \mathrm{s}^{-1} \nonumber \\
    \mathrm{and} \, Q_{\mathrm{MW \, e}} = 1.05 \times 10^{39} \mathrm{erg} \, \mathrm{s}^{-1} 
    \label{eq:SFR_to_PCR}
\end{align}  

\begin{equation}
    \log(L_\gamma \mathrm{[erg/s]}) = \beta \log{(P_{\mathrm{CR}} \mathrm{[erg/s]})} + \delta
    \label{eq:PCR_to_Lgamma}
\end{equation}

Where $\beta$ and $\delta$ are the best-fit parameters of the four models presented in Martin14 \cite{Martin:2014nia} and $Q_{\mathrm{MW \, p,e}}$ is the proportional factor that relates the $\mathcal{SFR}$ with the proton/electron Power $P_{\mathrm{CR \, p,e}}$. The uncertainties of the SFR emission model are divided in two parts: the specific uncertainty in Equation~\ref{eq:M_to_SFR} and the scatter in the $\beta$ and $\delta$ parameters given by the four models in Martin14 (Equation~\ref{eq:PCR_to_Lgamma}). With these, we define our benchmark luminosity as the mean value of the range of the four models of Martin14 with the $\mathcal{SFR}$ computed with Equation~\ref{eq:M_to_SFR} without uncertainties. Finally, the differential flux of the emission is modeled with a power law of spectral index 2.5 whose normalization is computed from $L_\gamma$.

\begin{figure}[t!] 
  \centering 
  \includegraphics[width=0.49\textwidth]{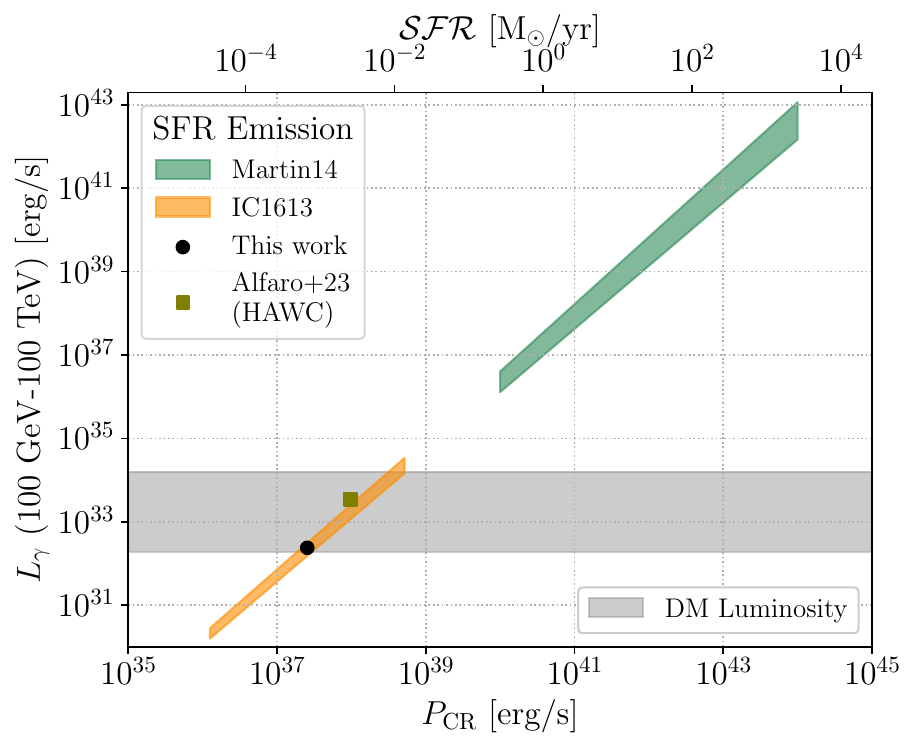}
  \includegraphics[width=0.49\textwidth]{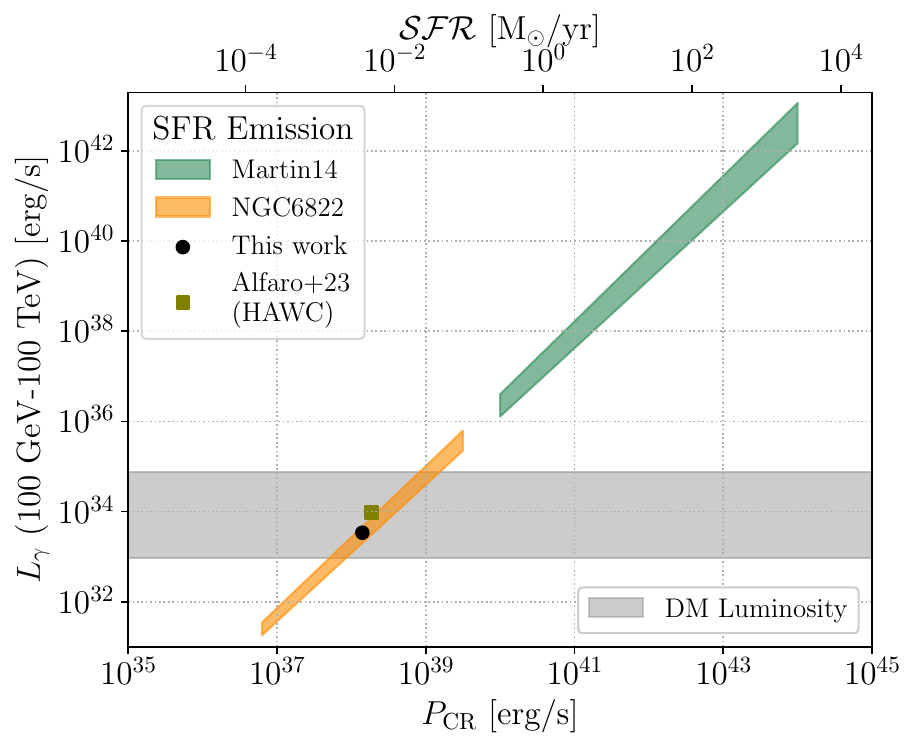}
    \caption{\footnotesize{Expected SFR luminosity following the Martin14 model \cite{Martin:2014nia} (green band), as a function of the $\mathcal{SFR}$ and CR power ($P_\mathrm{CR}$), for the galaxies IC1613 (left panel) and NGC6822 (right panel) compared with the DM luminosity (grey band, for the channels and range of DM masses $m_{\mathrm{DM}}$ used in this work). The yellow bands show the uncertainties of the SFR emission luminosity when estimating the $\mathcal{SFR}$ with \cite{2017ApJ...851...22M} and the gamma-ray emission extrapolating the Martin14 models. The black dot represents the benchmark value used in this work and the green square is the estimation made in HAWC 2023 \cite{HAWC:2023vtl}. See Appendix \ref{ap:rest_of_figs}, Figure~\ref{fig:SFR_vs_DM_appendix}, for the Figures corresponding to WLM and IC10.}}
\label{fig:SFR_vs_DM} 
\end{figure}

In Table~\ref{tab:dIrr_SFR_1} we list, for the four dIrrs targets, the stellar mass $M_*$, and the uncertatinty ranges (MIN, benchmark and MAX) for the $\mathcal{SFR}$\footnote{We note that in a recent work \cite{2025arXiv250910440H}, the listed $\mathcal{SFR}$ value for IC10 and NGC6822 are $0.3 \ \mathrm{M_\odot/yr}$ and $0.02 \ \mathrm{M_\odot/yr}$, respectively. While the value of NGC6822 lies in our uncertainty range, the updated IC10 value is slightly greater than the range of values presented in Table~\ref{tab:dIrr_SFR_1}. However, the corresponding luminosity in gamma rays ($1.56-4.96 \times 10^{36} \ \mathrm{erg/s}$) is not enough for a $3\sigma$ detection (see Figure~\ref{fig:Flux_TS_SFR_detection}).} and gamma-ray luminosity $L_\gamma$ integrated from 100 GeV to 100 TeV. In Figure~\ref{fig:SFR_vs_DM}, we show the expected luminosity and the corresponding $\mathcal{SFR}$ and $P_{\mathrm{CR}}$, with the uncertainties of the SFR gamma-ray luminosity from Equation~\ref{eq:M_to_SFR} and the four models from Martin14 \cite{Martin:2014nia} as the colored yellow band, and the benchmark SFR luminosity used in this work as the black point, for the case of the IC1613 galaxy (left panel), one of the target with lowest SFR emission, and NGC6822 galaxy (right panel), which together with IC10 are the brightest for this emission. Regarding the uncertainties, in the Figure, it can be seen that the most relevant one is from the estimation of the $\mathcal{SFR}$ (Equation~\ref{eq:M_to_SFR}). Therefore, using a better estimation of the $\mathcal{SFR}$ might reduce the uncertainties significantly. We also show, in grey, the expected DM luminosity for the channels and range of DM masses used in this work, and with the green band the Martin14 models \cite{Martin:2014nia}. Here it becomes clear that we had to extrapolate the Martin14 models to lower values of $\mathcal{SFR}$ to estimate the emission for our set of dIrrs. This further emphasizes that the SFR gamma-ray emission expected for the selected dIrrs is significantly lower with respect to the current detected values.

\begin{table}[t!]
    \begin{center}
    \begin{tabular}{|c|c|ccc|ccc|}
\hline \hline
\multirow{2}{*}{\textbf{Galaxy}} & \multirow{2}{*}{$M_*$ {[}$\mathrm{M}_\odot${]}} & \multicolumn{3}{c|}{$\mathcal{SFR}$ {[}$\times 10^{-3} \mathrm{M}_\odot$/yr{]}} & \multicolumn{3}{c|}{$L_{\gamma}$ {[}erg/s{]}}                                                                   \\ \cline{3-8} 
                                 &                                                 & \multicolumn{1}{c|}{MIN}       & \multicolumn{1}{c|}{Bench.}     & MAX       & \multicolumn{1}{c|}{MIN}                   & \multicolumn{1}{c|}{Bench.}             & MAX                   \\ \hline \hline
\textbf{IC10}                    & $1.18 \times 10^{8}$                            & \multicolumn{1}{c|}{0.20}      & \multicolumn{1}{c|}{4.42}          & 100     & \multicolumn{1}{c|}{$2.30 \times 10^{31}$} & \multicolumn{1}{c|}{$4.44 \times 10^{33}$} & $8.35 \times 10^{35}$ \\ \hline
\textbf{IC1613}                  & $1.94 \times 10^{7}$                            & \multicolumn{1}{c|}{0.03}      & \multicolumn{1}{c|}{0.68}          & 13.7     & \multicolumn{1}{c|}{$2.21 \times 10^{31}$} & \multicolumn{1}{c|}{$2.37 \times 10^{32}$} & $4.80 \times 10^{34}$ \\ \hline
\textbf{WLM}                     & $1.23 \times 10^{7}$                            & \multicolumn{1}{c|}{0.02}      & \multicolumn{1}{c|}{0.42}          & 8.30      & \multicolumn{1}{c|}{$7.97 \times 10^{29}$} & \multicolumn{1}{c|}{$1.13 \times 10^{32}$} & $1.55 \times 10^{34}$ \\ \hline
\textbf{NGC6822}                 & $1.00 \times 10^{8}$                            & \multicolumn{1}{c|}{0.17}      & \multicolumn{1}{c|}{3.72}          & 84.2     & \multicolumn{1}{c|}{$1.79 \times 10^{31}$} & \multicolumn{1}{c|}{$3.39 \times 10^{33}$} & $6.21 \times 10^{35}$ \\ \hline
        \end{tabular}
        \caption{\footnotesize{Parameters used to model the SFR spectral emission include the stellar mass $M_*$ and the uncertainty ranges (minimum (MIN), benchmark (Bench.), and maximum (MAX)) for the $\mathcal{SFR}$ and gamma-ray luminosity $L_\gamma$ (from 100 GeV to 100 TeV). The range of values for $\mathcal{SFR}$ reflects uncertainties computed with Equation~\ref{eq:M_to_SFR}. This corresponds to the yellow band in Figure~\ref{fig:SFR_vs_DM} for NGC6822. The median value is used as the benchmark $\mathcal{SFR}$. Benchmark $L_\gamma$ values correspond to the mean of the four models presented in Martin14 \cite{Martin:2014nia}. These are computed using the benchmark value of $\mathcal{SFR}$ from the table and are represented as the black dot in the Figure. References for the stellar mass: NGC6822~\cite{2012AJ....144....4M}  and \cite{Oh:2015xoa} for the rest. Note that the references do not provide uncertainties for stellar mass $M_*$.}}
        \label{tab:dIrr_SFR_1}
    \end{center}
\end{table}

We also need to estimate the spatial shape of such emissions. In this case, since this diffuse emission follows the star formation distribution, we can restrict the emission shape to the optical size of the galaxies, which we approximate to an ellipse with a given inclination $i$, eccentricity $e$, and semi-major axis $\theta_{\mathrm{opt}}$. We note that this is an approximation of the SFR emission, but given that the optical sizes $\theta_{\mathrm{opt}}$ of the galaxies are of the order $\sim 0.1^\circ$, this is a good approximation since it translates up to a few pixels for the estimated PSF of the CTAO. In Table~\ref{tab:dIrr_SFR_2}, we list the optical sizes $\theta_{\mathrm{opt}}$, inclinations $i$ and eccentricities $e$ of each galaxy. The inclination angle $i$ refers to the rotation angle of the major semiaxis, increasing counterclockwise from the North direction.

\begin{table}[t!]
    \begin{center}
\begin{tabular}{|c|c|c|c|}
\hline \hline
\textbf{Galaxy}  & $\theta_{\mathrm{opt}}$ {[}deg{]} & $i$ {[}deg{]} & $e$    \\ \hline \hline
\textbf{IC10}    & $0.04$                            & $-$           & $0.59$ \\ \hline
\textbf{IC1613}  & $0.11$                            & $-40$         & $0.46$ \\ \hline
\textbf{WLM}     & $0.13$                            & $-86$         & $0.89$ \\ \hline
\textbf{NGC6822} & $0.04$                            & $240$         & $0.65$ \\ \hline
\end{tabular}
        \caption{\footnotesize{Parameters used to model the SFR spatial emission: semi-major axis $\theta_{\mathrm{opt}}$, inclinations $i$ and eccentricities $e$. Note that we refer to the major semiaxis of the ellipse as the overall optical size $\theta_{\mathrm{opt}}$ of the galaxies. If a value is given as $-$, no data is available and the value set to 0 as the default. References: for the semi-major axis $\theta_{\mathrm{opt}}$, see~\cite{2012AJ....144....4M} and \cite{Oh:2015xoa} for NGC6822 and the rest, respectively; for inclinations $i$ and eccentricities $e$, see~\cite{2012AJ....144....4M}.}}
        \label{tab:dIrr_SFR_2}
    \end{center}
\end{table}

\section{Correlation matrices}
\label{ap:Correlation_matrix}

\begin{figure}[t!] 
  \centering 
    \includegraphics[width=0.49\textwidth]{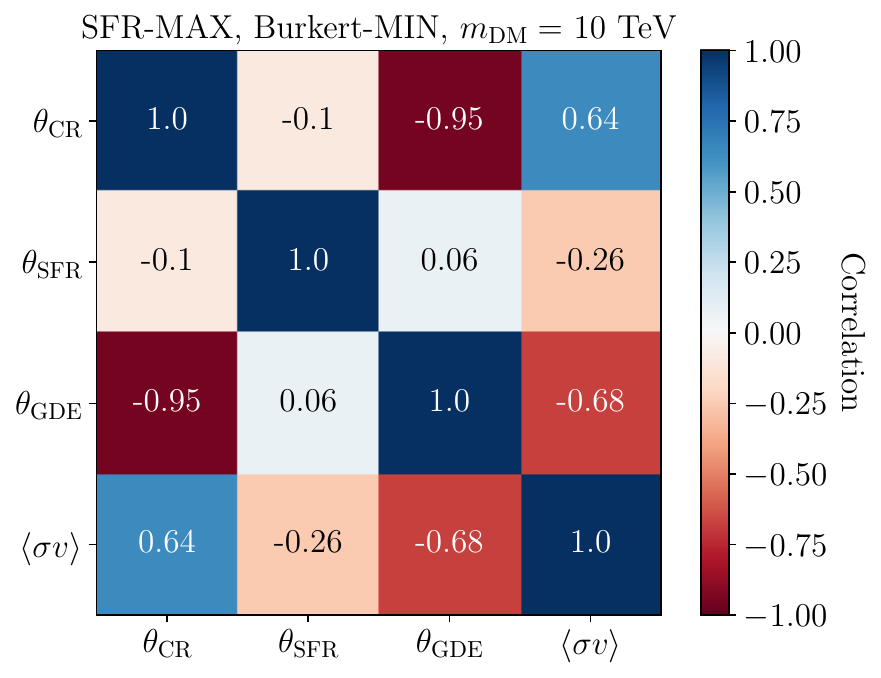}
    \includegraphics[width=0.49\textwidth]{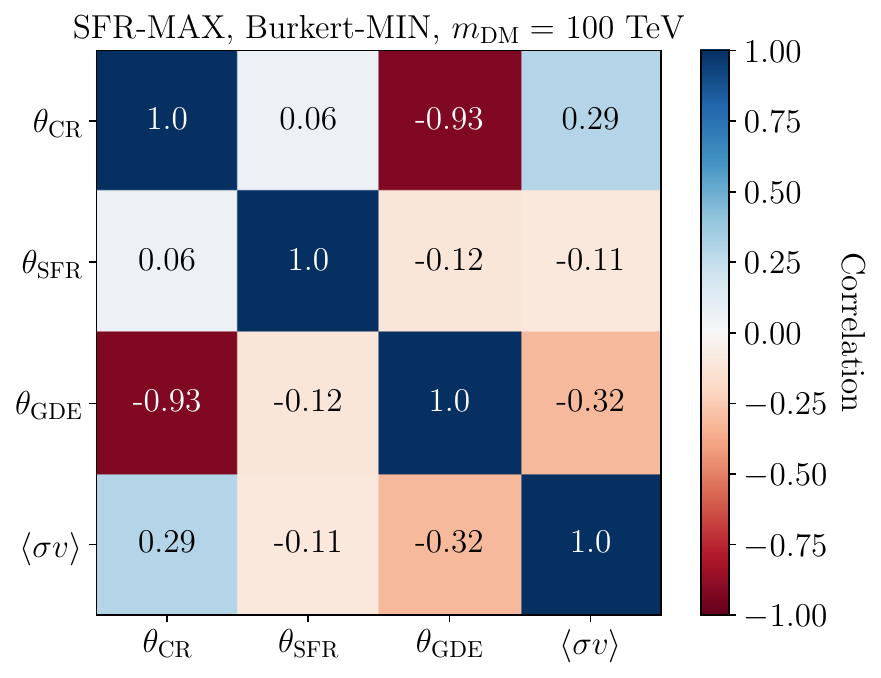}
    \includegraphics[width=0.49\textwidth]{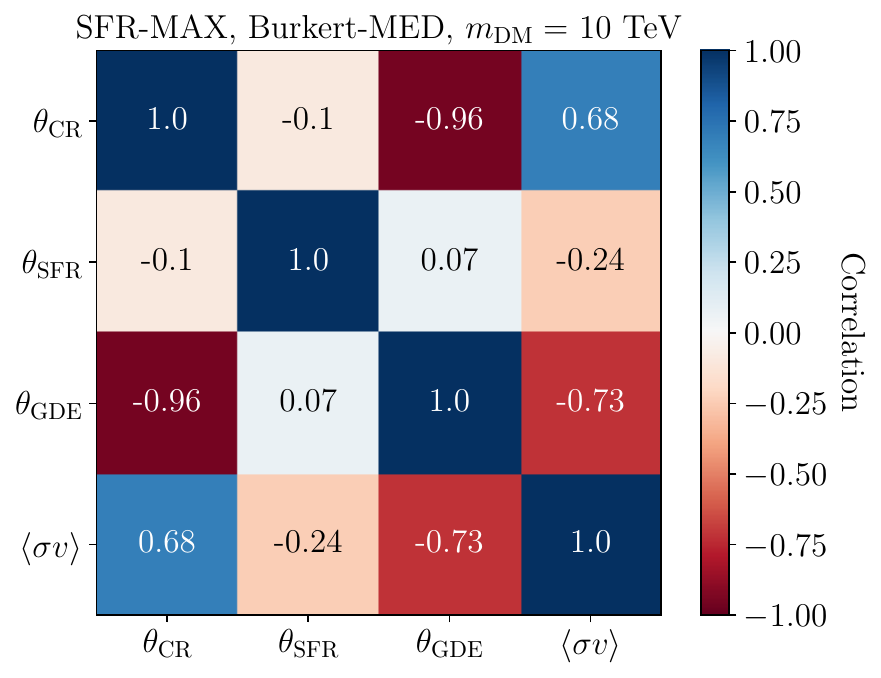}
    \includegraphics[width=0.49\textwidth]{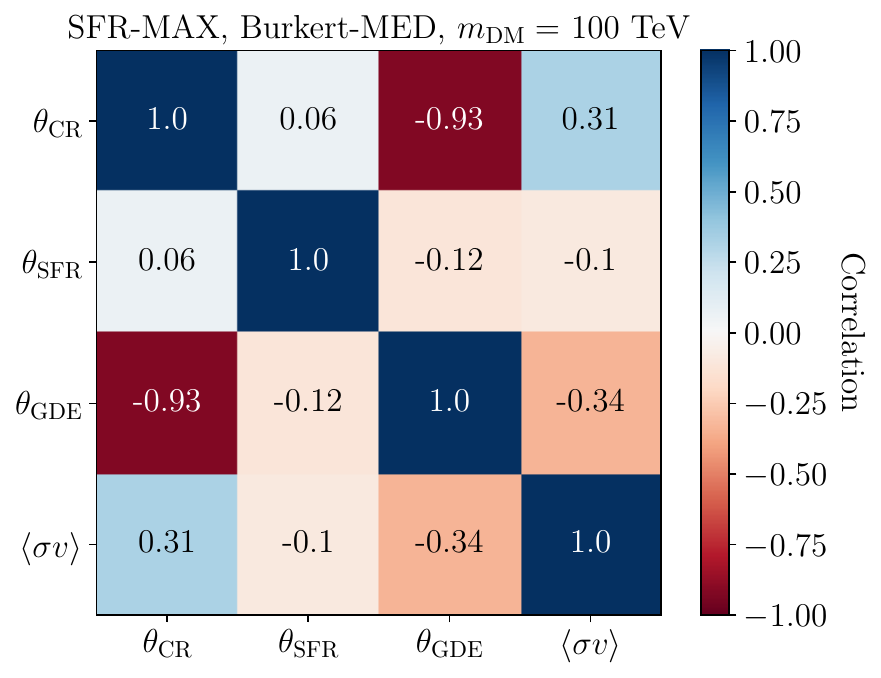}
    \includegraphics[width=0.49\textwidth]{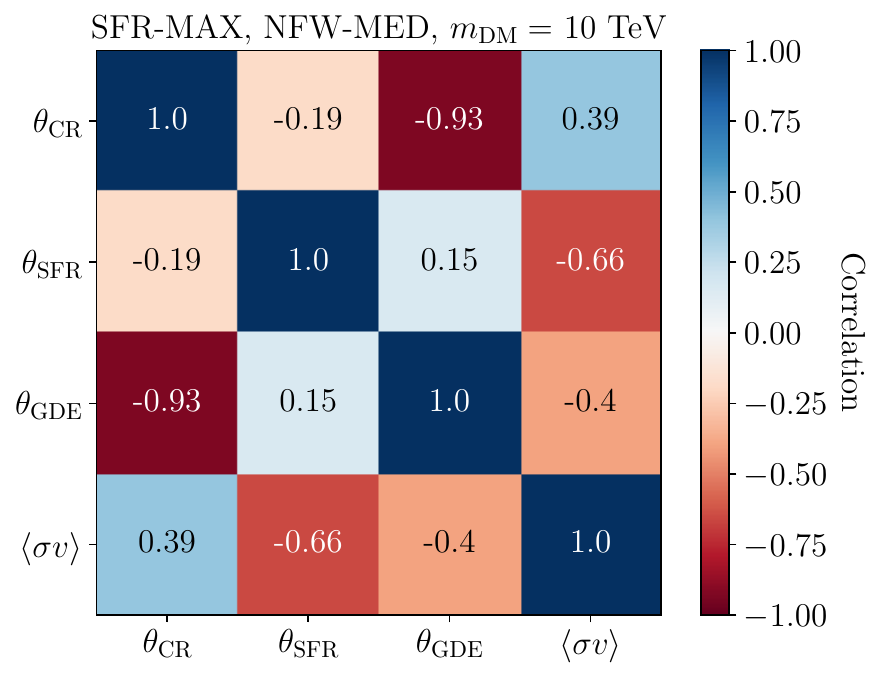}
    \includegraphics[width=0.49\textwidth]{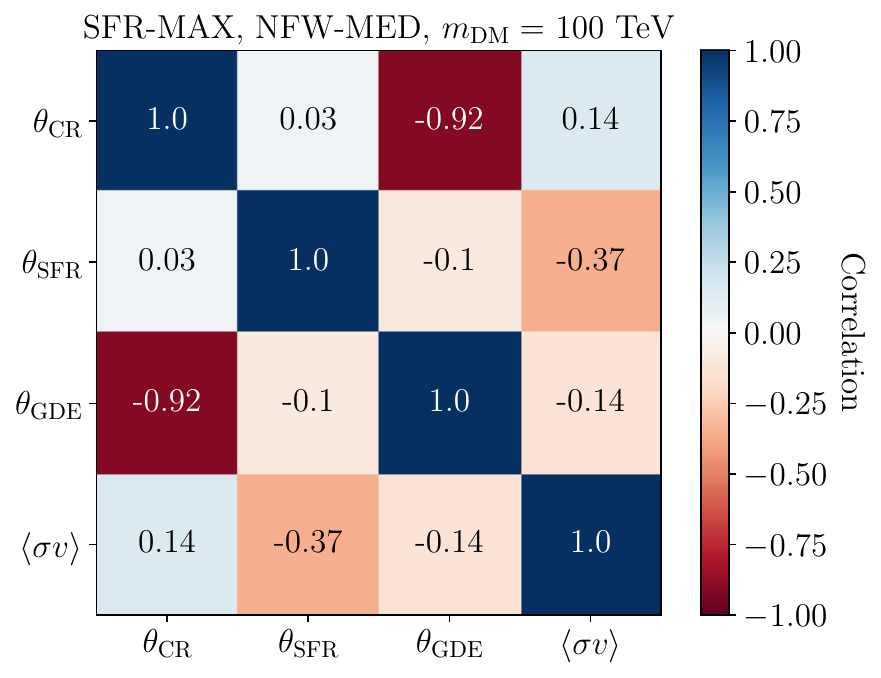}
    \caption{\footnotesize{Correlation matrices for different representative cases in the analysis. The analysis is made for the NGC6822 galaxy assuming the SFR-MAX emission (Table~\ref{tab:dIrr_SFR_1}) and the $\tau^+\tau^-$ annihilation channel. See text for more details on the choice of models and on the trends observed in these matrices.}}
\label{fig:correlation_matrices} 
\end{figure}

In this Appendix, we show the correlation matrices of different sky modelizations. To be representative of the sample, we focus on the galaxy NGC6822, as it is one of the galaxies with the greatest DM, SFR and GDE expected observed emission, together with IC10. We include all free parameters in this analysis: CR background $\theta_\mathrm{CR}$, SFR $\theta_\mathrm{SFR}$, GDE $\theta_\mathrm{GDE}$ and the DM $\langle \sigma v \rangle$ normalizations. For the DM flux, we show the correlation matrices for the $\tau^+\tau^-$ channel, as it is the one yielding the highest flux, but for different DM masses. The corresponding correlation matrices can be seen in Figure~\ref{fig:correlation_matrices}, for $m_\mathrm{DM} = 10$ (first column) and $100$ TeV (second column). As for the spatial emission, we compute the matrices for the Burkert-MIN (first row), Burkert-MED (second row) and NFW-MED models (last row). The chosen SFR emission normalization  corresponds to the SFR-MAX normalization value (see Table~\ref{tab:dIrr_SFR_1}), although we have checked that the results are similar in the SFR-MED case (but with an overall smaller correlation/anti-correlation). With these choices, we are left with the most general cases possible. Given that the results depend on Poissonian noise from the simulations, we present results computed with the Asimov dataset \cite{2011EPJC...71.1554C}, in which the observed gamma-ray counts are set to the predicted counts from the models. This noiseless dataset choice corresponds to the most idealistic observation possible, in the sense that all possible realizations of the simulations converge to it.

In all cases, the GDE normalization $\theta_\mathrm{GDE}$ and CR instrumental background normalization $\theta_\mathrm{CR}$ are largely anti-correlated (with values $\sim -0.9$). Intuitively, this anti-correlation is explained because both the GDE and CR backgrounds have a spatial distribution approximately homogeneous in the observations, and spectrally their counts are well described by a power law with a similar slope (see Figure~\ref{fig:dIrr_spectral_counts_simulation}). Thus, in order to respect the total amount of counts observed, the increase of one of the parameters will necessarily imply the decrease of the other. On the other hand, the normalization of the SFR emission $\theta_\mathrm{SFR}$ is barely correlated with the GDE and CR background. As the spatial extension of the SFR emission is very concentrated, contrarily to the extended emission of the other two components, low correlation values are expected. As for the DM annihilation cross-section $\langle \sigma v \rangle$,  a trend can be observed in which the correlation (or anti-correlation) decreases as the DM mass increases. The reason comes from the fact that the DM flux is proportional to $\propto 1/m_\mathrm{DM}^2$ (Equation~\ref{eq:flux_annih}) and, thus, the higher the mass, the lesser the expected flux, decreasing all possible correlations. Fixing the DM mass (i.e., focusing on a single column in the Figure), $\langle \sigma v \rangle$ shows different correlation values with the other two parameters. In the Burkert-MIN and Burkert-MED case (first and second row), $\langle \sigma v \rangle$ exhibits a high correlation with $\theta_\mathrm{CR}$ and anti-correlation with $\theta_\mathrm{GDE}$. This is because the Burkert profile is cored (Figures~\ref{fig:clumpy_maps} and \ref{fig:All_Jfac_integrated}), and so it has a greater resemblance to the spatial distribution of the other two models, yet with a slightly greater correlation/anti-correlation to the Burkert-MED, thanks to the increase in the extension of the DM signal after including the subhalos contribution. A similar case happens with SFR emission: being the most spatially concentrated emission among all, the expected correlation must be low between $\langle \sigma v \rangle$ and $\theta_\mathrm{SFR}$. On the other hand, the NFW-MED model (last row) is considerably cuspier than the Burkert-MIN and Burkert-MED so, as a consequence, the spatial distribution of the DM signal is more distinguishable from the CR background and GDE templates, therefore presenting a smaller correlation. For the same reason, the peak of the NFW profile makes the emission more degenerate with the SFR emission (also concentrated), thus creating a higher anti-correlation with $\theta_\mathrm{SFR}$.

\section{SFR flux bin-by-bin likelihood fit}
\label{ap:SFR_bin_by_bin_lklh}

\begin{figure}[t!] 
  \centering 
    \includegraphics[width=0.49\textwidth]{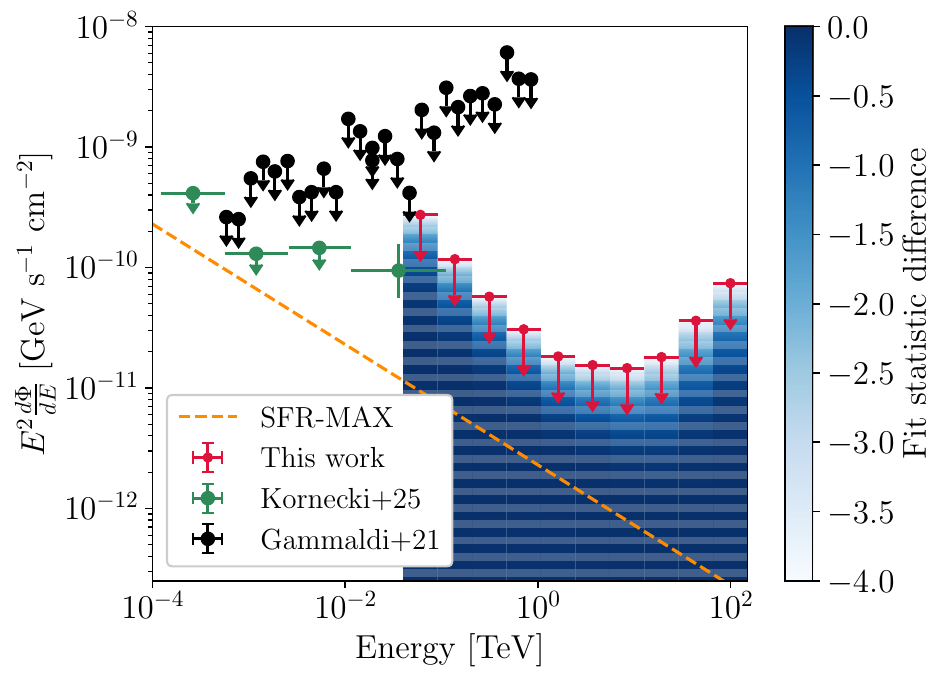}
    \includegraphics[width=0.49\textwidth]{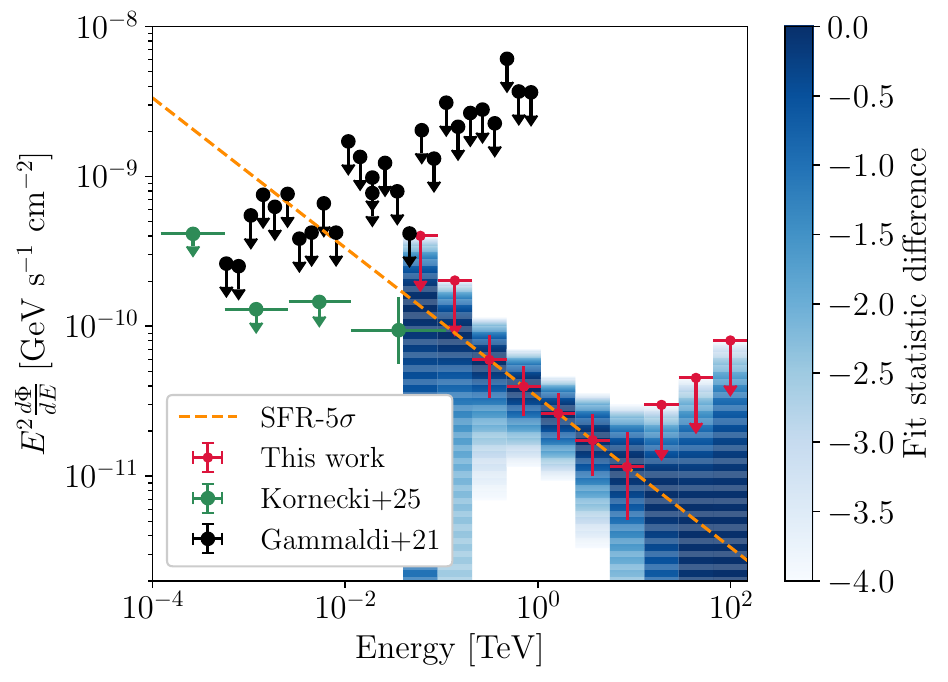}
    \caption{\footnotesize{Bin-by-bin likelihood fit of the SFR emission for NGC6822 (red points), compared to the flux upper limits obtained with \textit{Fermi}-LAT data (Kornecki+25 \cite{2025A&A...699A..43K}, in green; Gammaldi+21 \cite{2021PhRvD.104h3026G}, in black). The dashed yellow line corresponds to the SFR emission normalization used for the computation: left panel for SFR-MAX (see Table~\ref{tab:dIrr_SFR_1}) and right panel for a $5\sigma$ detection. As indicated by the color bar, the upper limits to the flux (arrows) indicate a 2$\sigma$ C.L., whereas error bars on detections indicate a 1$\sigma$ C.L., with a minimum of a $TS=4$ detection for that energy bin.}}
\label{fig:SFR_bin_by_bin_lklh} 
\end{figure}

This Appendix illustrates the expected CTAO sensitivity to the SFR emission produced in NGC6822, the best of the four candidates for this type of study. Figure~\ref{fig:SFR_bin_by_bin_lklh} shows the bin-by-bin likelihood fit for two different scenarios: a first one where we set the normalization of the flux to the SFR-MAX case (see Table~\ref{tab:dIrr_SFR_1}); and a second one for which we set the normalization of the flux such that a 5$\sigma$ detection is obtained. As in Appendix \ref{ap:Correlation_matrix}, we make use of the Asimov dataset \cite{2011EPJC...71.1554C} to ensure statistical consistency. For both cases, the SFR flux is depicted as a dashed yellow line in the Figure. We also compare to results obtained with \textit{Fermi}-LAT data (Kornecki+25 \cite{2025A&A...699A..43K}, in green; Gammaldi+21 \cite{2021PhRvD.104h3026G}, in black). Despite the SFR model apparently not being compatible with the \textit{Fermi}-LAT (green) data point in any of the two mentioned cases, we note that in \cite{2025A&A...699A..43K} the reported data point has a $\mathrm{TS}=8$, although it is not high enough to be linked to the SFR emission itself or just a background fluctuation. Also, at lower energies, the SFR emission is expected to have a lower spectral index ($\gamma \sim 2.3-2.4$) than the one adopted at TeV energies ($\gamma = 2.5$) \cite{Martin:2014nia}. Nonetheless, the case of a 
5$\sigma$ detection of the SFR flux is disfavored by the \textit{Fermi}-LAT upper limits.

\section{Individual constraints and Poissonian uncertainties}
\label{ap:poisson_uncertainties}

\begin{figure}[t!] 
  \centering 
    \includegraphics[width=0.6\textwidth]{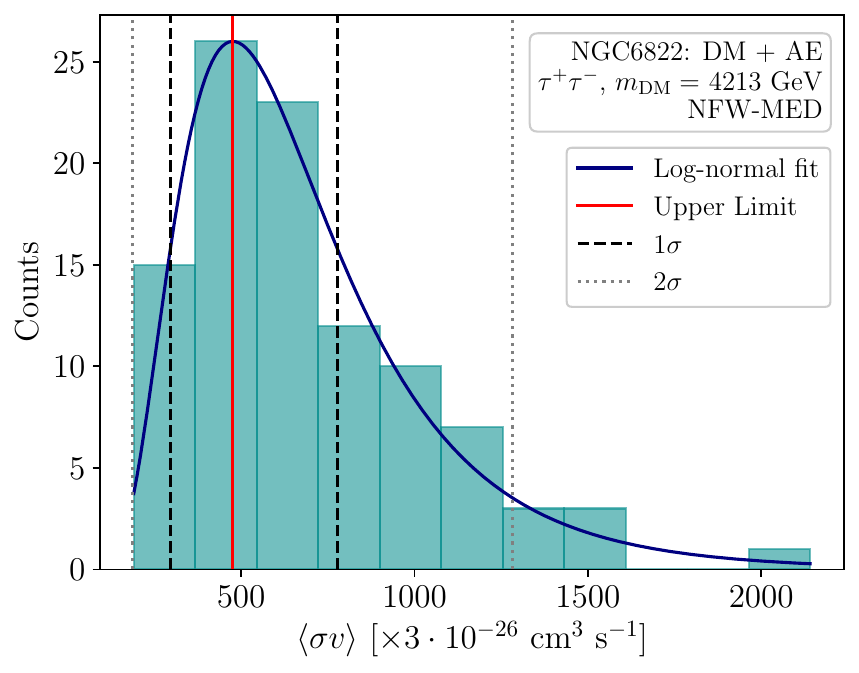}
    \caption{\footnotesize{Histogram of the $95\%$ C.L. projected constraints on the annihilation cross-section $\langle \sigma v \rangle$, for the 100 realizations made for this specific case in the analysis of the NGC6822 galaxy: full DM + AE and NFW-MED spatial modeling, DM mass $m_{\mathrm{DM}} = 4213$ GeV annihilating into the $\tau^+\tau^-$ channel. In dark blue we show the log-normal fit of the histogram, and the vertical lines show the final $95\%$ upper limit (solid red), with the 1$\sigma$ (dashed black) and 2$\sigma$ (dotted grey) statistical uncertainties.}}
\label{fig:UL_histogram} 
\end{figure}

Following \cite{CTAConsortium:2023yak}, we create 100 simulations of each combination of parameters (DM mass value $m_{\mathrm{DM}}$, DM annihilation channel, spatial DM density profile mode, AE modeling and galaxy target) to take into account the Poissonian nature of the simulations. With each of these 100 simulations, we directly obtain 100 values for the annihilation cross-section $\langle \sigma v \rangle$ $95\%$ C.L. projected upper limits. A distribution of the annihilation cross-section upper limits is obtained this way, its peak value being the final, reported constraint in each case  (i.e., the bin with the most values repeated). From such a distribution, we can also estimate the uncertainties of these upper limits by fitting the whole distribution to a log-normal distribution function. Once this distribution function is defined, we can easily obtain the final upper limit value and the $1\sigma$ and $2\sigma$ statistical uncertainties. In Figure~\ref{fig:UL_histogram}, we show the histogram for the example case of the full DM + AE NGC6822 galaxy and NFW-MED spatial modeling, with a DM mass $m_{\mathrm{DM}} = 4213$ GeV annihilating into the $\tau^+\tau^-$ channel (this mass value corresponds to the 8th mass considered in our sampling of 13 DM masses). The vertical solid red line corresponds to the final upper limit computed and reported (i.e., the peak of the log-normal distribution function), while the dashed black (dotted grey) vertical lines correspond to the $1\sigma$ ($2\sigma$) statistical uncertainties range. Also, in dark blue, we show the log-normal fit to the histogram.

\section{TS profiles and combined limits computation}
\label{ap:combined_analysis}

\begin{figure}[t!] 
  \centering 
    \includegraphics[width=0.49\textwidth]{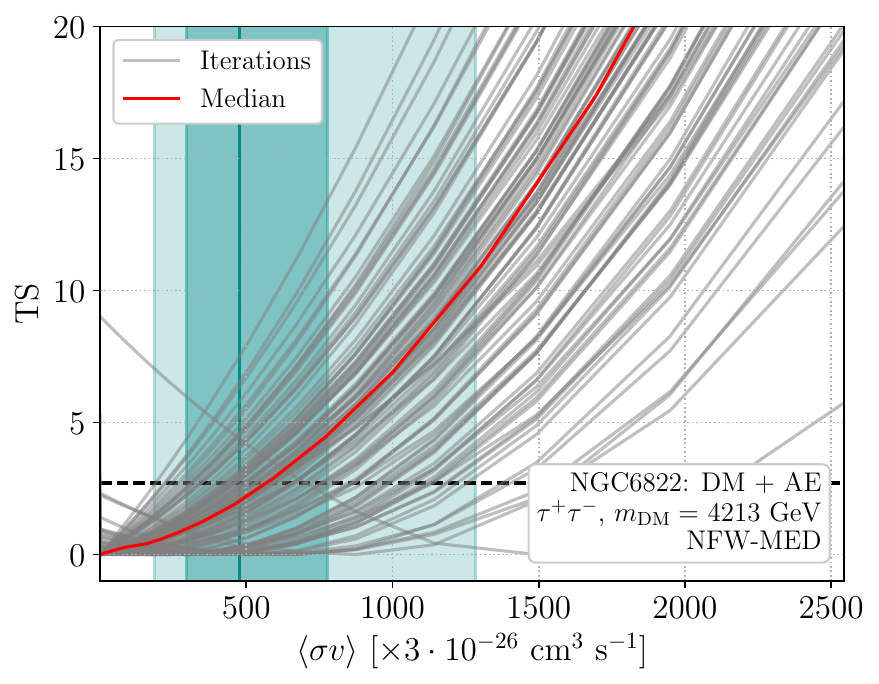}
    \includegraphics[width=0.49\textwidth]{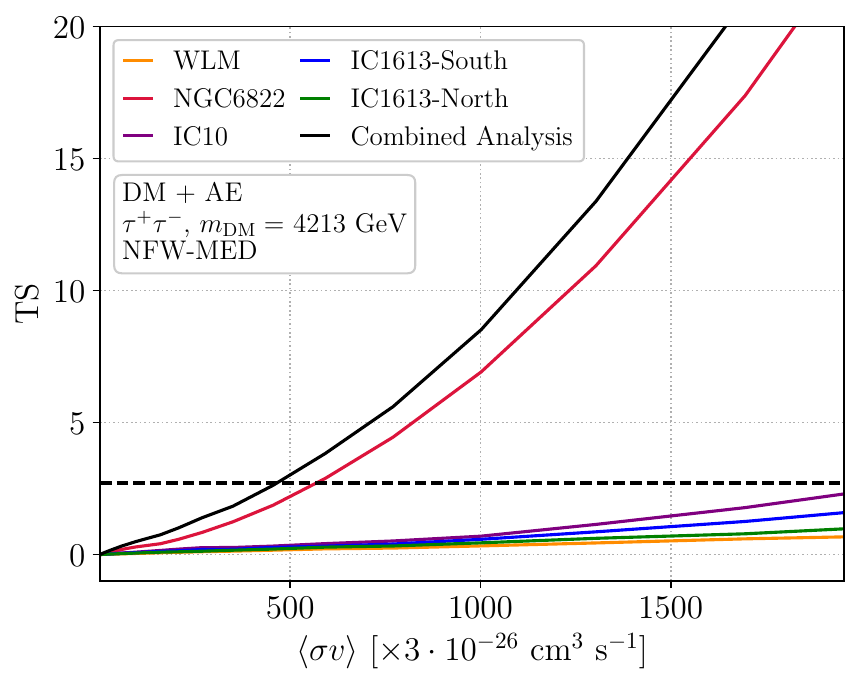}
    \caption{\footnotesize{TS profiles obtained for the computation of the $\langle \sigma v \rangle$ upper limits. Left panel: in grey the TS curves of the 100 realizations made for a single setup, in this case for the same example shown in Figure~\ref{fig:UL_histogram}. In red, the median of the profiles, which is the one adopted as the representative one to be used for the combined analysis (see text in this Appendix \ref{ap:combined_analysis} for more information). The vertical blue line corresponds to the $\langle \sigma v\rangle$ upper limit computed from Figure~\ref{fig:UL_histogram}, with the blue bands being the 1 and 2 $\sigma$ uncertainties range. Right panel: combined (black) TS profile and the rest of the individual representative likelihoods (colored lines) for all of the dIrrs in the sample, and for the same example setup depicted in Figure~\ref{fig:UL_histogram}.}}
\label{fig:combined_individual_likelihoods} 
\end{figure}

To compute the combined DM limit, we first need to define a way of obtaining a single TS profile representative of the 100 realizations done for each of the considered sets of parameters (DM mass value $m_{\mathrm{DM}}$, DM annihilation channel, spatial DM density profile mode, AE modeling and galaxy target). Indeed, for each of the 100 realizations of simulated observations, we compute one TS profile, thus we end up having 100 of these TS profiles, but there is no easy way to define a representative TS. In the left panel of Figure~\ref{fig:combined_individual_likelihoods}, we show the set of 100 TS profiles we have computed for the same example case as in Figure~\ref{fig:UL_histogram}. As expected, there is a significant spread of the TS curves in the figure. To solve this problem, we decide to take the median at each value of $\langle \sigma v \rangle$ considered, so that we can build a representative individual TS profile (red line in the figure) or the whole set of realizations in the considered setup. The same procedure is adopted to compute the combined TS profile of the rest of the galaxies in our sample. For consistency, we also show in the same figure the upper limit (vertical blue line) as well as the 1$\sigma$ and 2$\sigma$ uncertainties (blue vertical bands) depicted in Figure~\ref{fig:UL_histogram}. It can be seen that the median TS profile crosses the value of 2.71\footnote{We assume a one-sided distribution, thus the $95\%$ C.L. corresponds to a change of $\Delta$TS = 2.71, with respect to the best fit \cite{2015APh....62..165C}.} (horizontal black dashed line) well within the uncertainties.  

We show in the right panel of Figure~\ref{fig:combined_individual_likelihoods} the individual TS profiles for each dIrr in the sample (colored lines), together with the combined (black) TS profile for the whole sample, all of it for the same example setup as before. In this case, the $95\%$ C.L. combined constraint is computed when the combined likelihood crosses the $\mathrm{TS}=2.71$ black dashed line. As can be seen, the NGC6822 dIrr galaxy dominates the combined result, yielding the best individual constraints. We note, though, that this is only the case for the NFW-MED profile: in the other two spatial models (Burkert-MIN and Burkert-MED), IC10 turns out to be the most constraining galaxy.

\begin{figure}[t!] 
  \centering 
    \includegraphics[width=0.49\textwidth]{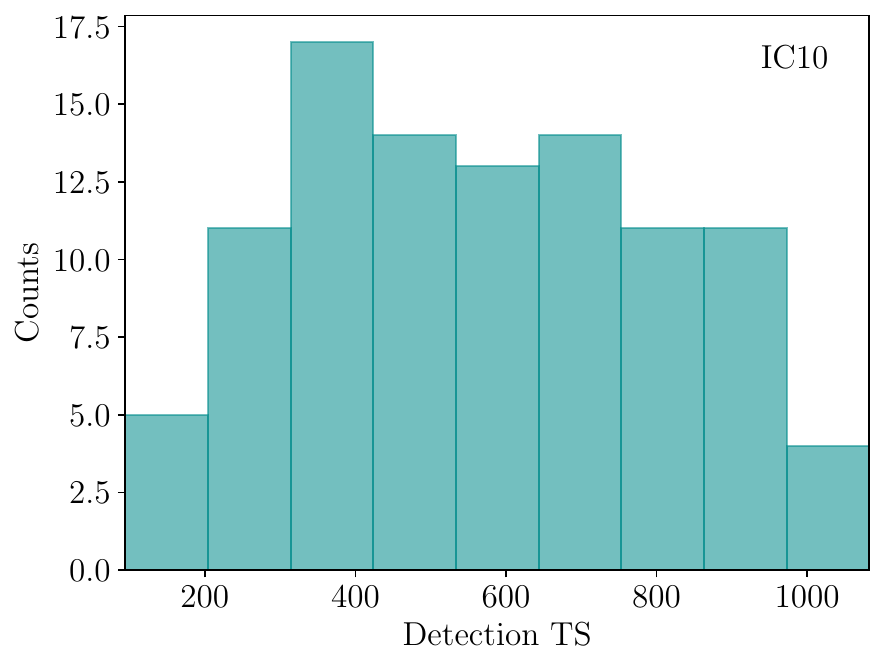}
    \includegraphics[width=0.49\textwidth]{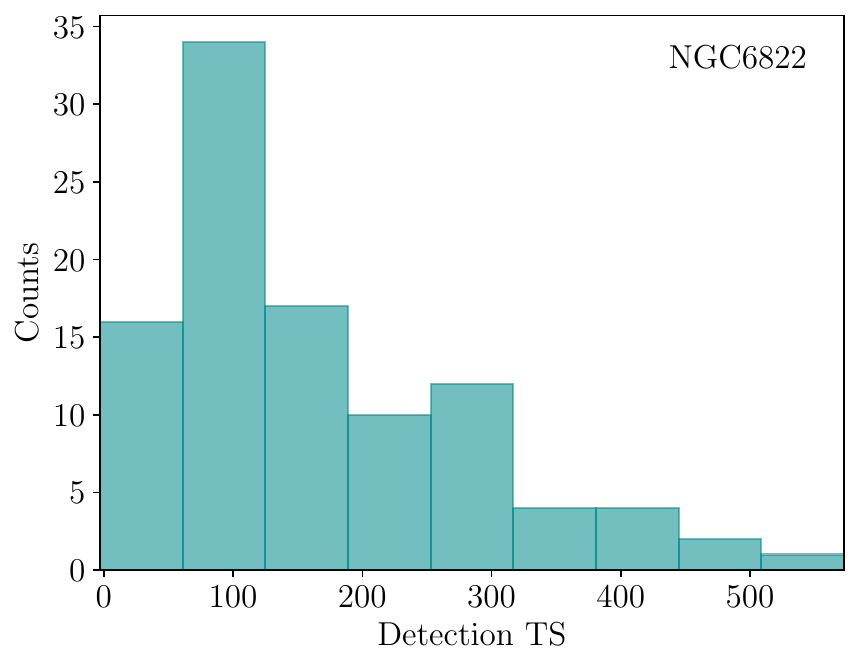}    
    \caption{\footnotesize{Histogram of the GDE detection TS for the 100 simulations made of IC10 (left panel) and NGC6822 (right panel), i.e., the two dIrrs where a GDE detection is consistently found. Given the Poissonian uncertainty, we show the results for 100 sky simulations, including the modeling of the DM signal, GDE, SFR emission and the instrumental background.}}
\label{fig:GDE_TS_histograms} 
\end{figure}

\section{GDE detection discussion}
\label{ap:GDE_detection}

In Figure~\ref{fig:GDE_TS_histograms}, we show the histogram of the GDE detection TS for the 100 simulations made of IC10 and NGC6822. The 100 sky simulations include the modeling of the GDE, SFR emission and the instrumental background. As we can see, most of the observations have their detection TS greater than 25 ($\sigma > 5$). This is due to their angular location being close to the GP. For the other two dIrrs, however, the GDE is not expected to be detected. For an in-depth discussion on the degeneracies and correlations between the free parameters of the full analysis, see Appendix \ref{ap:Correlation_matrix}.

\section{DM constraints mismodeling the emission}
\label{ap:mismodeling_constraints}
In this Appendix, we illustrate how the projected upper limits depend on whether the GDE emission, the SFR emission and DM-induced emission are correctly modeled in the mock dataset and fitted models. We show the cases of a DM-only PL analysis, the spatially extended DM-only analysis and the final case of the full simulation (GDE, SFR emission and DM) without including SFR emission in the fitted models. As a final remark, for an in-depth discussion on the degeneracies and correlations between the free parameters of the full analysis, see Appendix \ref{ap:Correlation_matrix}.

\subsection{DM-only point-like analysis}
\label{ap:Point_Like_results}

As a first approximation, we study the CTAO sensitivity to a DM annihilation signal by modeling the DM target as a PL source. With this approximation, we aim to quantify the impact on the results when simplifying the spatial analysis, i.e., when considering a PL analysis instead of using the capabilities of the CTAO to perform an extended template analysis, where the signal is more spread. We include only the PL DM source and the instrumental background in the model, with the normalization of the instrumental CR background and the DM annihilation cross-section $\langle \sigma v \rangle$ as the only free parameters in this approach.

\begin{figure}[t!] 
  \centering 
  \includegraphics[width=0.99\textwidth]{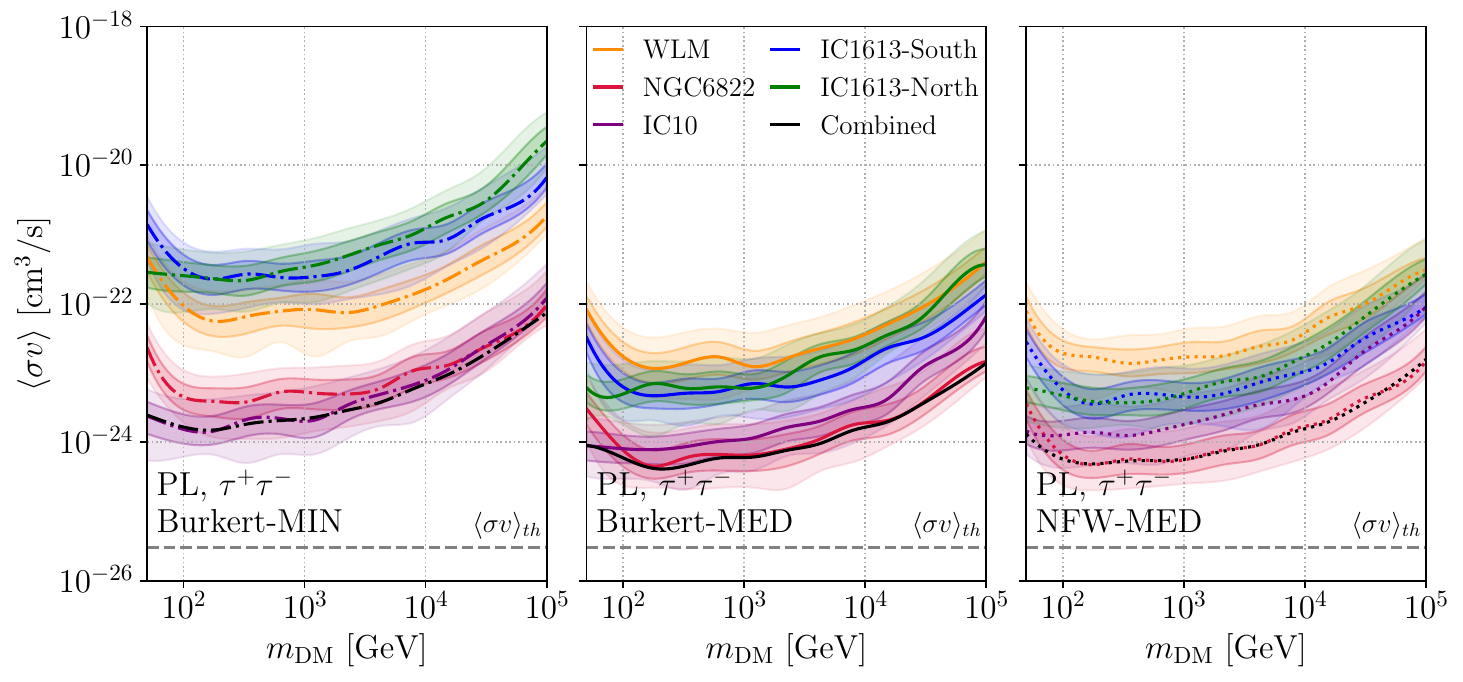}
  \includegraphics[width=0.99\textwidth]{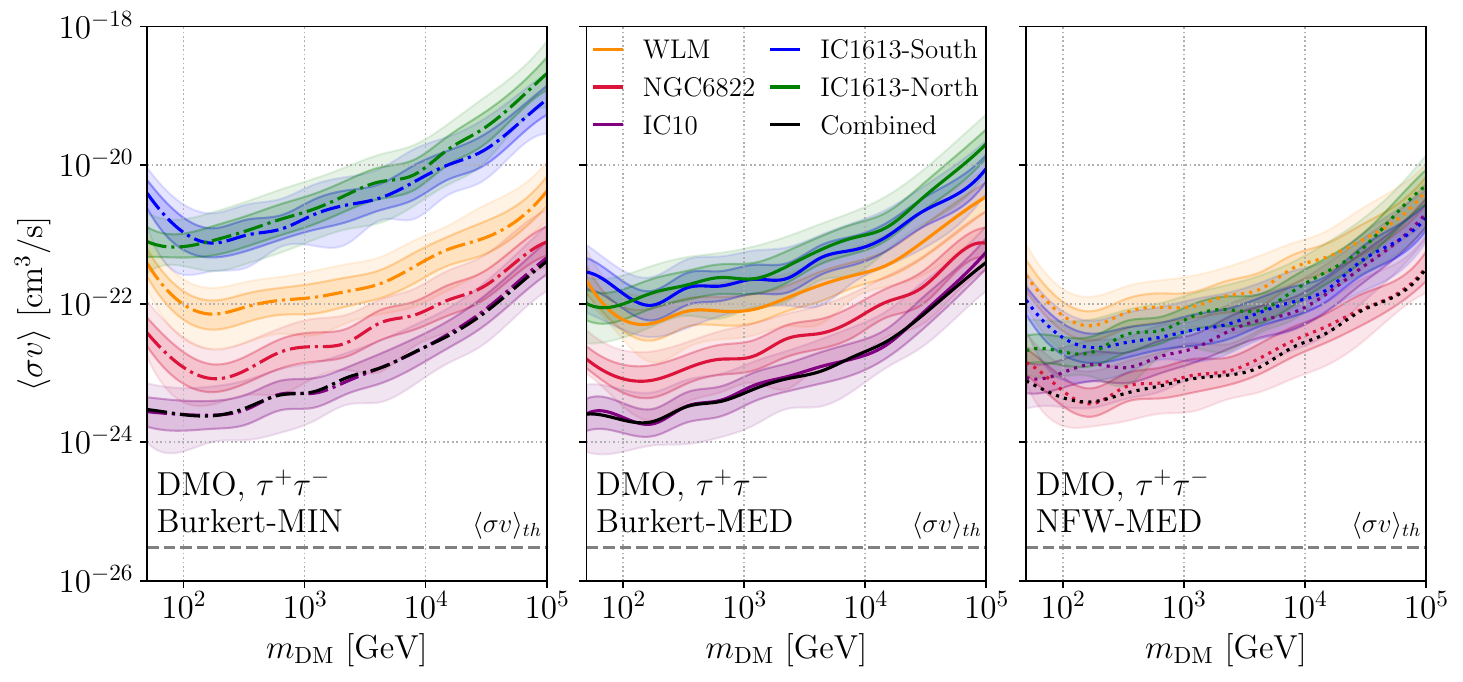}
    \caption{\footnotesize{DM annihilation cross-section $\langle \sigma v \rangle$ projected upper limits for individual dIrrs (colored lines) and the combined sample (black lines, see Appendix \ref{ap:combined_analysis}) for the $\tau^+\tau^-$ channel, in the case of the PL (first row; Section \ref{ap:Point_Like_results}) and the extended DM-only scenario (DMO, second row; Section \ref{ap:DM_only_results}). Each column refers to a different DM density profile: Burkert-MIN (first column), Burkert-MED (second column) and NFW-MED (third column); as described in Section \ref{sec:DM_spatial_modeling}. We also show the 1$\sigma$ and 2$\sigma$ uncertainties (colored bands) for each of the individual upper limits due to the Poissonian noise related to the simulation of the counts (Appendix \ref{ap:poisson_uncertainties}). As a comparison, the horizontal grey dashed line represents the thermal relic cross-section $\langle \sigma v \rangle_{\mathrm{th}}$.}}
\label{fig:tau_channel_UL_DMOnly} 
\end{figure}

In the first row of Figure~\ref{fig:tau_channel_UL_DMOnly}, we show the individual (colored lines) and combined (black lines) projected upper limits computed with this approach. In the figure, we restrict ourselves to the case of the $b\bar{b}$ channel, and, in each column, show one of the three different DM density profiles adopted (Burkert-MIN, Burkert-MED and NFW-MED). We also show in the Figure the statistical uncertainties related to the Poissonian noise of the simulations (colored bands), in which we present the 1$\sigma$ and 2$\sigma$ regions, as explained in Appendix \ref{ap:poisson_uncertainties}. For this particular approach, the most constraining galaxies are given by the highest J-factor: IC10 for the Burkert-MIN case, and the other two DM density profiles modeling IC10 (for lower masses) and NGC6822 for higher DM masses. And, in all cases, with the corresponding individual constraints almost coinciding with the combined analysis line. Overall, the best constraint is given by the Burkert-MED analysis at $m_\mathrm{DM} \simeq 200$ GeV, reaching $\langle \sigma v \rangle \sim 4\times 10^{-25} \ \mathrm{cm^3/s}$ for the $\tau^+\tau^-$ channel. In general, compared to Figure~\ref{fig:tau_channel_UL_all}, the constraints are overestimated by an order of a few better than the full DM+AE analysis, given that with a PL treatment all the signal is concentrated in the center of the observation and, therefore, the limits are more constraining.

\subsection{Spatially extended DM-only analysis}
\label{ap:DM_only_results}

In a more realistic scenario, we consider each dIrr to be an extended target for the CTAO (with the spatial templates shown in Section \ref{sec:DM_spatial_modeling}). With this analysis, we aim to estimate the error when not including the AE models to the analysis, assuming the astrophysical background to be negligible in this approach. This, in principle, is a good approximation for the low DM masses, given that, as introduced in Section \ref{sec:SFR_IE}, the expected SFR flux is lower than the DM fluxes. However, as can be seen in Figure~\ref{fig:dIrr_DMFlux_SFR}, when evaluating higher DM masses, both fluxes are of the same order. Therefore, we only simulate the instrumental background and the prospective DM signal. As before, we only have two free parameters in the fit: the normalization of the instrumental CR background and the DM annihilation cross-section $\langle \sigma v \rangle$. We show in the second row of Figure~\ref{fig:tau_channel_UL_DMOnly} the individual and combined upper limits for the case of DM annihilating into the $\tau^+\tau^-$ channel, with the corresponding 1 and 2 $\sigma$ uncertainties estimated from the Poissonian noise of the simulations (see Appendix \ref{ap:poisson_uncertainties}). Only two galaxies dominate the constraints. For the Burkert-MIN and Burkert-MED, IC10 is the most constraining galaxy, and for the NFW-MED modeling, NGC6822. In all cases, the individual constraints are very similar to the combined results, with the best constraints reaching $m_\mathrm{DM} \simeq 200$ GeV, reaching $\langle \sigma v \rangle \sim 2\times 10^{-24} \ \mathrm{cm^3/s}$ for the $\tau^+\tau^-$ channel in the Burkert-MED case. Compared to the previous PL results, these constraints are a factor of a few worse than before, although in the case of Burkert-MIN, the differences are smaller. This can be explained by the fact that the signal now is treated as extended, and the DM signal, therefore, is not as concentrated. When including subhalos in the profiles (Burkert-MED and NFW-MED), the corresponding J-factors are more extended (Figure~\ref{fig:clumpy_maps}), thus worsening the constraints more with respect to the PL case. On the other hand, we do not find any noticeable difference between the DM-only and full DM+AE modelization in the projected constraints (Figure~\ref{fig:tau_channel_UL_all}).

\subsection{Full simulation without fitting SFR}
\label{ap:UL_no_SFR_fitted}

\begin{figure}[t!] 
  \centering 
  \includegraphics[width=0.99\textwidth]{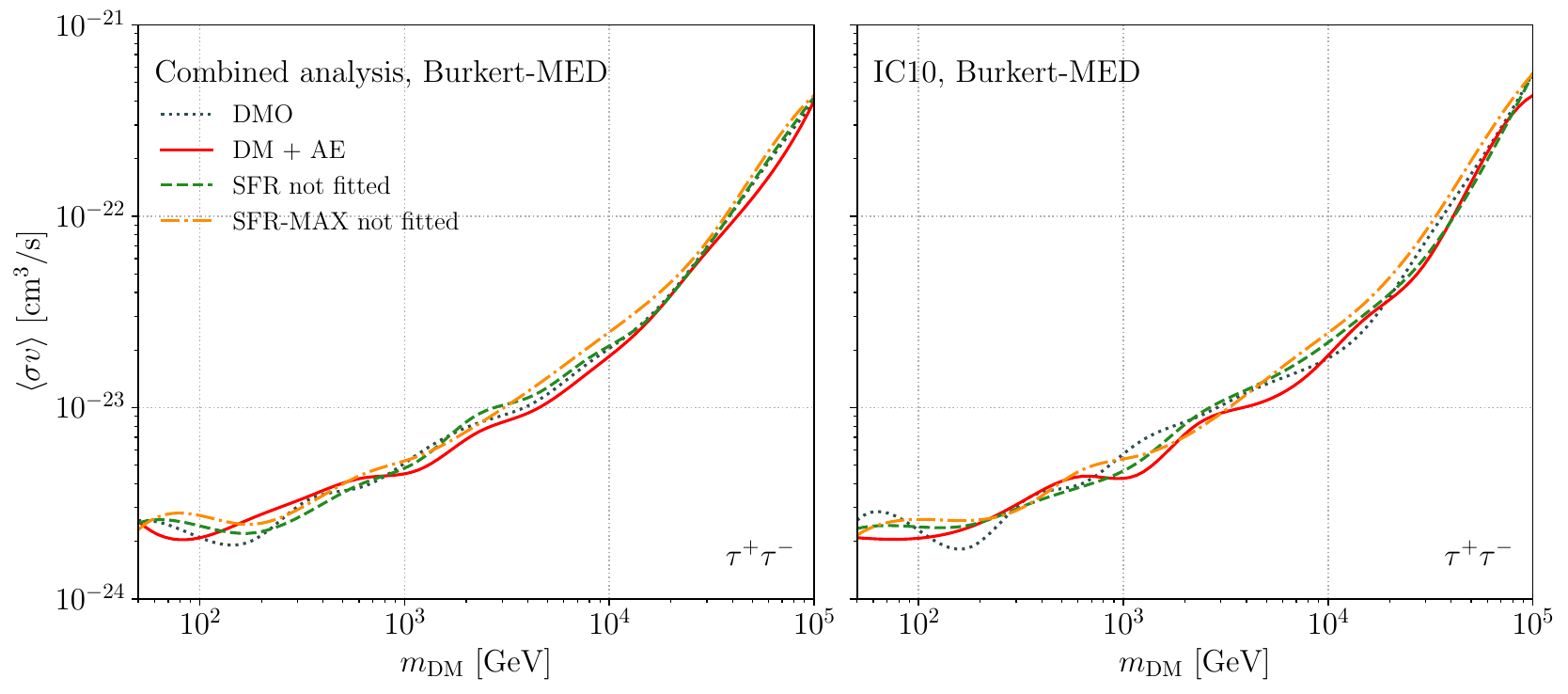}
    \caption{\footnotesize{DM annihilation cross-section $\langle \sigma v \rangle$ projected combined (left panel) and IC10 individual (right panel) upper limits for the $\tau^+\tau^-$ channel and Burkert-MED modeling. The dotted grey line corresponds to the DM-only extended analysis (DMO, Appendix \ref{ap:DM_only_results}), the solid red line to the full DM + AE modeling and fitting (Section \ref{sec:results_DM}), the dashed green line is the full modeling of the mock data (following the SFR emission benchmark emission) without fitting SFR emission and the dot-dashed yellow line is for the SFR-MAX case (see Table~\ref{tab:dIrr_SFR_1}), where we set the SFR emission to the maximum flux allowed by the uncertainties.}}
\label{fig:UL_no_SFR_fitted} 
\end{figure}

In the previous section, we demonstrated that excluding the SFR emission from the mock data and fitting templates yields constraints on $\langle \sigma v \rangle$ that are similar to those obtained from the full AE modeling. As a final check, we now show how ignoring the SFR emission in the fitting templates (but still included in the mock data sky simulation) affects the constraints. We do so for two cases: the case in which mock data includes the SFR emission modeled with our benchmark value, and an extreme case where the SFR emission is set to the maximum allowed by the uncertainties of the modeling (which we call SFR-MAX, see Table~\ref{tab:dIrr_SFR_1}). Note that the GDE is treated as a background emission. The results, for the $\tau^+\tau^-$ channel and Burkert-MED modeling, are in Figure~\ref{fig:UL_no_SFR_fitted}, where we represent the projected combined (left panel) upper limits for the DM-only extended simulation (DMO, dotted grey line, Appendix \ref{ap:DM_only_results}), the full DM + AE modeling and fitting (solid red line, Section \ref{sec:results_DM}), and the full modeling of the mock data (following the SFR emission becnhmark emisison) without fitting SFR emission (dashed green line) and the SFR-MAX case (dot-dashed yellow line). Since the IC10 galaxy yields the best constraints for the Burkert profile, almost coinciding with the combined analysis, we also show in the right panel the related individual projected constraints. As can be seen, there are no differences in any of the projected results, so the AE intrinsic emission can be neglected for future dIrrs studies, as in the case of dSphs. We have also checked that the same conclusions can be obtained for the rest of the DM density profile models.

\section{Sommerfeld enhancement results for NGC6822 and WLM}
\label{ap:Sommerfeld_rest_galaxies}

\begin{figure}[t!] 
  \centering 
    \includegraphics[width=0.85\textwidth]{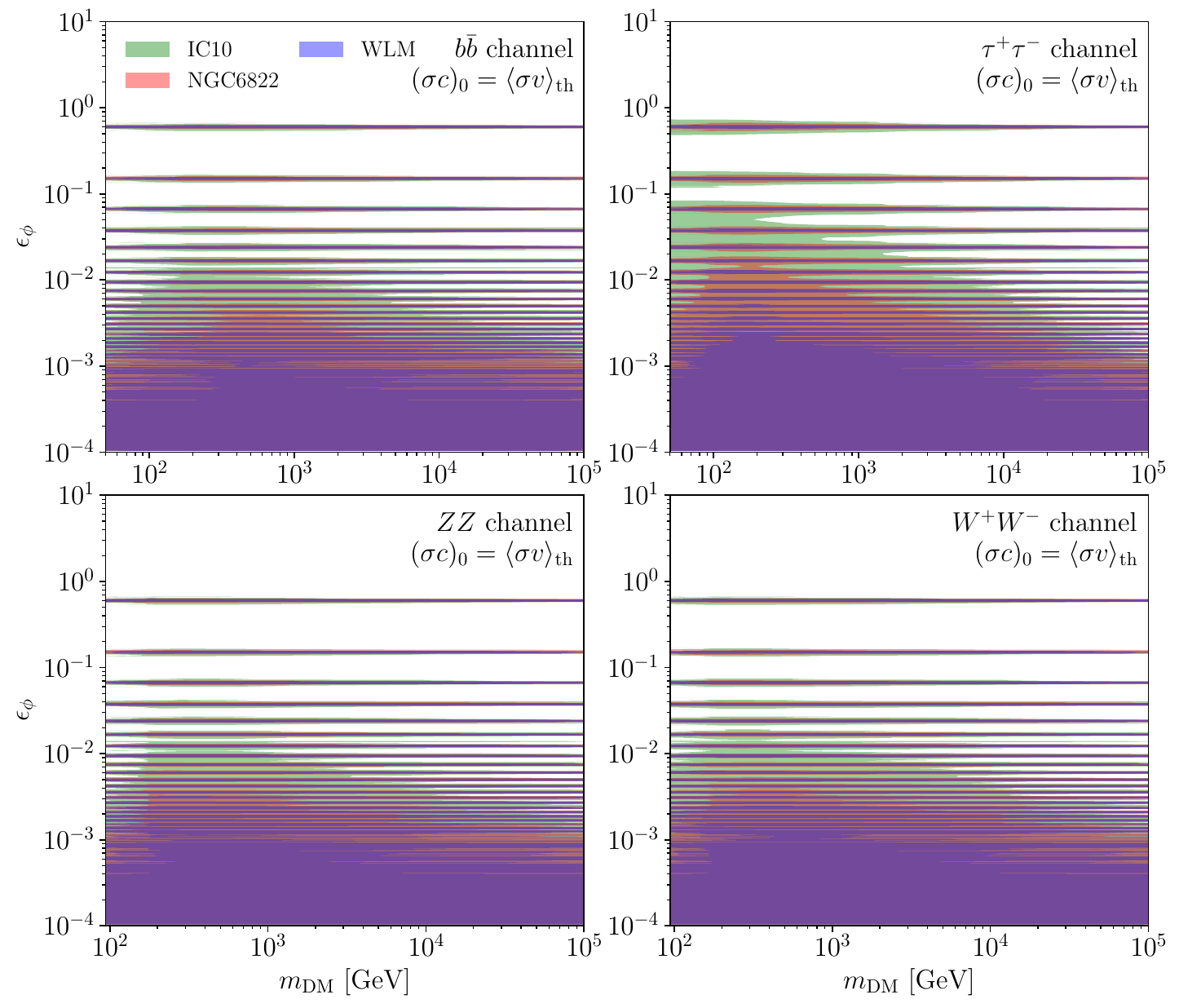}
    \caption{\footnotesize{CTAO sensitivity predictions for dIrrs for the $\epsilon_\phi - m_\mathrm{DM}$ parameter space (in green for IC10, red for NGC6822, and blue for WLM). The value of the thermal relic cross-section $\langle \sigma v \rangle_{\mathrm{th}} = 3 \times 10^{-26} \ \mathrm{cm}^3 \mathrm{s}^{-1}$ has been adopted as the normalization parameter $(\sigma c)_0$. We show results for the $b\bar{b}$ (upper left panel), $\tau^+\tau^-$ (upper right panel), $ZZ$ (lower left) and $W^-W^+$ (lower right) annihilation channel. See Section \ref{sec:sommerfeld} for details.}}
\label{fig:velocity_dependent_epsilonlimits_all_galaxies} 
\end{figure}

In Figure~\ref{fig:velocity_dependent_epsilonlimits_all_galaxies} we show how the projected constraints for IC10 (in green) compare with NGC6822 (red) and WLM (blue) for the case of the $\tau^+\tau^-$ (upper left panel), $b\bar{b}$ (upper right), $ZZ$ (lower left) and $W^-W^+$ (lower right) annihilation channels, assuming $(\sigma c)_0 = \langle \sigma v \rangle_{\mathrm{th}} = 3 \times 10^{-26} \mathrm{cm}^3 \mathrm{s}^{-1}$. From the Figure, we can see that the best constraints are given by IC10, followed by NGC6822 and with the worst constraints given by WLM.

\section{Extra figures}
\label{ap:rest_of_figs}

\begin{figure}[h!] 
  \centering 
  \includegraphics[width=0.49\textwidth]{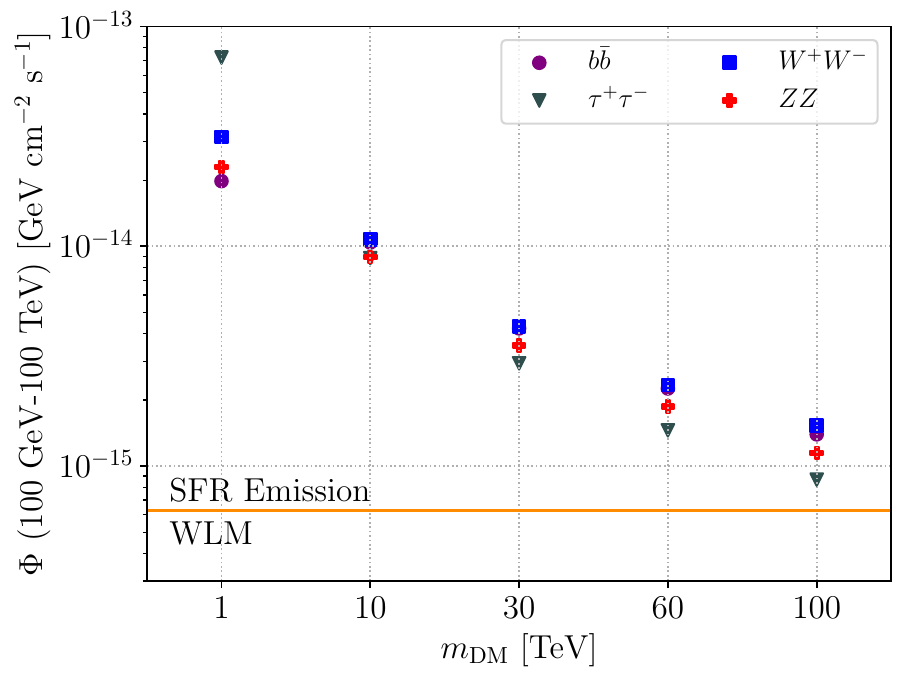}
  \includegraphics[width=0.49\textwidth]{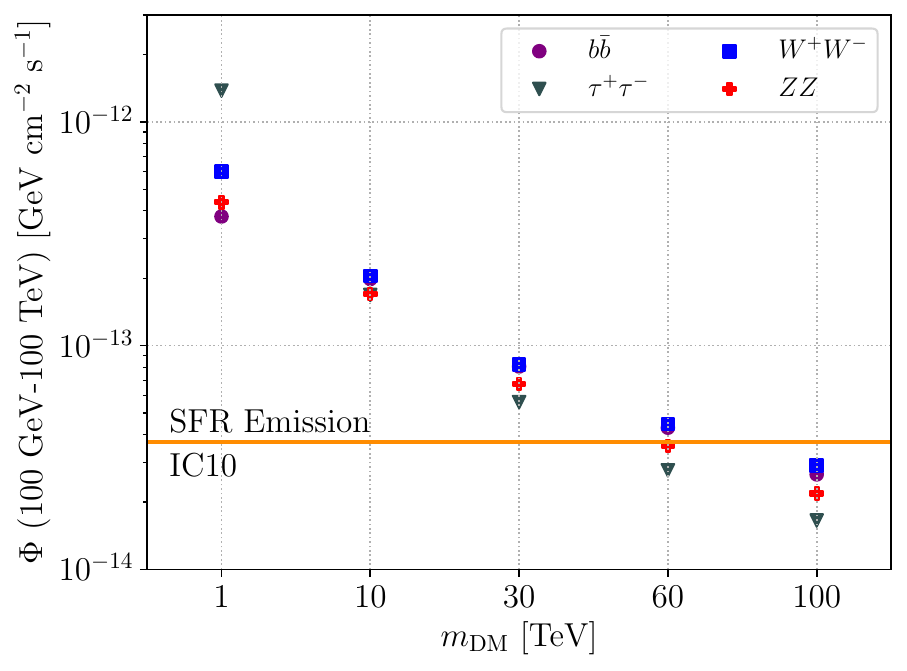}
    \caption{\footnotesize{Same as Figure~\ref{fig:dIrr_DMFlux_SFR} but for the WLM (left panel) and IC10 (right panel) galaxies.}}
\label{fig:dIrr_DMFlux_SFR_appendix} 
\end{figure}

\begin{figure}[h!] 
  \centering 
  \includegraphics[width=0.49\textwidth]{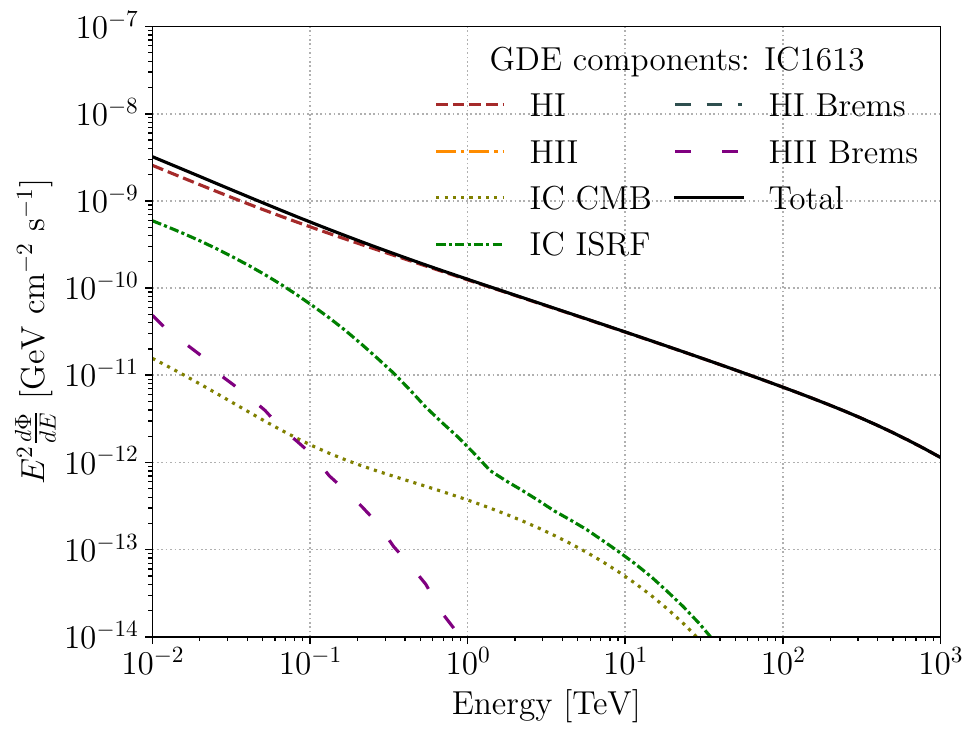}
  \includegraphics[width=0.49\textwidth]{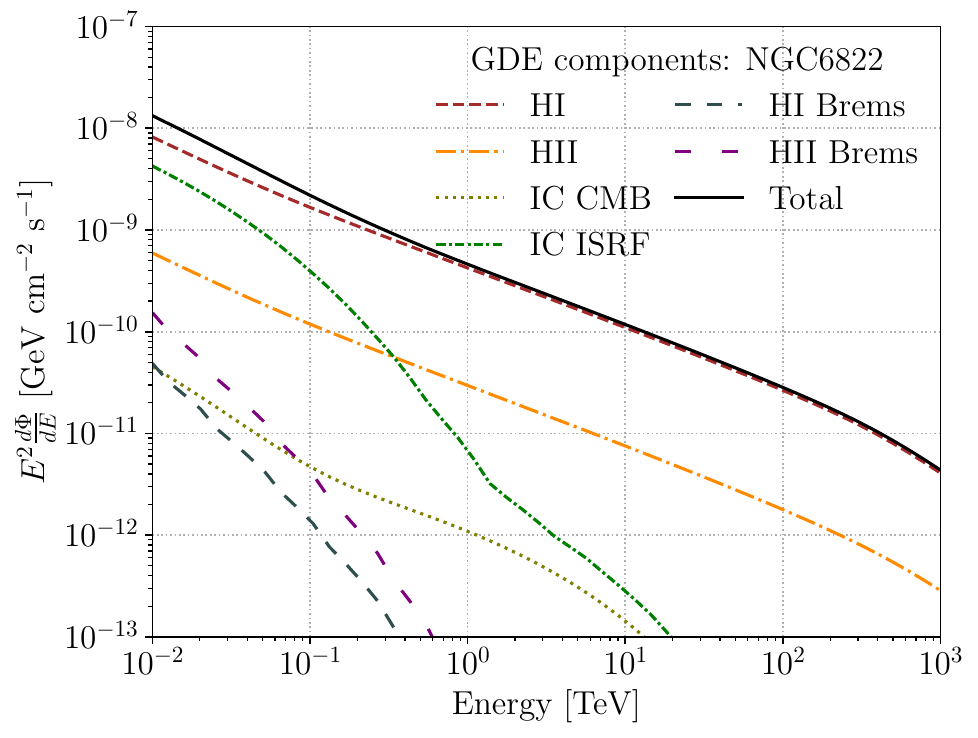}
    \caption{\footnotesize{Same as Figure~\ref{fig:GDE_components} but for the IC1613 (left panel) and NGC6822 (right panel) galaxies.}}
\label{fig:GDE_components_appendix} 
\end{figure}

\begin{figure}[t!] 
  \centering 
  \includegraphics[width=0.49\textwidth]{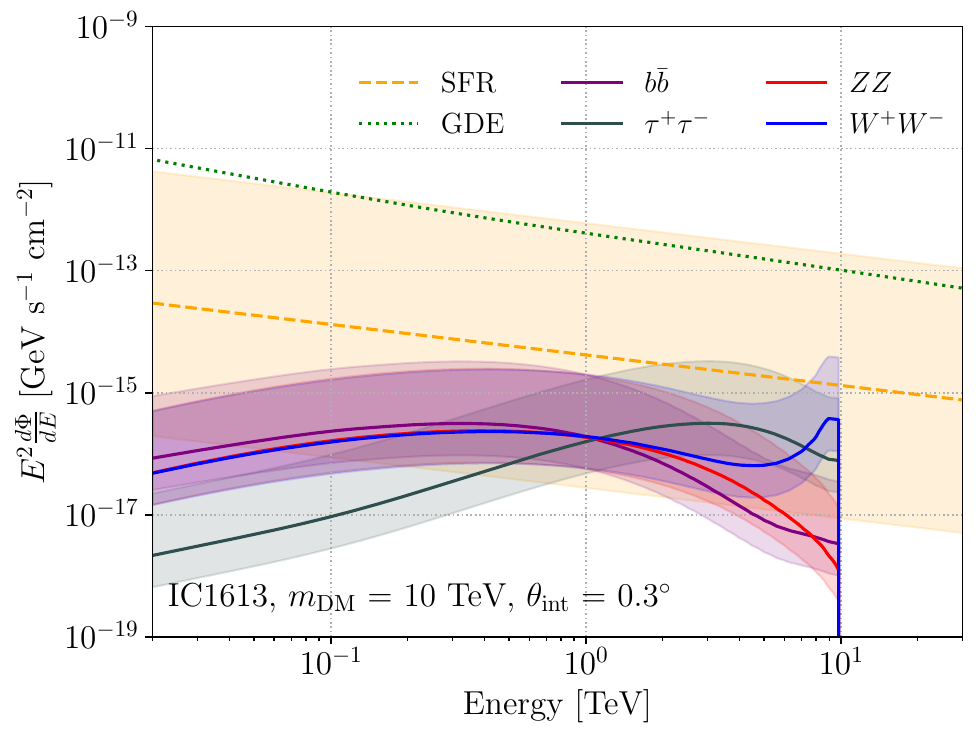}
  \includegraphics[width=0.49\textwidth]{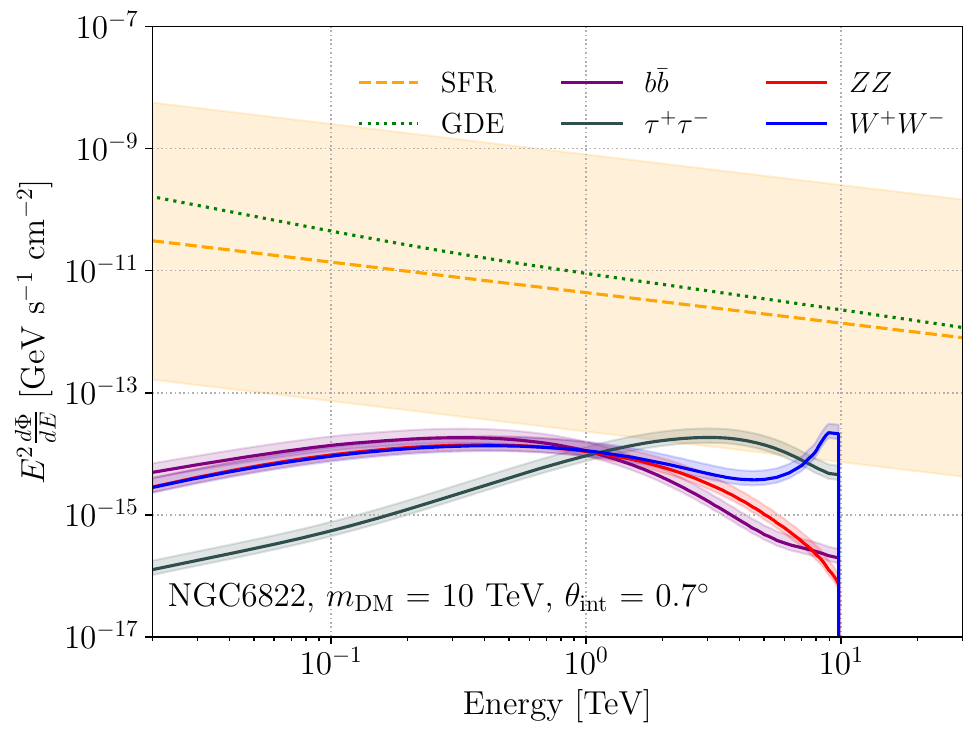}
    \caption{\footnotesize{Same as Figure~\ref{fig:dIrr_All_spectral_fluxes} but for the IC1613 (left panel) and NGC6822 (right panel) galaxies.}}
\label{fig:dIrr_All_spectral_fluxes_appendix} 
\end{figure}

\begin{figure}[t!] 
  \centering 
  \includegraphics[width=0.49\textwidth]{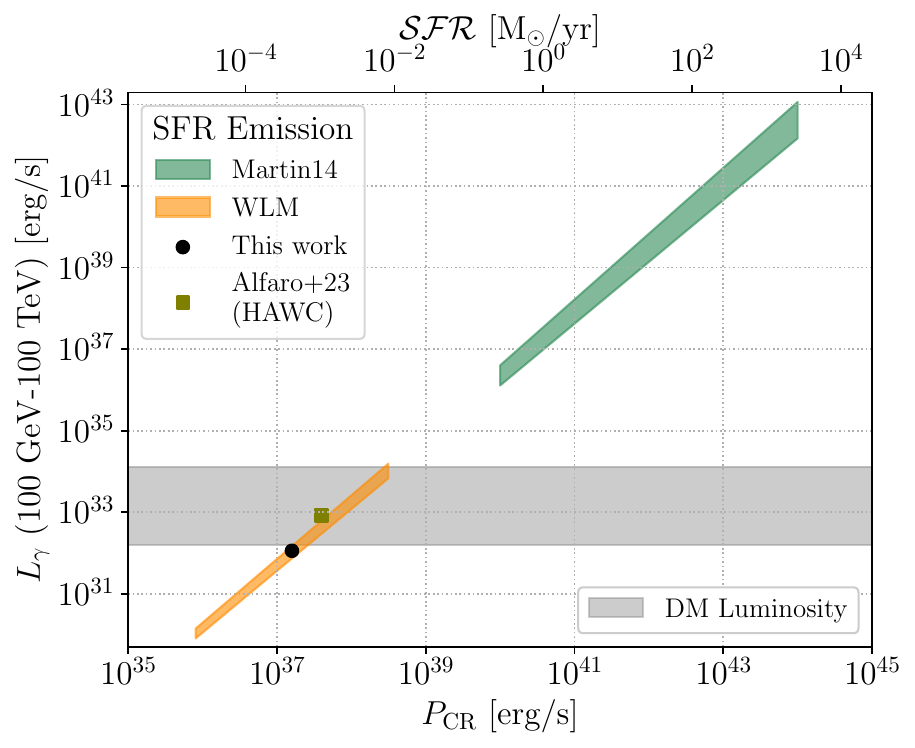}
  \includegraphics[width=0.49\textwidth]{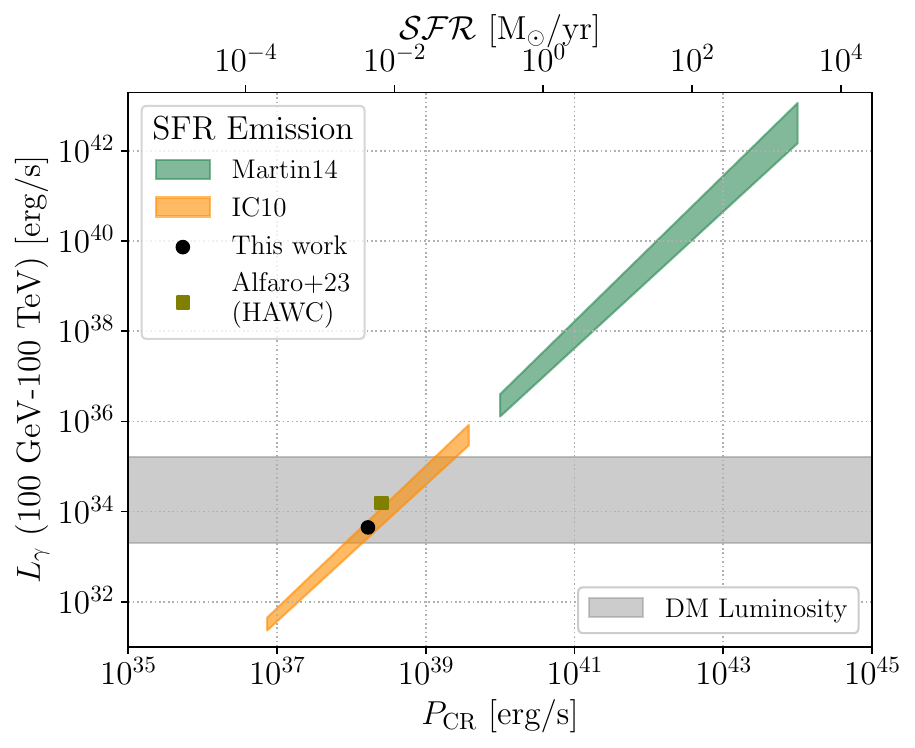}
    \caption{\footnotesize{Same as Figure~\ref{fig:SFR_vs_DM} but for the WLM (left panel) and IC10 (right panel) galaxies.}}
\label{fig:SFR_vs_DM_appendix} 
\end{figure}

In this Appendix, we show in Figures~\ref{fig:dIrr_DMFlux_SFR_appendix}, \ref{fig:GDE_components_appendix}, \ref{fig:dIrr_All_spectral_fluxes_appendix}, \ref{fig:SFR_vs_DM_appendix} the same as in Figures~\ref{fig:dIrr_DMFlux_SFR},  \ref{fig:GDE_components}, \ref{fig:dIrr_All_spectral_fluxes}, \ref{fig:SFR_vs_DM}, but for the other two galaxies not shown in the main text in each of the cases.

\bibliographystyle{JHEP}
\bibliography{bibliography}

\end{document}